\documentclass{aa}
\usepackage{graphicx}
\usepackage{natbib}
\usepackage{longtable,lscape}
\usepackage[figuresright]{rotating}
\bibpunct{(}{)}{;}{a}{}{,}
\usepackage{txfonts}
\begin{document}
   \title{Automated supervised classification of variable stars}
   \subtitle{I. Methodology\thanks{The documented classification software codes as well
   as the light curves and the set of classification parameters for the definition stars, are
   available electronically from the $\aap$ anonymous ftp site.}\fnmsep\thanks{Figures 5 to 12 are only available in
   the electronic form of the paper.} }

   \author{J. Debosscher
          \inst{1}
          \and
          L. M. Sarro\inst{3,7}
          \and
          C. Aerts\inst{1,2}
          \and
          J. Cuypers\inst{4}
          \and
          B. Vandenbussche\inst{1}
          \and
          R. Garrido\inst{5}
          \and
          E. Solano\inst{6,7} 
          }
         
   \institute{Instituut voor Sterrenkunde, KU Leuven, Celestijnenlaan 200B, 3001 Leuven, Belgium
         \and
            Department of Astrophysics, Radbout University Nijmegen, POBox 9010, 6500 GL Nijmegen, the Netherlands
          \and 
           Dpt. de Inteligencia Artificial , UNED, Juan del Rosal, 16, 28040 Madrid, Spain
          \and
          Royal Observatory of Belgium, Ringlaan 3, B-1180 Brussel, Belgium
           \and
           Instituto de Astrof{\'{i}}sica de Andaluc{\'{i}}a-CSIC, Apdo 3004, 18080 Granada, Spain
          \and 
           Laboratorio de Astrof{\'{i}}sica Espacial y F{\'{i}}sica Fundamental, INSA, Apartado de Correos 50727, 28080 Madrid, Spain
	   \and
	   Spanish Virtual Observatory, INTA, Apartado de Correos 50727, 28080 Madrid, Spain
}

 \date{}

 
  \abstract
{The fast classification of new variable stars is an important step
in making them available for further research. Selection of science targets from
large databases is much more efficient if they have been classified
first. Defining the classes in terms of physical parameters is also important
to get an unbiased statistical view on the variability
mechanisms and the borders of instability strips.}
{Our goal is twofold: provide an overview of the stellar variability
classes that are presently known, in terms of some relevant stellar
parameters; use the class descriptions obtained as the basis for an
  automated 
`supervised classification' of large databases. Such automated
classification will compare and assign new objects to a set of pre-defined
variability training classes.}
{For every variability class, a literature search was performed to find as many
well-known member stars as possible, or a considerable subset if too many were
present. Next, we searched on-line and private databases for their light curves
in the visible band and performed period analysis and harmonic fitting. The
derived light curve parameters are used to describe the classes and define the
training classifiers.}
{We compared the performance of different classifiers in terms of percentage of
correct identification, of confusion among classes and of computation time. We
describe how well the classes can be separated using the proposed set of
parameters and how future improvements can be made, based on new large 
databases such as the light curves to be assembled by the CoRoT and Kepler 
space missions.}
{The derived classifiers' performances are so good in terms of success rate
and computational speed that we will evaluate them in practice from the
application of our methodology to a large subset of variable stars in the OGLE
database and from comparison of the results 
with published OGLE variable star classifications based on human
  intervention. These results will be published
in a subsequent paper.}

   \maketitle
%
%
\section{Introduction}

The current rapid progress in astronomical instrumentation provides us with a
torrent of new data. For example, the large scale photometric monitoring of
stars with ground-based automated telescopes and space telescopes delivers us large numbers of high quality light curves. The HIPPARCOS space mission is an example of this and led to a large number of new variable stars discovered in
the huge set of light curves. In the near future, new space missions will
deliver even larger numbers of light curves of much higher quality (in terms of
sampling and photometric precision).  The CoRoT mission (Convection Rotation and
planetary Transits, launched on $27$ December $2006$) has two main scientific
goals: asteroseismology and the search for exoplanets using the transit
method. The latter purpose requires the photometric monitoring of a large number
of stars with high precision. As a consequence, this mission will produce
excellent time resolved light curves for up to $60000$ stars with a sampling
rate better than $10$ minutes during $5$ months. Even higher numbers of stars
($>100000$) will be measured for similar purposes and with comparable sampling
rate by NASA's Kepler mission (launch end $2008$, duration $4$ years). The ESA
Gaia mission (launch foreseen in $2011$) will map our Galaxy in three
dimensions. About one billion stars will be monitored for this purpose, with
about $80$ measurements over $5$ years for each star.

 Among these large samples, many new variable stars of known and unknown
 type will be present. Extracting them, and making their characteristics and
 data available to the scientific community within a reasonable timescale, will
 make these catalogues really useful. It is clear that automated methods have to
 be used here. Mining techniques for large databases are more and more frequently used in
 astronomy. Although we are far from reproducing capabilities of the human
 brain, a lot of work can be done efficiently using intelligent
 computer codes.

In this paper, we present automated supervised classification methods for
variable stars. Special attention is paid to computational speed and robustness,
with the intention to apply the methods to the huge datasets expected to come
from the CoRoT, Kepler and Gaia satellite missions. We tackle this problem with
two parallel strategies. In the first, we construct a Gaussian mixture
model. Here, the main goals are to optimize speed, simplicity and
interpretability of the model rather than optimizing the classifiers'
performance. In the second approach, a battery of state-of-the-art pattern
recognition techniques is applied to the same training set in order to select
the best performing algorithm by minimizing the misclassification rate. The
latter methods are more sophisticated and will be discussed in more detail in a
subsequent paper (Sarro et al., in preparation).

For a supervised classification scheme, we need to predefine the classes. Every
new object in a database to be classified will then be assigned to one of those
classes (called definition or training classes) with a certain probability. The
construction of the definition classes for stellar variability is, therefore, an
important part of this paper. Not only are these classes necessary for this type
of classification method, they also provide us with physical parameters
describing the different variability types. They allow us to attain a good view
on the separation and overlap of the classes in parameter space. For every
variability class, we derive descriptive parameters using the light curves of
their known member stars. We use exclusively 
light curve information for the basic
methodology we present here, 
because additional information is not always available and we want
to see how well the classes can be described (and separated) using only this
minimal amount of information. This way, the method is broadly applicable. It is
easy to adapt the methods when more information such as colors, radial
velocities, etc. is available.

The first part of this paper is devoted to the description of the stellar
variability classes and the parameter derivation. The classes are visualized in
parameter space. In the second part, a supervised classifier based on
multivariate statistics is presented in detail. We also summarize the results of
a detailed statistical study on Machine Learning methods such as Bayesian Neural
Networks. Our variability classes are used to train the classifiers and the
performance is discussed. In a subsequent paper, the methods will be applied to
a large selection of OGLE (Optical Gravitational Lensing Experiment) light
curves, while we plan to update the training classes from the CoRoT exoplanet
light curves in the coming two years.

\section{Description of stellar variability classes from photometric time 
series}
\label{attributes}

We provide an astrophysical description of the stellar variability classes by
means of a fixed set of parameters. These parameters are derived using the light curves of known
member stars. An extensive literature search provided us with the object identifiers of well-known class
members. We retrieved their available light curves from
different sources. The main sources are the HIPPARCOS space data
\citep{ESA:1997,Perryman:1997} and the Geneva and OGLE ground-based data \citep{Udalski:1999d,Wyrzykowski:2004,Soszynski:2002}. Other
sources include ULTRACAM data (ULTRA-fast, triple-beam CCD CAMera), see
\cite{Dhillon:2001}, MOST data (Microvariability and Oscillations of STars, see
\textit{http://www.astro.ubc.ca/MOST/}), WET data (Whole Earth Telescope, see
\textit{http://wet.physics.iastate.edu/}), ROTOR data \citep{ROTOR},
Lapoune/CFHT data (Fontaine, private communication), and ESO-LTPV data (European
Southern Observatory Long-Term Photometry of Variables project), see
\cite{Sterken:1995}. Table\,\ref{tabel1} lists the number of light curves used from each instrument, together with their average total time span and their average number of measurements. For every considered class (see Table\,\ref{tabel2}), we have tried to find the best available light curves, allowing recovery of the class' typical variability. Moreover, in order to be consistent in our description of the classes, we tried, as much as possible, to use light curves in the visible band (V-mag). This was not possible for all the classes however, due to a lack of light curves in the V-band, or an inadequate temporal sampling of the available V-band light curves. The temporal sampling (total time span and size of the time steps) is of primordial importance when seeking a reliable description of the variability present in the light curves.
While HIPPARCOS light curves, for example, are adequate in describing the long term variability of Mira stars, they do not allow recovery of the rapid photometric variations seen in some classes such as rapidly oscillating Ap stars. We used WET or ULTRACAM data in this case, dedicated to this type of object. For the double-mode Cepheids, the RR-Lyrae stars of type RRd and the eclipsing binary classes, we used OGLE light curves, since they both have an adequate total time span and a better sampling than the HIPPARCOS light curves.
\begin{table}
\caption{The sources and numbers of light curves $NLC$ used to define the classes, their average total time span $<T_{tot}>$ and their average number of measurements $<N_{points}>$}
\begin{tabular}{|c|c|c|c|}
\hline
Instrument&$NLC$&$<T_{tot}>$ (days)&$<N_{points}>$\\
\hline
HIPPARCOS&$1044$&$1097$&$103$\\
GENEVA&$118$&$3809$&$175$\\
OGLE&$527$&$1067$&$329$\\
ULTRACAM&$19$&$22$&$15820$\\
MOST&$3$&$34$&$59170$\\
WET&$3$&$5.6$&$11643$\\
ROTOR&$3$&$5066$&$881$\\
CFHT&$3$&$0.18$&$1520$\\
ESO-LTPV&$20$&$2198$&$209$\\

\hline
\end{tabular}

\label{tabel1}
\end{table}

For every definition class, mean parameter values and variances are
calculated. Every variability class thus corresponds to a region in a
multi-dimensional parameter space. We investigate how well the classes are
separated with our description and point out where additional information is
needed to make a clear distinction. Classes showing a large overlap will have a
high probability of resulting in misclassifications when using them in the training
set.

The classes considered are listed in Table\,\ref{tabel2}, together with the code
we assigned to them and the number of light curves we used to define the class. We
use this coding further in this paper, and in the reference list, to indicate
which reference relates to which variability type. For completeness, we also
list the ranges for $T_{\rm eff}$, $\log g$, and the range for the dominant frequencies and their amplitudes present in the light curves. The first two physical parameters cannot be measured directly, but are calculated from modeling. We do not use
these parameters for classification purposes here because they are in general
not available for newly measured stars. Also, for some classes, such as those
with non-periodic variability or outbursts, it is not possible to define a
reliable range for these parameters. The ranges for the light curve parameters result from our analysis, as described in Sect. $2.1$.

\begin{figure}
   \centering
   \includegraphics[angle=360,scale=0.36]{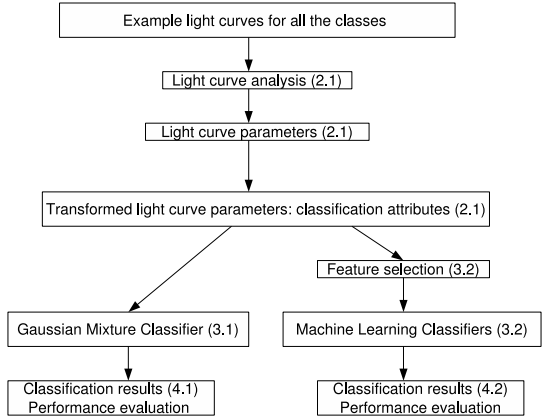}
   \caption{Schematic overview of the different steps (sections indicated) in the development and comparison of the classification methods presented in this paper.}
   \label{fig1}
\end{figure}

We stress that the classes considered in Table\,\ref{tabel2} constitute the vast
majority of known stellar variability classes, but certainly not all of them. In
particular, we considered only those classes whose members show clear and
well-understood visual photometric variability. Several additional classes exist
which were defined dominantly on the basis of spectroscopic diagnostics or
photometry at wavelengths outside the visible range. For some classes, we were
unable to find good consistent light curves.  Examples of omitted classes are
hydrogen-deficient carbon stars, extreme helium stars, $\gamma$ or $X$-ray
bursts, pulsars, etc. Given that we do not use diagnostics besides light curves at or around visible wavelengths in our methods presently, these classes are not considered
here. In the following we describe our methods in detail. A summary of the different steps is shown in Fig. \ref{fig1}

\begin{table*}
\caption{Stellar variability classes considered in this study, their code, the
number of light curves we used ($NLC$) and their source. Also listed (when relevant for
the class) are the ranges for the parameters $T_{\rm eff}$ and $\log g$ if they
could be determined from the literature. The last two columns list the range for the dominant frequencies ($f_1$) and their amplitudes ($A_{11}$) present in the light curves, resulting from our analysis (Sect. $2.1$).}  \centering
\scriptsize
\begin{tabular}{|c|c|c|c|c|c|c|}
  \hline
  Class&NLC& Instrument&Range $T_{\rm eff}$&Range $\log g$& Range $f_1$ (c/d)&Range $A_{11}$ (mag)\\
  \hline
  Periodically variable supergiants (PVSG)&76&HIPPARCOS/GENEVA/ESO&$3890-56234$K&$1.0-4.5$&$0.0004-14.1668$&$0.0027-0.4689$\\
  Pulsating Be-stars (BE)&57&HIPPARCOS/GENEVA&$17100-23850$K&$3.30-4.33$&$0.0003-14.0196$&$0.0023-2.9385$\\
  $\beta$-Cephei stars (BCEP) & 58&HIPPARCOS/GENEVA&$18238-36813$K&$3.18-4.30$&$0.0180-11.3618$&$0.0030-0.1344$\\
  Classical Cepheids (CLCEP)&195&HIPPARCOS/GENEVA&$4800-6648$K&$1.45-2.6$&$0.0222-0.4954$&$0.0493-1.1895$\\
  Beat (double-mode)-Cepheids (DMCEP)&95&OGLE&$5000-7000$K&$2-3.5$&$0.5836-1.7756$&$0.0544-0.1878$\\
  Population II Cepheids (PTCEP)&24&HIPPARCOS&$5200-6550$K&&$0.0038-1.5377$&$0.1561-0.6364$\\
  Chemically peculiar stars (CP)&63&HIPPARCOS/GENEVA&$6500-18900$K&$3.2-4.6$&$0.0076-33.4158$&$0.0027-0.0604$\\
  $\delta$-Scuti stars (DSCUT) & 139&HIPPARCOS/GENEVA&$6550-9126$K&$3.5-4.25$&$0.0109-26.9967$&$0.0043-0.3841$\\
  $\lambda$-Bootis stars (LBOO)&13&HIPPARCOS&$6637-9290$K&$3.4-4.1$&$7.0865-14.5035$&$0.0036-0.0143$\\ 
  SX-Phe stars (SXPHE)&7&HIPPARCOS/GENEVA&$6940-8690$K&$3.34-4.3$&$6.2281-16.2625$&$0.0138-0.3373$\\
  $\gamma$-Doradus stars (GDOR) & 35&HIPPARCOS/GENEVA&$5980-7375$K&$3.32-4.58$&$0.3803-13.7933$&$0.0048-0.0325$\\
  Luminous Blue Variables (LBV)&21&HIPPARCOS/GENEVA/ESO&$8000-30000$K&&$0.0004-2.0036$&$0.0296-0.9877$\\
  Mira stars (MIRA)& 144&HIPPARCOS&$2500-3500$K&&$0.0020-0.6630$&$0.2828-3.9132$\\
  Semi-Regular stars (SR) &42&HIPPARCOS&$2500-3500$K&&$0.0012-11.2496$&$0.0216-1.9163$\\
  RR-Lyrae, type RRab (RRAB) & 129&HIPPARCOS/GENEVA&$6100-7400$K&$2.5-3.0$&$1.2150-9.6197$&$0.0745-0.5507$\\
  RR-Lyrae, type RRc (RRC)&29&HIPPARCOS&&&$2.2289-4.5177$&$0.0313-0.2983$\\
  RR-Lyrae, type RRd (RRD)&57&OGLE&&&$2.0397-2.8177$&$0.0899-0.2173$\\
  RV-Tauri stars (RVTAU) & 13&HIPPARCOS/GENEVA&$4250-7300$K&$-0.9-2.0$&$0.0011-1.0280$&$0.2851-2.3831$\\
  Slowly-pulsating B stars (SPB) & 47&HIPPARCOS/GENEVA/MOST&$12000-18450$K&$3.8-4.4$&$0.1394-3.7625$&$0.0036-0.0982$\\
  Solar-like oscillations in red giants (SLR) &1&MOST&&&$0.0352$&$0.0014$\\ 
  Pulsating subdwarf B stars (SDBV)&16&ULTRACAM&$23000-32000$K&$4.5-5.6$&$242.5726-612.7225$&$0.0038-0.0739$\\
  Pulsating DA white dwarfs (DAV)&2&WET&$10350-11850$K&$7.73-8.74$&$149.2038-401.5197$&$0.0020-0.0226$\\
  Pulsating DB white dwarfs (DBV)&1&WET/CFHT&$11000-30000$K&$\sim8$&$150.5844$&$0.0401$\\
  GW-Virginis stars (GWVIR)&2&CFHT&$70000-170000$K&&$192.9965-215.3986$&$0.0141-0.0216$\\
  Rapidly oscillating Ap stars (ROAP)&4&WET/ESO&$6800-8400$K&$3.77-4.52$&$123.0299-235.0878$&$0.0013-0.0022$\\
  T-Tauri stars (TTAU)&17&HIPPARCOS/GENEVA&$3660-4920$K&$3.8-4.5$&$0.0009-11.0231$&$0.0092-0.8925$\\
  Herbig-Ae/Be stars (HAEBE)&21&HIPPARCOS/GENEVA&$5900-16000$K&$3.5-5$&$0.0009-10.9516$&$0.0053-0.8925$\\
  FU-Ori stars (FUORI)&3&ROTOR&$13000-15000$K&&$0.0002-0.0006$&$0.0432-0.2181$\\
  Wolf-Rayet stars (WR)&63&HIPPARCOS/GENEVA/ESO/MOST&$14800-91000$K&&$0.0003-15.9092$&$0.0016-0.3546$\\
  X-Ray binaries (XB)&9&HIPPARCOS/GENEVA&&&$0.0057-11.2272$&$0.0063-0.0813$\\
  Cataclysmic variables (CV)&3&ULTRACAM&&&$27.5243-36.9521$&$0.1838-0.5540$\\
  Eclipsing binary, type EA (EA) &169&OGLE&&&$0.0127-3.1006$&$0.0371-0.2621$\\
  Eclipsing binary, type EB (EB)& 147&OGLE&&&$0.0175-4.5895$&$0.0454-0.7074$\\
  Eclipsing binary, type EW (EW)& 59&OGLE&&&$0.2232-8.3018$&$0.0376-0.4002$\\
  Ellipsoidal binaries (ELL)& 16&HIPPARCOS/GENEVA&&&$0.1071-3.5003$&$0.0136-0.0629$\\
  \hline
\end{tabular}

 \label{tabel2}
\end{table*}

\subsection{Light curve analysis and parameter selection}

After removal of bad quality measurements, the photometric time series of the
definition stars were subjected to analysis. First, we checked for possible
linear trends of the form $a+bT$, with $a$ the intercept, $b$ the slope and $T$
the time. These were subtracted, as they can have a large influence on the
frequency spectrum. The larger the trend is for pulsating stars, the more the
frequency values we find can deviate from the stars' real pulsation frequencies.

Subsequently, we performed a Fourier analysis to find periodicities in the light
curves. We used the well-known Lomb-Scargle method
\citep{Lomb:1976,Scargle:1982}. The computer code to calculate the periodograms
was based on an algorithm written by J. Cuypers. It followed outlines given by \cite{Ponman:1981} and \cite{Kurtz:1985} focussed on speedy
calculations. As is the case with all frequency determination methods, we
needed to specify a search range for frequencies ($f_0$, $f_N$ and $\Delta
f$). Since we were dealing with data coming from different instruments, it was
inappropriate to use the same search range for all the light curves. We adapted
it to each light curve's sampling, and took the starting frequency as
$f_0=1/T_{tot}$, with $T_{tot}$ the total time span of the observations. A
frequency step $\Delta f=0.1/T_{tot}$ was taken. For the highest frequency, we
used the average of the inverse time intervals between the measurements:
$f_N=0.5<1/\Delta T>$ as a pseudo Nyquist frequency. Note that $f_N$ is equal to
the Nyquist frequency in the case of equidistant sampling. For particular
cases, an even higher upper limit can be used \citep[see][]{Eyer:1999}. Our
upper limit should be seen as a compromise between the required resolution to
allow a good fitting, and computation time.

We searched for up to a maximum of three independent frequencies for every
star. The procedure was as follows: the Lomb-Scargle periodogram was calculated
and the highest peak was selected. The corresponding frequency value $f_1$ was
then used to calculate a harmonic fit to the light curve of the form:
\begin{equation}
y(t)=\sum_{j=1}^{4} (a_j \sin{2\pi f_1 j t}+b_j \cos{2\pi f_1 j t})+b_0, 
\end{equation}
with $y(t)$ the magnitude as a function of time. Next, this curve was subtracted
from the time series (prewhitening) and a new Lomb-Scargle periodogram was
computed. The same procedure was repeated until three frequencies were found. Finally, the three frequencies were used to make a harmonic best-fit to the
original (trend subtracted) time series:
\begin{equation}
y(t)=\sum_{i=1}^3\sum_{j=1}^{4} (a_{ij} \sin{2\pi f_i j t}+b_{ij} 
\cos{2\pi f_i j t})+b_0. 
\end{equation}
The parameter $b_0$ is the mean magnitude value of the light curve. The
frequency values $f_i$ and the Fourier coefficients $a_{ij}$ and $b_{ij}$
provide us with an overall good description of light curves, if the latter are
periodic and do not show large outbursts. It is important to note, in the
context of classification, that the set of Fourier coefficients obtained here is
not unique: identical light curves can have different coefficients, just because
the zero-point of their measurements is different. The Fourier coefficients are
thus not invariant under time-translation of the light curve. Since we want to
classify light curves, this is inconvenient. We ideally want all light curves,
identical apart from a time-translation, to have the same set of parameters (called attributes when used for classifying). On the other hand, we want different parameter sets to correspond to different
light curves as much as possible. To obtain this, one can first transform the
Fourier coefficient into a set of amplitudes $A_{ij}$ and phases $PH_{ij}$ as
follows:
\begin{eqnarray}
A_{ij}=\sqrt{a_{ij}^2+b_{ij}^2},\\
PH_{ij}=\arctan(b_{ij},a_{ij}),
\end{eqnarray}
with the arctangent function returning phase angles 
in the interval ]$-\pi$,$+\pi$].
This provides us with a completely equivalent description of the light curve:
\begin{equation}
y(t)=\sum_{i=1}^3\sum_{j=1}^{4} A_{ij}\sin(2\pi f_i j t+PH_{ij})+b_0.
\end{equation}
The positive amplitudes $A_{ij}$ are already time-translation invariant, but the
phases $PH_{ij}$ are not. This invariance can be obtained for the phases as
well, by putting $PH_{11}$ equal to zero and changing the other phases
accordingly (equivalent to a suitable time-translation, depending on the
zero-point of the light curve). Although arbitrary, it is preferable to choose
$PH_{11}$ as the reference, since this is the phase of the most significant
component in the light curve. The new phases now become:
\begin{eqnarray}
PH'_{ij}=\arctan(b_{ij},a_{ij})-(\frac{jf_i}{f_1})\arctan(b_{11},a_{11}),
\end{eqnarray}
with $PH'_{11}=0$. The factor $(jf_i/f_1)$ in this expression is the ratio of
the frequency of the $j$-th harmonic of $f_i$ to the frequency $f_1$, because
the first harmonic of $f_1$ has been chosen as the reference.  Note that these
new phases can have values between $-\infty$ and $+\infty$. We can now constrain
the values to the interval $]-\pi,+\pi]$, since all phases differing with
an integer multiple of $2\pi$ are equivalent. This can be done using the same
arctangent function:
\begin{eqnarray}
PH''_{ij}=\arctan(\sin(PH'_{ij}),\cos(PH'_{ij})).
\end{eqnarray}    
The parameters $A_{ij}$ and $PH''_{ij}$ now provide us with a time-translation
invariant description of the light curves and are suitable for classification
purposes. Note that this translation invariance strictly only holds for
monoperiodic light curves, and is not valid for multiperiodic light
curves. Alternate transformations are being investigated to extend the
translation invariance to multiperiodic light curves as well. For ease of
notation, we drop the apostrophes when referring to the phases
$PH''_{ij}$.

Another important parameter, which is also calculated during the fitting
procedure, is the ratio of the variances $v_{f1}/v$ in the light curve, after
and before subtraction of a harmonic fit with only the frequency $f_1$. This
parameter is very useful for discriminating between multi- and monoperiodic
pulsators. Its value is much smaller for monoperiodic pulsators, where most of
the variance in the light curve can be explained with a harmonic fit with only
$f_1$.

In total, we calculated $28$ parameters starting from the original time series:
the slope $b$ of the linear trend, $3$ frequencies, $12$ amplitudes, $11$ phases
($PH_{11}$ is always zero and can be dropped) and $1$ variance ratio. This way,
the original time series, which can vary in length and number of measurements,
were transformed into an equal number of descriptive parameters for every star.

We calculated the same parameter set for each star, irrespective of the
variability class they belong to. This set provided us with an overall good
description of the light curves for pulsating stars, and even did well for
eclipsing binaries. It is clear, however, that the whole parameter set might not
be needed for distinguishing, say, between class A and class B. The distinction
between a Classical Cepheid and a Mira star is easily made with only the
parameters $f_1$ and $A_{11}$, other parameters are thus not necessary and might
even be completely irrelevant for this example. For other classes, we have to
use more parameters to reach a clear distinction. 

With these $28$ selected parameters, we found a good compromise between maximum
separability of all the classes and a minimum number of descriptive
parameters. Our class definitions are based on the entire parameter set
described above. A more detailed study on statistical attribute selection methods is
presented in Sect. 3.2.1, as this is closely related to the performance of a
classifier.

\subsection{Stellar variability classes in parameter space}

\begin{figure*}
   \centering
   \includegraphics[width=17cm,angle=180,scale=0.94]{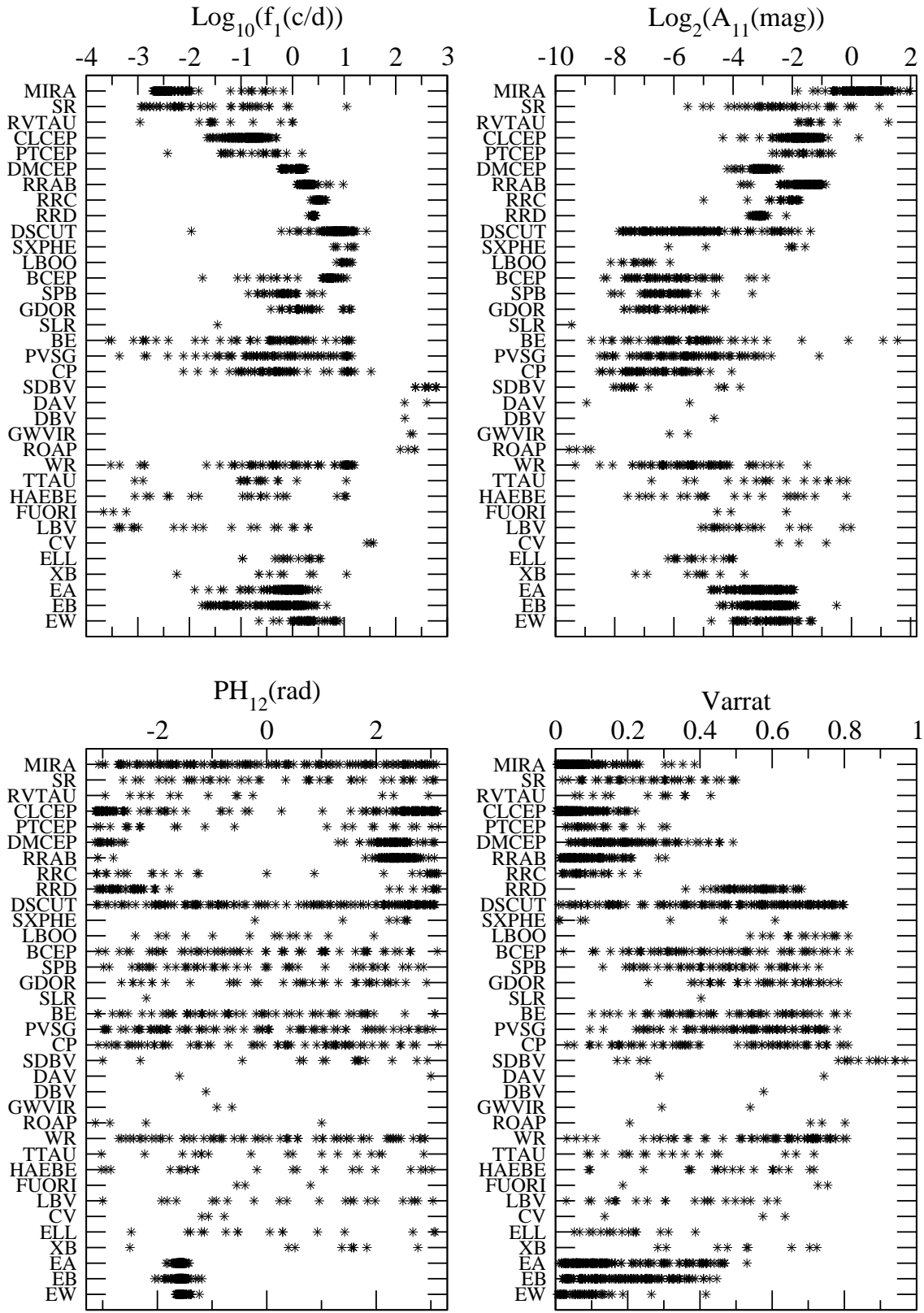}
   \caption{The range for the frequency $f_1$ (in cycles/day), its first
   harmonic amplitude $A_{11}$ (in magnitude), the phases $PH_{12}$ (in radians)
   and the variance ratio $v_{f1}/v$ (varrat) for all the 35 considered
   variability classes listed in Table\,\ref{tabel1}. For visibility
   reasons, we have plotted the logarithm of the frequency and amplitude
   values. Every symbol in the plots corresponds to the parameter value of
   exactly one light curve. In this way, we attempt to visualize the
   distribution of the light curve parameters, in addition to their mere range.}
   \label{fig2}
\end{figure*}

The different variability classes can now be represented as sets of points in
multi-dimensional parameter space. Each point in every set corresponds to the
light curve parameters of one of the class' member stars. The more the clouds
are separated from each other, the better the classes are defined, and the fewer
the misclassifications which will occur in the case of a supervised classification, using
these class definitions. As an external check for the quality of our class
definitions, we performed a visual inspection of phase plots made with only
$f_1$, for the complete set. If these were of dubious quality (or the wrong variability type), the objects were deleted from the class
definition. It turned out to be very important to retain only definition stars
with high-quality light curves. This quality is much more important than the
number of stars to define the class, provided that enough stars are available for a good sampling of the class' typical parameter ranges.
Visualizing the classes in multi-dimensional space is difficult. Therefore we
plot only one parameter at a time for every class. Figures \ref{fig2},
\ref{fig3}, \ref{fig4}-\ref{fig8} show the spread of the derived light
curve parameters for all the classes considered. Because the range can be quite
large for frequencies and amplitudes, we have plotted the logarithm of the
values (base $10$ for the frequencies and base $2$ for the amplitudes). As can
be seen from Fig. \ref{fig2}, using only $f_1$ and $A_{11}$, we already attain a
good distinction between monoperiodically pulsating stars such as Miras,
RR-Lyrae and Cepheids. For the multiperiodic pulsators, a lot of overlap is
present and more parameters ($f_2$, $f_3$, the $A_{2j}$ and the $A_{3j}$) are
needed to distinguish between those classes. If we look at the frequencies and
amplitudes, we see that clustering is less apparent for the non-periodic
variables such as Wolf-Rayet stars, T-Tauri stars and Herbig Ae/Be stars. For
some of those classes, we only have a small number of light curves, i.e. we do
not have a good `sampling' of the distribution (selection effect). The main
reason for their broad distribution is, however, the frequency spectrum: for the
non-periodic variables, the periodogram will show a lot of peaks over a large
frequency range, and selecting three of them is not adequate for
describing the light curve. Selecting more than three, however, entails the danger
of picking non-significant peaks. The phase values $PH_{1i}$ corresponding to
the harmonics of $f_1$ cluster especially well for the eclipsing binary classes,
as can be expected from the nature of their light curves. These parameters are
valuable for separating eclipsing binaries from other variables. The phase
values for the harmonics of $f_2$ and $f_3$ do not show significant clustering
structure. On the contrary, they tend to be rather uniformly distributed for
every class and thus, they will likely not constitute very informative
attributes for classification. This is not surprising, since these phases belong
to less significant signal components and will vary more randomly for the
majority of the stars in our training set. In the next section, we discuss
more precise methods for assessing the separation and overlap of the classes in
parameter space.

\begin{figure*}
   \centering
   \includegraphics[angle=-90,scale=0.6]{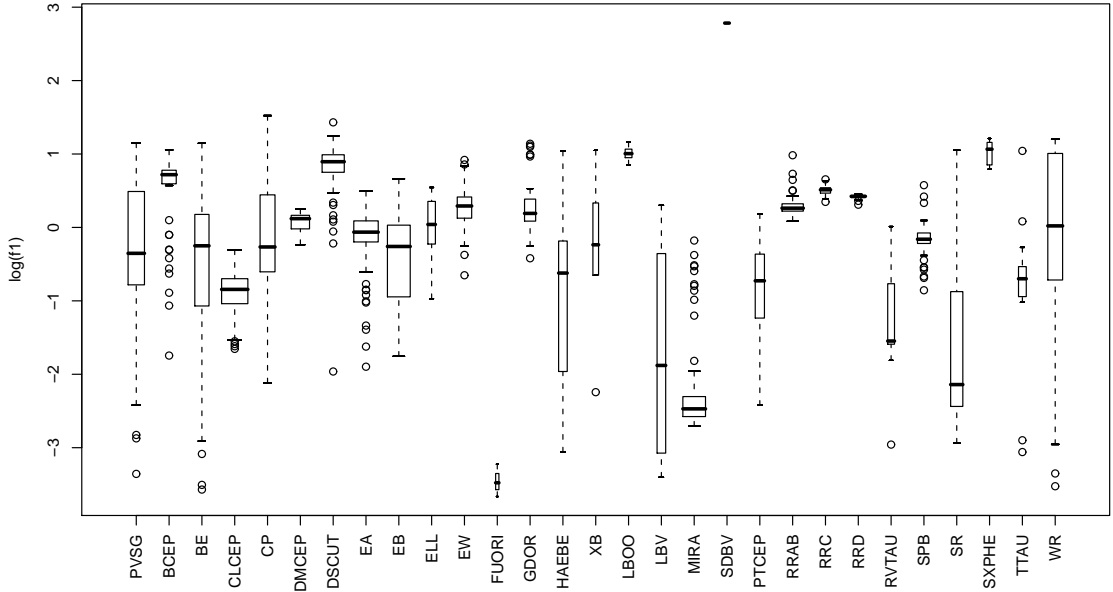}
   \caption{Box-and-whiskers plot of the logarithm of $f_1$ for 29 classes with
   sufficient members to define such tools in the training set. Central boxes
   represent the median and interquantile ranges (25 to 75\%) and the outer
   whiskers represent rule-of-thumb boundaries for the definition of outliers
   (1.5 the quartile range). The box widths are proportional to the number of
   examples in the class.}
   \label{baw1}
\end{figure*}

\begin{figure*}
   \centering
   \includegraphics[angle=-90,scale=0.5]{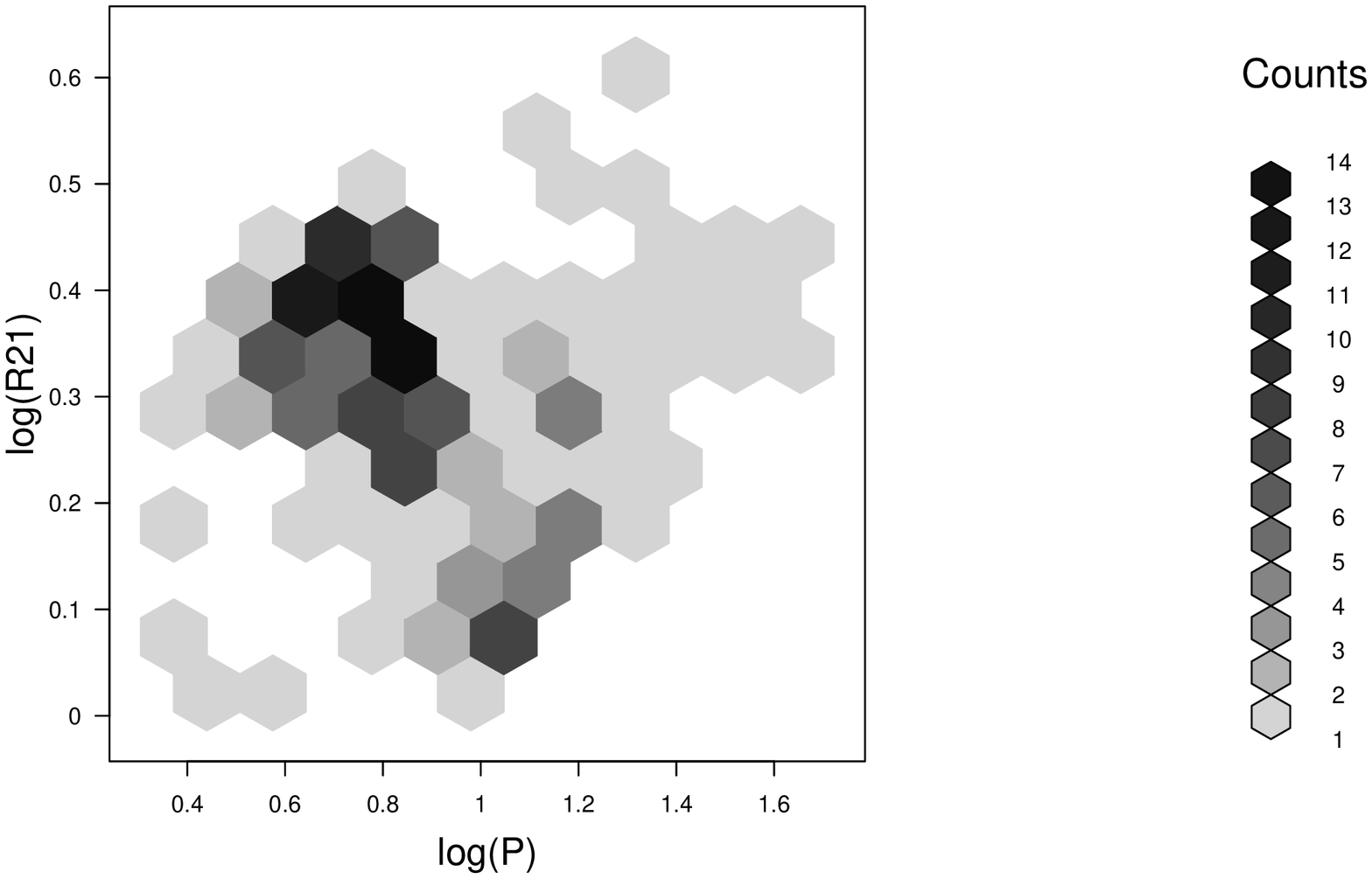}
   \caption{Hexagonal representation of the two dimensional density of examples
   of the Classical Cepheids class in the $\log(P)$-$\log(R_{21})$ space. The
   quantity $R_{21}\equiv A_{12}/A_{11}$ represents the ratio of the second to
   the first harmonic amplitude of $f_1$. This plot is comparable to Fig.\,3
   of \citet{Udalski:1999b}.}
   \label{2d-densities}
\end{figure*}

\begin{figure*}
   \centering
   \includegraphics[width=17cm,angle=180,scale=0.94]{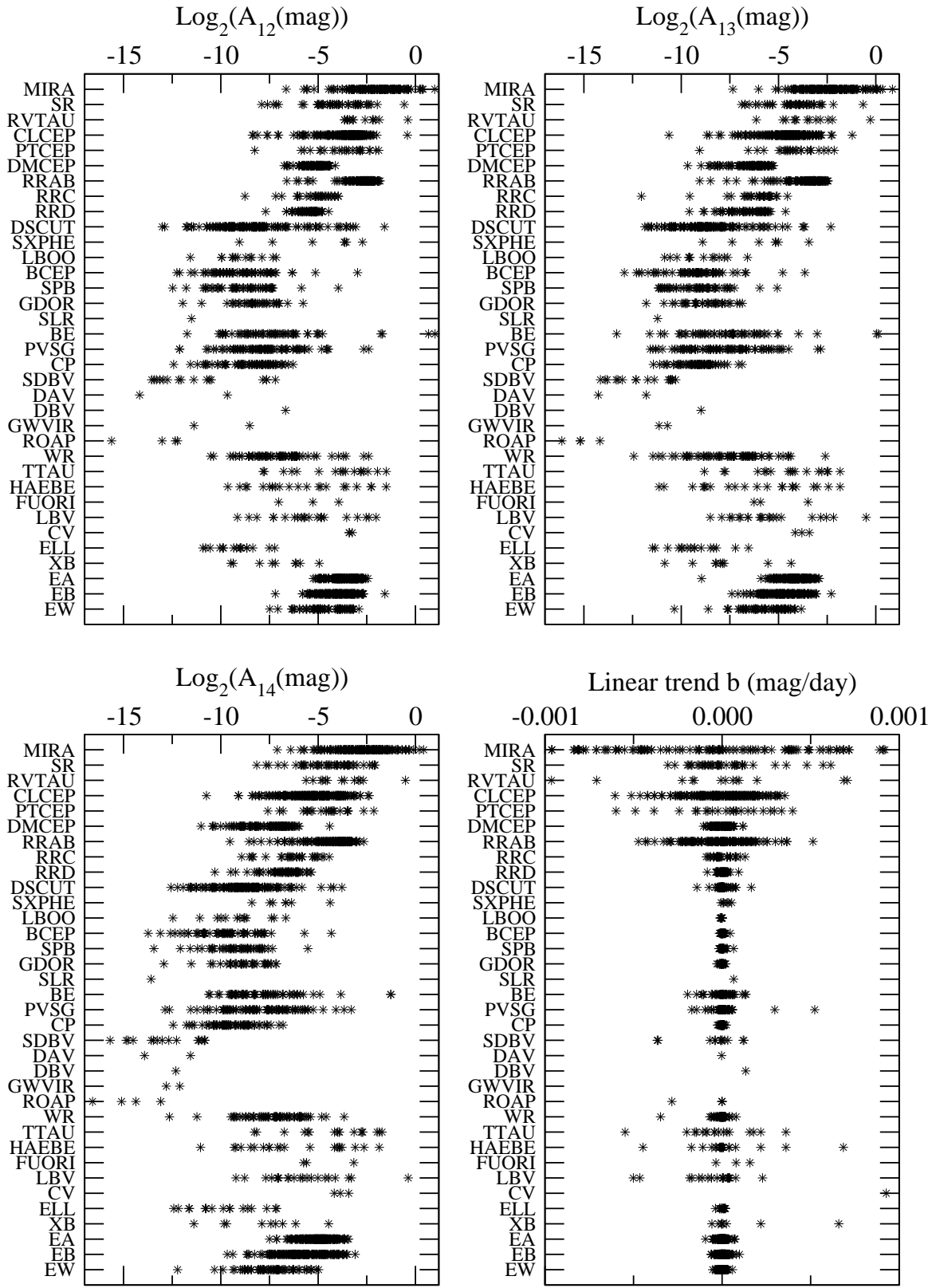}
   \caption{The range in amplitudes $A_{1j}$ for the $3$ higher harmonics of $f_1$, and the linear trend $b$. For visibility reasons, we have plotted the logarithm of the amplitude values.}
   \label{fig3}
\end{figure*}

\begin{figure*}
   \centering
   \includegraphics[width=17cm,angle=180,scale=0.94]{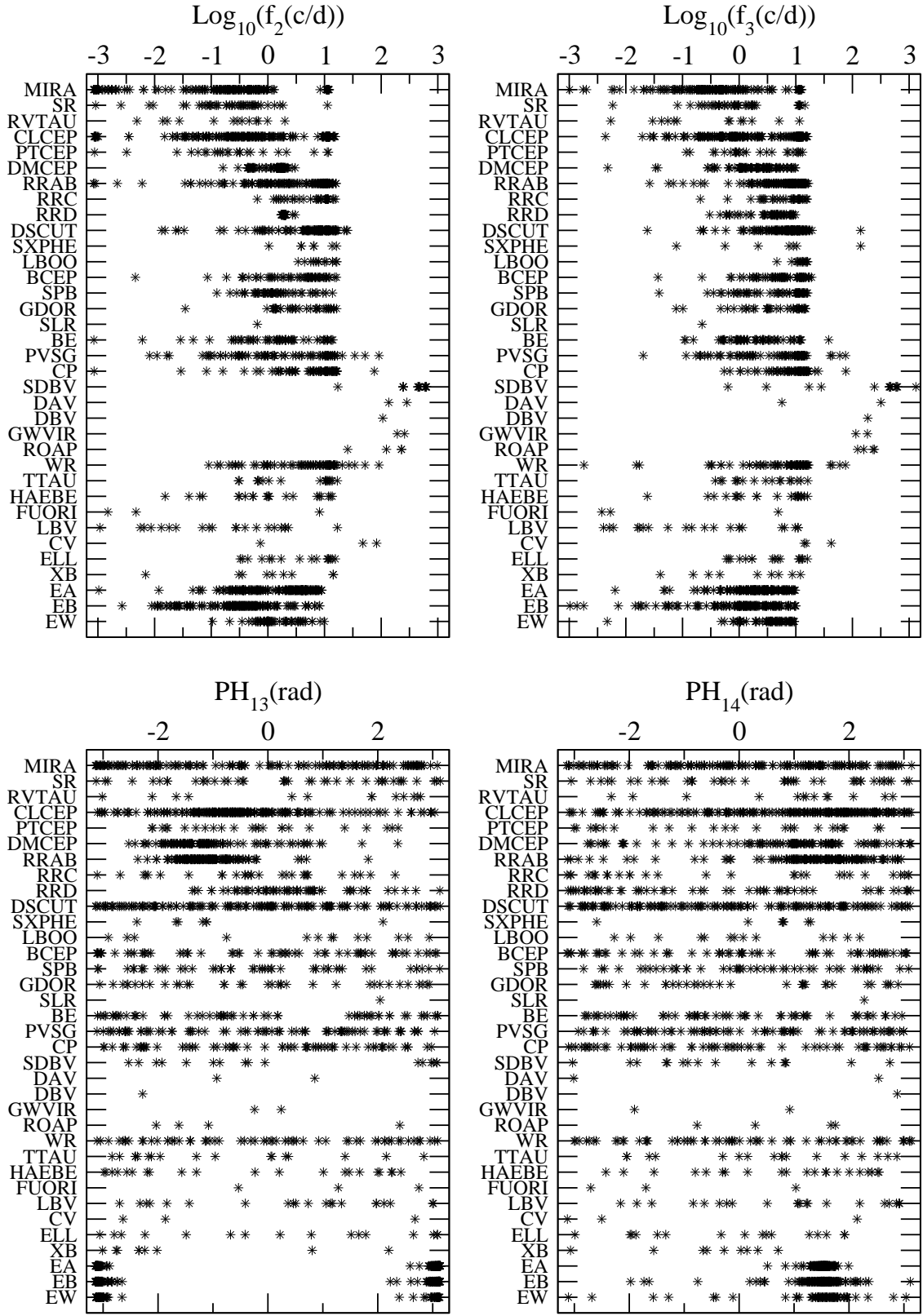}
   \caption{The range for the frequencies  $f_2$ and $f_3$ and the phases $PH_{1j}$ of the higher harmonics of $f_1$. For visibility reasons, we have plotted the logarithm of the frequency values. Note the split into two clouds of the phase values $PH_{13}$ for the eclipsing binary classes. This is a computational artefact: phase values close to $-\pi$ are equivalent to values close to $+\pi$, so the clouds actually represent a single cloud.}
   \label{fig4}
\end{figure*}

\begin{figure*}
   \centering
   \includegraphics[width=17cm,angle=180,scale=0.94]{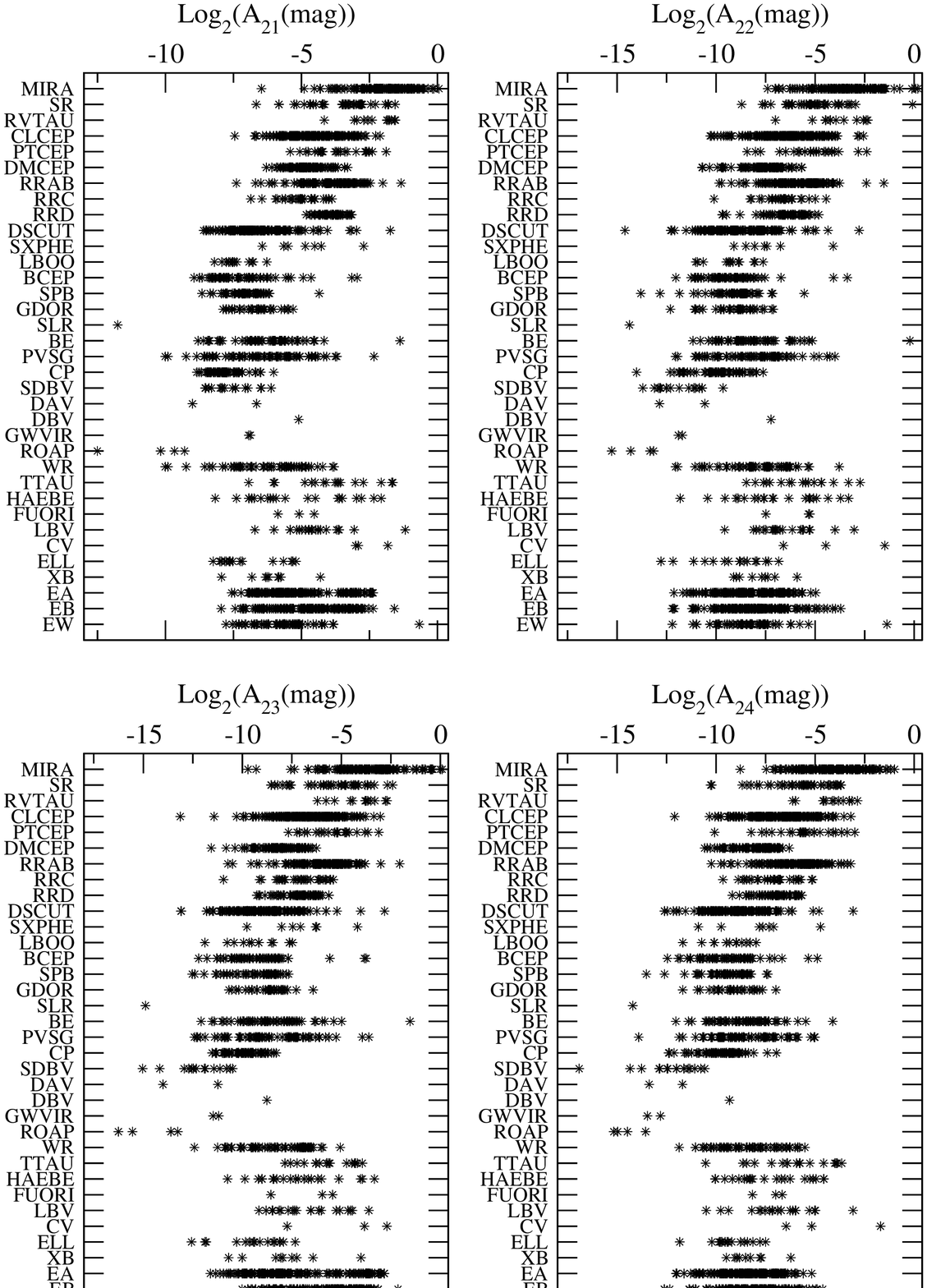}
   \caption{The range in amplitudes $A_{2j}$ for the $4$ harmonics of $f_2$. For visibility reasons, we have plotted the logarithm of the amplitude values.}
   \label{fig5}
\end{figure*}

\begin{figure*}
   \centering
   \includegraphics[width=17cm,angle=180,scale=0.94]{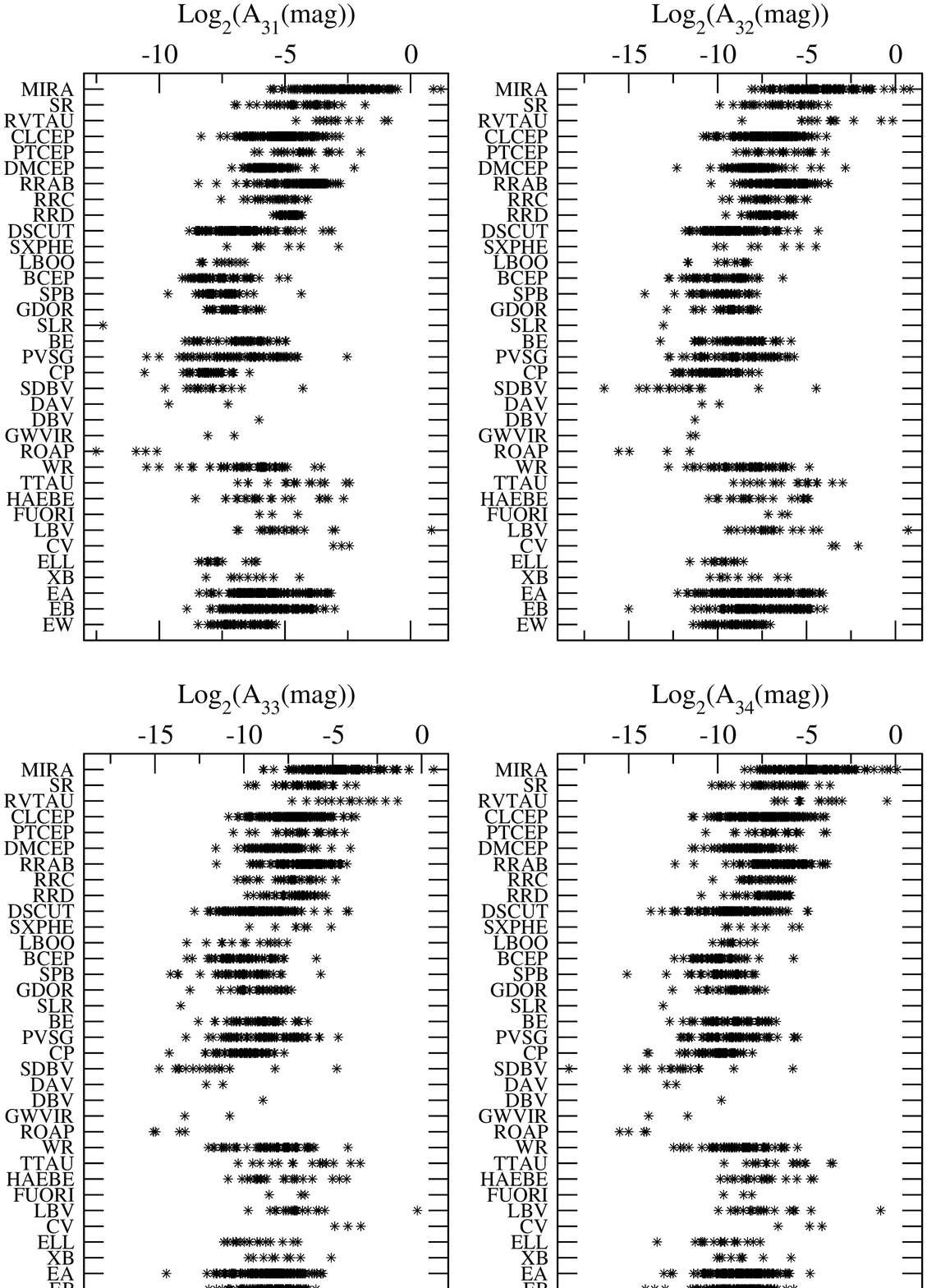}
   \caption{The range in amplitudes $A_{3j}$ for the $4$ harmonics of $f_3$. For visibility reasons, we have plotted the logarithm of the amplitude values.}
   \label{fig6}
\end{figure*}

\begin{figure*}
   \centering
   \includegraphics[width=17cm,angle=180,scale=0.94]{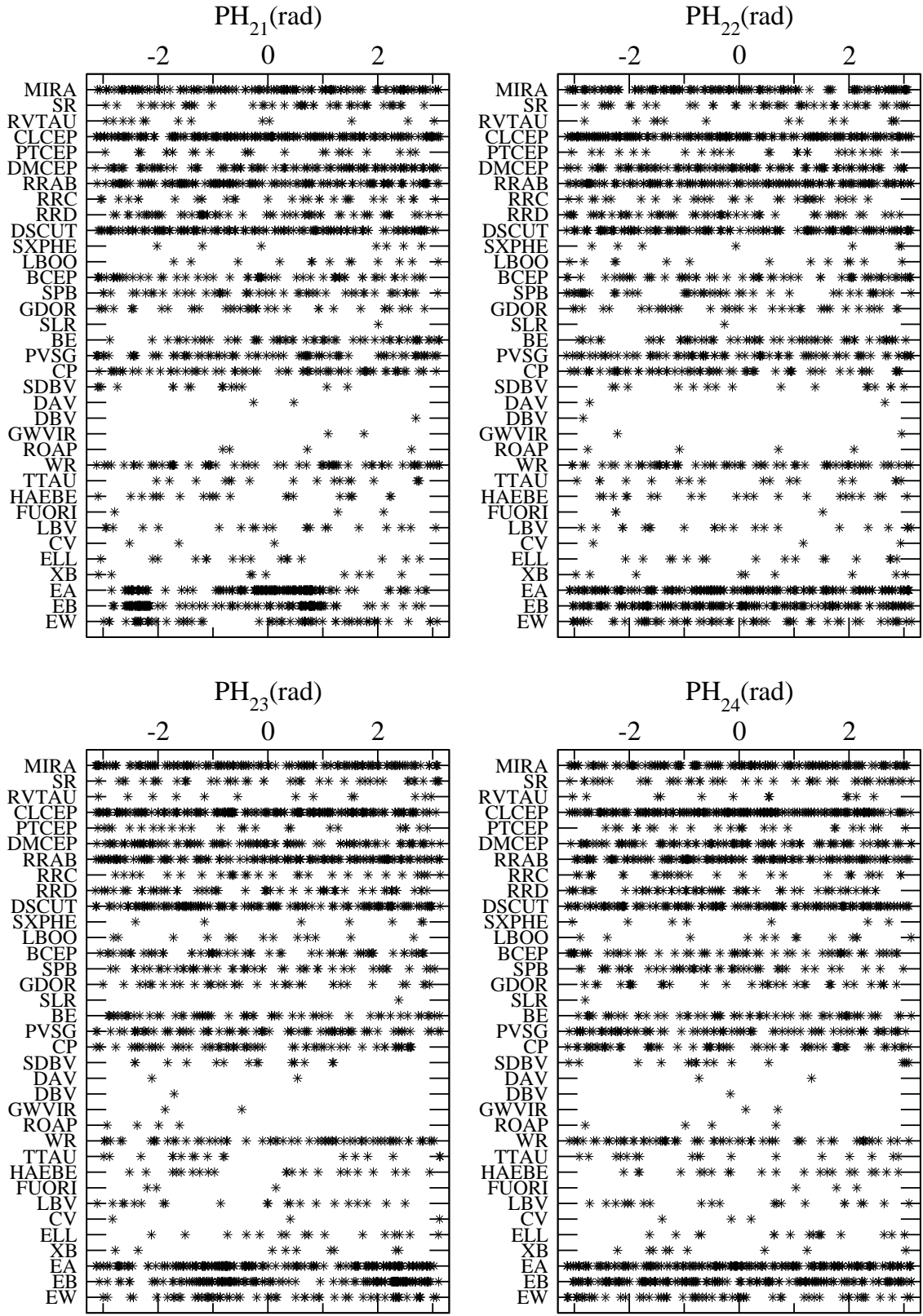}
   \caption{The range in phases $PH_{2j}$ for the $4$ harmonics of $f_2$. As can be seen from the plots, the distribution of these parameters is rather uniform for every class. They are unlikely to be good classification parameters, since for none of the classes, clear clustering of the phase values is present.} 
   \label{fig7}
\end{figure*}

\begin{figure*}
   \centering
   \includegraphics[width=17cm,angle=180,scale=0.94]{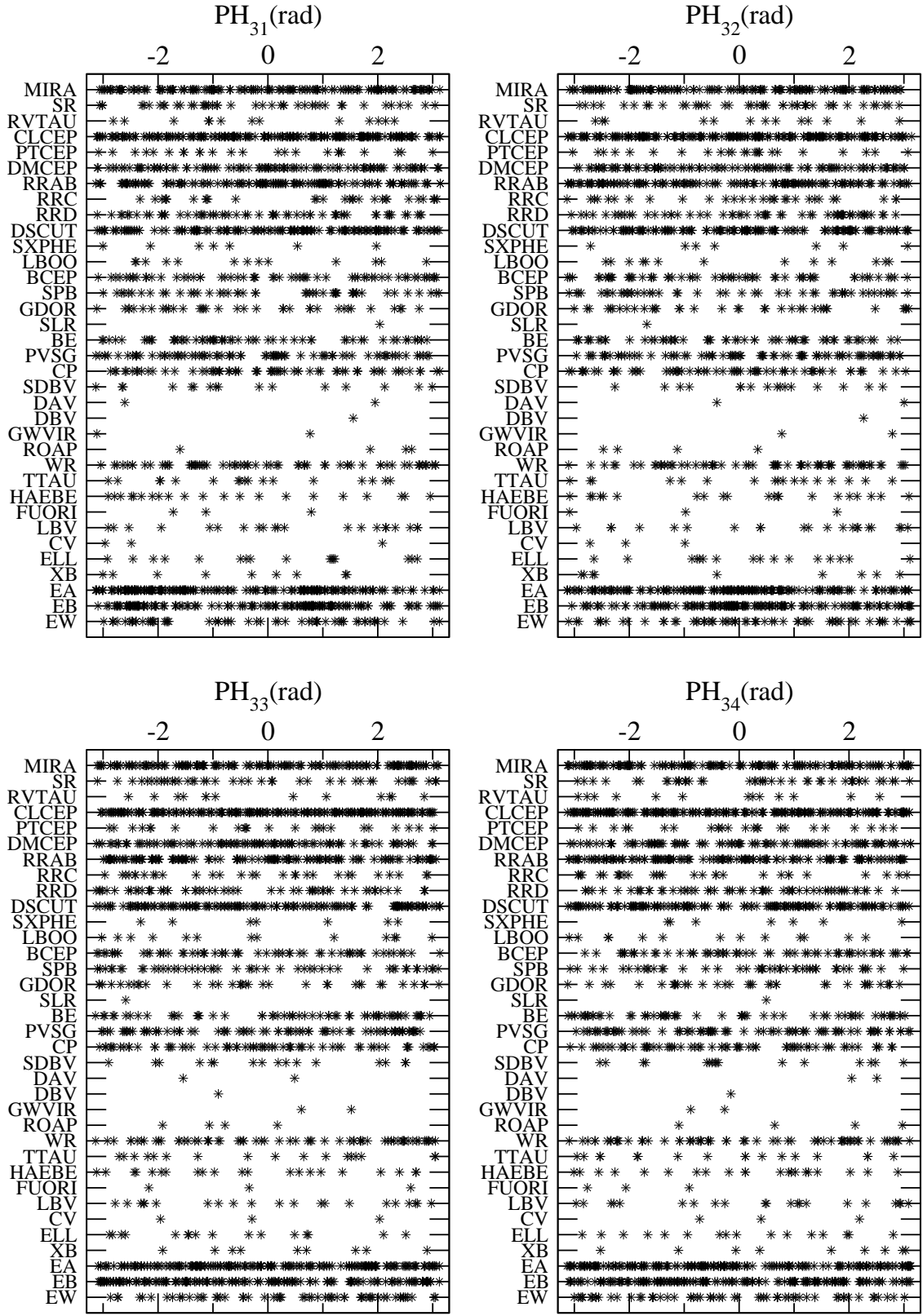}
   \caption{The range in phases $PH_{3j}$ for the $4$ harmonics of $f_3$. The same comments as for Fig. \ref{fig7} apply here.}
   \label{fig8}
\end{figure*}

\begin{figure*}
   \centering
   \includegraphics[angle=-90,scale=0.6]{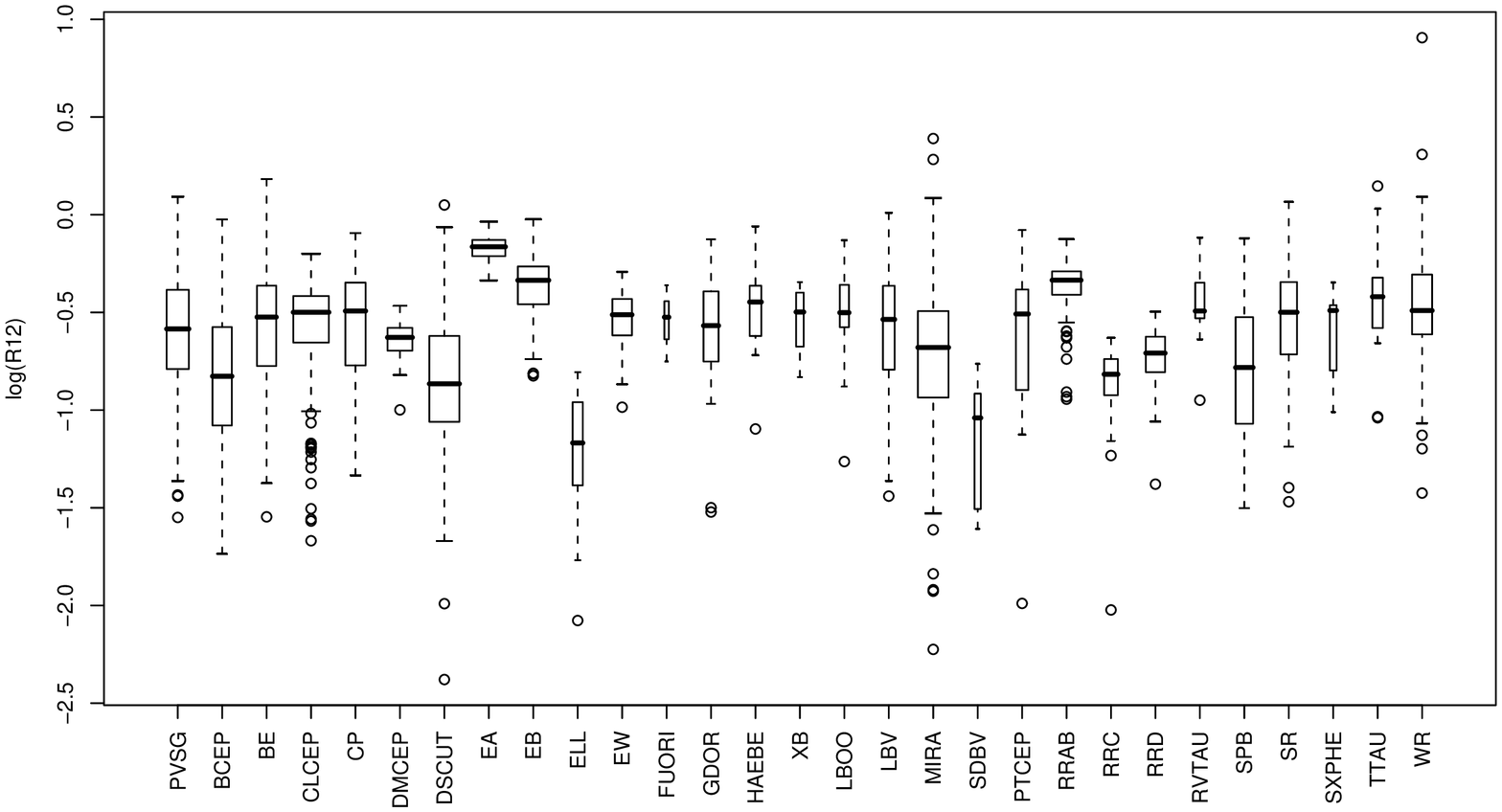}
   \caption{Box-and-whiskers plot of the logarithm of $R_{21}$ for all
   classes in the training set. Central boxes represent the median and
   interquantile ranges (25 to 75\%) and the outer whiskers represent
   rule-of-thumb boundaries for the definition of outliers (1.5 the
   quartile range). The boxes widths are proportional to the number of
   examples in the class.}
   \label{baw2}
\end{figure*}

\begin{figure*}
   \centering
   \includegraphics[angle=-90,scale=0.6]{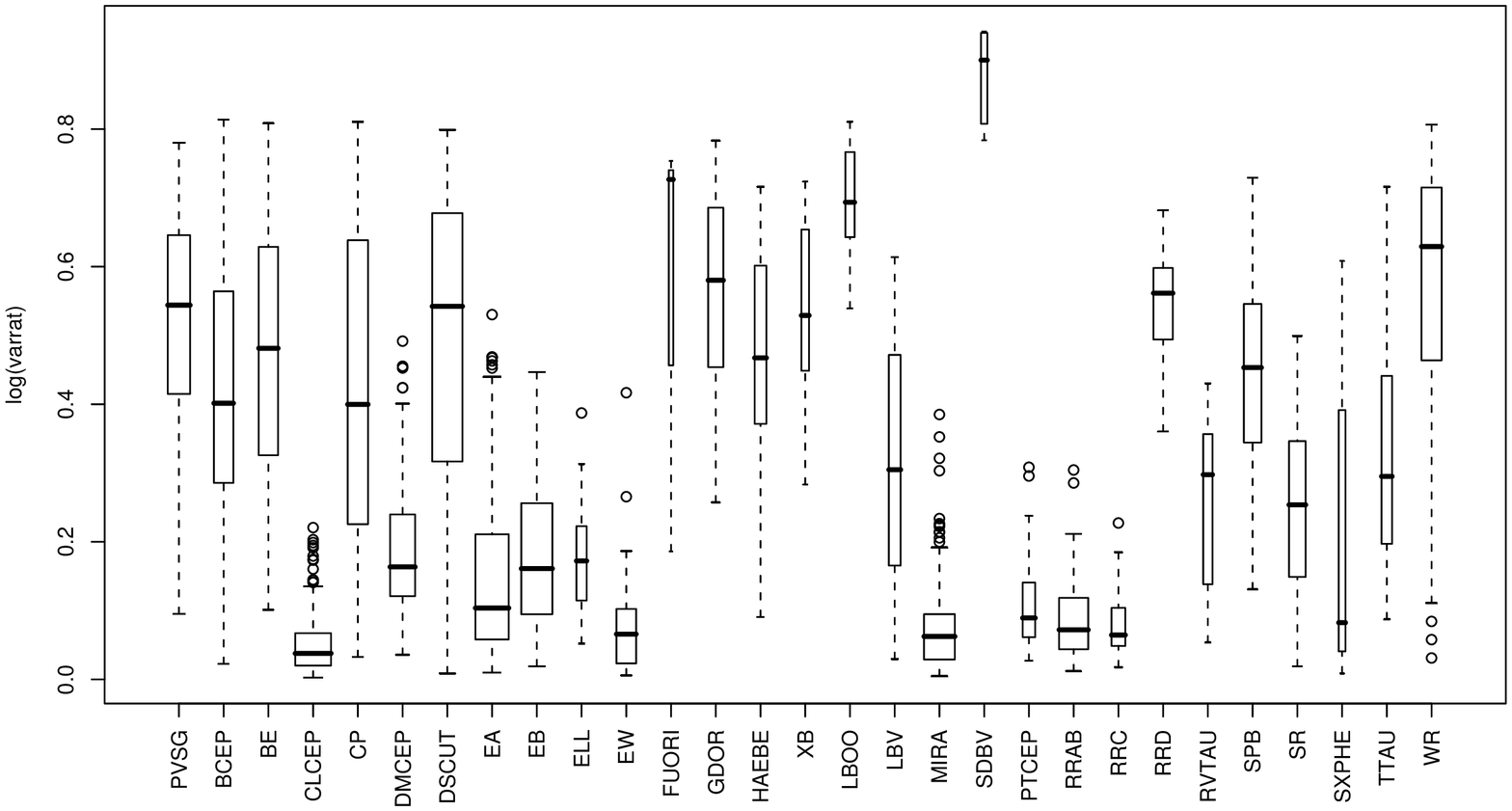}
   \caption{Box-and-whiskers plot of the logarithm of the variance ratio $v_{f1}/v$ (varrat) for all classes in
   the training set. Central boxes represent the median and
   interquantile ranges (25 to 75\%) and the outer whiskers represent
   rule-of-thumb boundaries for the definition of outliers (1.5 the
   quartile range). The boxes widths are proportional to the number of
   examples in the class.}
   \label{baw3}
\end{figure*}

Complementary to these plots, we have conducted a more detailed analysis of the
statistical properties of the training set. This analysis is of importance for a
sensible interpretation of the class assignments obtained for unknown objects, since the class boundaries of the classifiers depend critically on the
densities of examples of each class as functions of the classification
parameters. This analysis comprises i) the computation of box-and-whiskers plots
for all the attributes used in classification (see Fig. \ref{baw1}, \ref{baw2}, and \ref{baw3} for example); ii) the search for correlations between the different
parameters; iii) the computation of 1d, 2d and 3d nonparametric density
estimates (see Fig.\,\ref{2d-densities} for an easily interpretable hexagonal
histogram); iv) clustering analysis of each class separately and for the
complete training set. The results of this analysis are especially useful for
guiding the extension of the training set as new examples become available
to users, such as those from CoRoT and Gaia.


\section{Supervised Classification}

The class descriptions we attained, form the basis of the so-called `Supervised
Classification'. This classification method assigns every new
object to one of a set of pre-defined classes (called `training classes'), meaning that, given the time series characteristics described above, the system
gives a set of probabilities that the source of the time series belongs to one
of the set of classes listed in Table\,\ref{tabel1}.

A supervised classification can be done in many ways. The most suitable method
depends on the kind of data to be classified, the required performance and the
available computational power. We focus here on a statistical method based on
multivariate analysis, also known as the `Gaussian Mixture Model'. We have chosen
for a fast and easily adaptable code written in {\tt FORTRAN}. We also summarize
the results of a detailed study of other supervised classification methods,
based on Artificial Intelligence techniques.

\subsection{Multivariate Gaussian mixture classifier}

We assume that the descriptive parameters for every class have a multinormal
distribution. This is a reasonable assumption for a first approach. There is no
reason to assume a more complicated distribution function, unless there is clear evidence. The added advantages of the multinormal distribution are the well-known properties and the relatively simple calculations. We use our derived
light curve parameters to estimate the mean and the variance of the multinormal
distributions. If the vector $X_{ij}$ represents the parameters of light curve
number $j$ belonging to class $i$, the following quantities are calculated for
every variability class. The class mean vector of length P (number of light
curve parameters, $P=28$ in our method for example):
\begin{equation}
\overline{X_i}=\frac{1}{N_i}\sum_{j=1}^{N_i} X_{ij} 
\end{equation}
and the class variance-covariance matrix of dimension P$\times$P:
\begin{equation}
S_i=\frac{1}{N_i-1}\sum_{j=1}^{N_i}(X_{ij}-\overline{X_i})(X_{ij}-\overline{X_i})'.
\end{equation}
Every class is now defined by a mean vector $\overline{X_i}$ and a
variance-covariance matrix $S_i$, which corresponds to the mean and the variance
of a normal distribution in the one-dimensional case.

If we want to classify a new object, we first have to calculate the same light
curve parameters as described in Sect. $2$. We can then derive the statistical
distance of this object with respect to the different classes, and assign the
object to the nearest (most probable) class. If $X$ denotes the parameters for
the new object, we calculate the following statistical distance for every class:
\begin{equation} 
 D_i=(X-X_i)'S_i^{-1}(X-X_i)+\ln|S_i|,
\label{dist} 
\end{equation}
with $|S_i|$ the determinant of the variance-covariance matrix
\citep[e.g.][]{Sharma}. The first term of $D_i$ is known as the squared Mahalanobis
distance. The object is now assigned to class $i$ for which $D_i$ is
minimal. This minimum of $D_i$ is equivalent to the maximum of the corresponding
density function (under the assumption of a multinormal distribution):
\begin{equation}
f_i(X)=\frac{1}{(2\pi)^{P/2}|S_i|^{1/2}}\exp{-\frac{1}{2}(X-\overline{X_i})'S_i^{-1}(X-\overline{X_i})}.  
\end{equation}
This statistical class assignment method can cause an object to be assigned to a
certain class even if its light curve parameters deviate from the class' typical
parameter values. This is a drawback, and can cause contamination in the
classification results. It does, however, has an important advantage: objects near
the border of the class can still be correctly assigned to the class. If one is
only interested in objects that are very similar to the objects used to define
the class, one can define a cutoff value for the Mahalanobis distance. Objects
that are too far from the class centers will then not be assigned to any of the
classes. To illustrate this, consider a classifier where only $f_1$ would be used as a classification attribute, and suppose we are interested in $\beta$-Cephei stars. If we do not want a star to be classified as $\beta$-Cephei if the value of $f_1$ is larger than $15$ c/d, we have to take a cutoff value for the Mahalanobis distance equal to $4$ in frequency space (this value only holds for our definition of the $\beta$-Cephei class). In terms of probabilities: objects with a Mahalanobis distance larger than $4$ are more than $4\sigma$ away from the class center (the class' mean value for $f_1$), and are therefore very unlikely to belong to the class.

We emphasize the difference between a supervised classification method as
defined here and an extraction method: a supervised classifier assigns new
objects to one of a set of definition classes with a certain probability, given
the object's derived parameters. An extractor, on the other hand, will only
select those objects in the database for which the derived parameters fall
within a certain range. Extractor methods are typically used by scientists only
interested in one class of objects. The specified parameter range for an
extractor (based on the knowledge of the variability type) can be chosen as to
minimize the number of contaminating objects, to make sure that the majority of
the selected objects will indeed be of the correct type. Of course, extraction
methods can also be applied to our derived parameter set. The goal of our
supervised classifier is much broader, however: we consider all the known
variability classes at once and also get a better view on the differences and
similarities between the classes. Moreover, our method does not need visual
inspection of the light curves, while this was always needed in practice with
extraction.  On top of this, our class definitions can also be used to specify
parameter ranges for extraction methods.

\subsection{Machine learning classifiers}

Following standard practice in the field of pattern recognition or
statistical learning, we have adopted a parallel approach where we
allow for more flexibility in the definition of the models used to
classify the data. The Gaussian mixtures model presented in the
previous section induces hyperquadratic boundaries
between classes (with hyperspheres/hyperellipses as special
cases). This has the advantage of providing a fast method for the
detection of outliers (objects at large Mahalanobis distances from all
centers) and easy interpretation of results. On the other hand, more
sophisticated methods offer the flexibility to reproduce more
complicated boundaries between classes, at the expense of more complex
models with varying degrees of interpretability. 

A common problem presented in the development of supervised
classification applications based on statistical learning methods, is
the search for the optimal trade-off between the two components of the
classifier error. In general, this error is composed of two
elements: the bias and the variance. The former is due to the
inability of our models to reproduce the real decision boundaries
between classes. To illustrate this kind of error, we can imagine a
set of training examples such that any point above the $y=\sin(x)$
curve belongs to class A and any point below it, to class B. Here,
classes A and B are separable (unless we add noise to the class
assignment), and the decision boundary is precisely the curve
$y=\sin(x)$. Obviously, if we try to solve this toy classification
problem with a classifier inducing linear boundaries we will
inevitably have a large bias component to the total error. The second
component (the variance) is due to the finite nature of the training
set and the fact that it is only one realization of the random
process of drawing samples from the true (but unknown) probability
density of having an object at a given point in the hyperspace of
attributes.

If the model used to separate the classes in the classification problem is
parametric, then we can always reduce the bias term by adding more and more
degrees of freedom. In the Gaussian mixtures case, where we model the
probability densities with multivariate Gaussians, this would be equivalent to
describing each class by the sum of several components (i.e. several
multivariate Gaussians). It has to be kept in mind, however, that there is an
optimal number of components beyond which the decrease in the bias term is more
than offset by an increase of the variance, due to the data being overfitted by
the complexity of the model.  The natural consequence is the loss of
generalization capacities of the classifier, where the generalization ability is
understood as the capacity of the model to correctly predict the class of unseen
examples based on the inferences drawn from the training set.

We computed models allowing for more complex decision boundaries where the
bias-variance trade-off is sought, using standard procedures. Here we present brief outlines of the methods and a summary of the results, while a more detailed analysis will be published in a forthcoming paper (Sarro et al., in
preparation). We made use of what is widely known as Feature Selection
Methods. These methods can be of several types and are used to counteract the
pernicious effect of irrelevant and/or correlated attributes for the performance
of classifiers. The robustness of a classifier to the degradation produced by
irrelevance and correlation depends on the theoretical grounds on which the
learning algorithms are based. Thus, detailed studies have to be conducted to find the optimal subset of attributes for a given problem.  The interested reader can find a good introduction to the field and references to
the methods used in this paper in \cite{FS-JMLR}.

We adopted two strategies: training a unique classifier for the 29 classes with
sufficient stars for a reliable estimate of the errors, or adopting a multistage
approach where several large groups with vast numbers of examples and well
identified subgroups (eclipsing binaries, Cepheids, RR-Lyrae and Long Period
Variables) are classified first by specialized modules in a sequential approach
and then, objects not belonging to any of these classes are passed to a final
classifier of reduced complexity.

\subsubsection{Feature selection}

Feature selection methods fall into one of two categories: filter and
wrapper methods. Filter methods rank the attributes (or subsets of
them) based on some criterion independent of the model to be
implemented for classification (e.g., the mutual information between
the attribute and the class or between attributes, or the statistical
correlation between them). Wrapper methods, on the other hand, explore
the space of possible attribute subsets and score each combination
according to some assessment of the performance of the classifier
trained only on the attributes included in the subset. The exhaustive
search for an optimal subset in the space of all possible combinations
rapidly becomes unfeasible as the total number of attributes in the
original set increases. Therefore, some sort of heuristic search, based
on expected properties of the problem, has to be used in order to
accomplish the selection stage in reasonable times. 

We applied several filtering techniques based on Information Theory (Information
Gain, Gain Ratio and symmetrical uncertainty) and statistical correlations to
the set of attributes described in Sect.\,2.1, extended with peak-to-peak
amplitudes, harmonic amplitude ratios (within and across frequencies) and
frequency ratios. These techniques were combined with appropriate search
heuristics in the space of feature subsets. Furthermore, we also investigated
feature relevance by means of wrapper techniques applied to Bayesian networks
and decision trees, but not to the Bayesian combination of neural networks or to
Support Vector Machines due to the excessive computational cost of combining the
search for the optimal feature subset and the search for the classifier's
optimal set of parameters. The Bayesian model averaging of neural networks in
the implementation used here, incorporates automatic relevance determination by
means of hyperparameters. For this reason, we have not performed any feature selection.

In general, there is no well-founded way to combine the results of
these methods. Each approach conveys a different perspective of the
data and it is only by careful analysis of the rankings and selected
subsets that particular choices can be made. As a general rule, we
have combined the rankings of the different methodologies when dealing
with single stage classifiers, whereas for sequential classifiers, each
stage had its own feature selection process. When feasible in terms of
computation time (e.g. for Bayesian networks), the attribute subsets
were scored in the wrapper approach. Otherwise, several filter methods
were applied and the best results used.

\subsubsection{Bayesian networks classifier}

Bayesian networks are probabilistic graphical models where the uncertainty
inherent to an expert system is encoded into two basic structures: a graphical
structure $S$ representing the conditional independence relations amongst the
different attributes, and a joint probability distribution for its nodes
\citep{BN}. The nodes of the graph represent the variables (attributes) used to
describe the examples (instances). There is one special node corresponding to
the class attribute. Here, we have constructed models of the family known as
$k$-dependent Bayesian classifier \citep{sahami96learning} with $k$, the maximum
number of parents allowed for a node in the graph, set to a maximum of 3 (it was
checked that higher degrees of dependency did not produce improvements in the
classifier performance).

The induction of Bayesian classifiers implies finding an optimal structure and
probability distribution according to it. We have opted for a score and search
approach, where the score is based on the marginal likelihood of the structure
as implemented in the K2 algorithm by \cite{K2-CH92}. Although there are
implementations of the $k$-dependent Bayesian classifiers for continuous
variables, also known as Gaussian networks, we have obtained significantly
better results with discretized variables. The discretization process is based
on the Minimum Description Length principle as proposed in
\cite{Fayyadirani93}. It is carried out as part of the cross validation
experiments to avoid overfitting the training set.

\subsubsection{Bayesian average of artificial neural networks classifier}

Artificial neural networks are probably the most popular methodology for
classification and clustering. They are taken from the world of Artificial
Intelligence. In its most frequent implementation, it is defined as a
feedforward network made up of several layers of interconnected units or
neurons. With appropriate choices for the computations carried out by the
neurons, we have the well-known multilayer perceptron. \cite{Bishop} has written
an excellent introductory text to the world of neural networks, statistical
learning and pattern recognition.

We do not deviate here from this widely accepted architecture but use a training
approach other than the popular error backpropagation algorithm. Instead of the
maximum likelihood estimate provided by it, we use Bayesian Model Averaging
(BMA). BMA combines the predictions of several models (networks) weighting each
by the {\it a posterior\/} probability of its parameters (the weights of network
synapses) given the training data. For a more in-depth description of the
methods, see \cite{Neal} or \cite{Sarro:2006}. In the following, we use the
acronym BAANN to refer to the averaging of artificial neural networks.

\subsubsection{Support vector machines classifier}

Support vector machines (SVM) are based on the minimization of the
structural risk \citep{SRM}. The structural risk can be proven to be upper-bounded by the sum of the empirical risk and the optimism, a quantity dependent on the Vapnik-Chervonenkis dimension of the chosen set of classifier
functions. For linear discriminant functions, minimizing the optimism amounts to
finding the hyperplane separating the training data with the largest margin
(distance to the closest examples called support vectors). For nonlinearly
separable problems, the input space can be mapped into a higher dimensional
space using kernels, in the hope that the examples in the new hyperspace are
linearly separable. A good presentation of the foundations of the method can be
found in \cite{Vapnik95}. Common choices for the kernels are $n$-th degree
polynomial and Gaussian radial basis functions. The method can easily
incorporate noisy boundaries by introducing regularization terms. We used radial
basis functions kernels. The parameters of the method (the complexity or
regularization parameter and the kernel scale) are optimized by grid search and
10-fold cross validation.

\section{Classifier performance}

One of the central problems of statistical learning from samples is that of
estimating the expected error of the developed classifiers.  The final goal of
automatic classification, as mentioned above, is to facilitate the analysis of
large amounts of data which would otherwise be left unexplored because the amount
of time needed for humans to undertake such an analysis is incommensurably large. This
necessity cannot mask the fact that classifiers have errors and these need to be
quantified if scientific hypotheses are to be drawn from their products.

When developing a classifier, the goal is to maximize the number of correct
classifications of new cases. Given the classification method, the performance
of a supervised classification depends, amongst other factors that measure the
faithfulness of the representation of the real probability densities by the
training set, on the quality of the descriptive parameters used for
classifying. We seek a set of classification attributes which describes most
light curves well and provides a good separation of the classes in
attribute space.

Several methodologies exist to evaluate classifiers.  A common way of testing a
classifier's performance is feeding it with objects with a known member class and
derive how many of them are correctly classified. This method is called
`cross validation' in the case that the complete training set is split up into two
disjointed sets: a training set and a set that will be classified, called the
validation set. It is also possible to use the complete set for both training
and classifying. This is known as `resampling'. This is no longer a cross validation
experiment, since the objects used for training and for classifying are
the same. For a real cross validation experiment, the objects to classify
must be different from the objects in the training set, in order to have
statistical independence. The resampling method thus has a bias towards
optimistic assessment of the misclassification rate, compared to a
cross validation method.  Another possibility (called holdout procedure)
consists of training the classifier with a subset of the set of examples and
evaluating its error rates with the remainder. Depending on the percentage split
it can be biased as well, but this time in the opposite (pessimistic)
direction. Finally, the most common approach to validating classification models
is called $k$-fold cross validation. This consists of dividing the set of
examples into $k$ folds, repeating $k$ times the process of training the
model with $k-1$ folds and evaluating it with the $k$-th fold not used for
training. Several improvements to this method can be implemented to reduce the
variance of its estimates, e.g. by assuring a proportional representation of
classes in the folds (stratified cross validation). Recent proposals include
bolstered resubstitution and several variants. Good and recent overviews of the
problem with references to relevant bibliography can be found in \cite{Demsar06}
and \cite{Bouckaert04}.

\subsection{Gaussian mixture classifier}

For the simplest classifier, we also considered the simplest performance test by
adopting the resampling approach. Using this method, we already get an idea of
the overlap and separability of the classes in parameter space.

The total number of correct classifications expressed as a percentage, can be
rather misleading. For example, if our training set contains many example
light curves for the well-separated classes, we will have a high rate of correct
classifications, even if the classifier performs very badly for some classes
with only a small number of training objects. Therefore, it is better to judge
the classifier's performance by looking at the `confusion matrix'. This is a
square matrix with rows and columns having a class label. It lists the numbers
of objects assigned to every class in the training set after
cross validation. The diagonal elements represent the correct classifications,
and their sum (trace of the matrix) divided by the total number of objects in
the training set, equals the total correct classification rate. The off-diagonal
elements show the number of misclassified (confused) objects and the classes
they were assigned to. In this way, we get a clear view on the classifier's
performance for every class separately. We can see which classes show high
misclassification rates and are thus not very well separated using our set of
classification attributes.

Table\,\ref{table3} shows the confusion matrix for a subset of $25$ variability
classes. These are the classes with more than $13$ member stars each. We have
chosen not to take the classes with fewer member stars into account here,
because the number of classification attributes is limited by the number of
member stars in the class. This is merely a numerical limitation of the
multivariate Gaussian mixture classifier: if the number of defining class
members is equal to or lower than the number of classification attributes, the
determinant of the variance-covariance matrix will become equal to zero. This
makes it impossible to calculate the statistical distance with respect to the
class using Eq. (\ref{dist}). We used $12$ light curve parameters to perform the
classification (the smallest class in this set contains $13$ member stars): the
three frequencies $f_i$, the four amplitudes $A_{1j}$, the phases $PH_{1j}$, the
linear trend $b$ and the variance ratio. The average correct classification rate
is about $69\%$ for this experiment. As can be seen from the matrix in
Table\,\ref{table3}, the monoperiodic pulsators such as MIRA, CLCEP, DMCEP,
RRAB, RRC and RRD are well separated. Some of the multiperiodic pulsators are
also well identified (SPB, GDOR). A lot of misclassifications (fewer than
$50\%$ correct classifications) occur for the following multiperiodic pulsators:
BE, PVSG, DSCUT. Also, some of the irregular and semi-regular variables show
poor results (SR, WR, HAEBE, TTAU) as was emphasized in Sect.\,2.2.

Depending on the intended goal, it can be better to take fewer classes into
account. For example, when the interest is focused on a few classes only, using
fewer classes will decrease the risk of misclassifying members of those
classes. To illustrate this, Table\,\ref{table4} shows the confusion matrix for
only $14$ classes using the complete set of $28$ light curve parameters defined
in Sect.\,2.1 to perform the classification. We did not include the classes
with very irregular light curves or the less well-defined classes such as BE,
CP, WR and PVSG.

The average correct classification rate amounts to $92\%$ for this
experiment. It is clear that the monoperiodically pulsating stars are again very
well separated (MIRA, CLCEP, DMCEP, RRAB, RRC and RRD). Most of the classes with
multiperiodic variables also show high correct classification rates now (SPB,
GDOR). Confusion is still present for the DSCUT and the BCEP classes.  This is
normal, as these stars have non-radial oscillations with similar amplitudes and
with overlap in frequencies.  For those classes in particular, additional (or
different) attributes are necessary to distinguish them, e.g.\ the use of a
color index as we will discuss extensively in our future application of the
methods to the OGLE database. Parameter overlap (similarity) with other classes
is the main reason for misclassifications if only light curves in a single
photometric band are available. Note the high correct classification rate for
the three classes of eclipsing binaries (EA, EB and EW). Some of their light
curves (mainly from the EA class) are highly non-sinusoidal, but they are
nevertheless well described with our set of attributes.

The higher correct classification rates for this classification experiment with
$14$ classes is caused by the removal of the most `confusing' classes compared
to the experiment with $25$ classes, and the increased number of discriminating
attributes (this was tested separately). Note that the use of fewer classes for
classifying also implies more contamination of objects which actually belong to
none of the classes in the training set. This can effectively be solved by
imposing limits on the Mahalanobis distance to the class centers.
Objects with a Mahalanobis distance larger than a certain user-defined
value, will then not be assigned to any class.

\begin{sidewaystable*}
\begin{minipage}[c][220mm]{\textwidth}
\caption{The Confusion Matrix for the Gaussian Mixture method, using $25$
variability classes and $12$ classification attributes. The last but one line
lists the total number of light curves (TOT) to define every class. The last
line lists the correct classification rate (CC) for every class separately. The
average correct classification rate is about $69\%$.}
\label{table3}
\tiny
\renewcommand{\tabcolsep}{1.3mm}
\renewcommand{\arraystretch}{1.7}

\begin{tabular}{l||c|c|c|c|c|c|c|c|c|c|c|c|c|c|c|c|c|c|c|c|c|c|c|c|c}

             &MIRA&SR&RVTAU&CLCEP&PTCEP&DMCEP&RRAB&RRC&RRD&DSCUT&LBOO&BCEP&SPB&GDOR&BE&PVSG&CP&WR&TTAU&HAEBE&LBV&ELL&EA &EB&EW\\ 
\hline \hline MIRA  &139 &2 &0     &3    &0     &0     &0    &0   &0   &0     &0    &0   &0   &0    &3 &0    &0  &1 &1   &1    &0   &0                &0   &0   &0  \\ 
\hline SR    &2   &19&0     &2    &0     &0     &2   &0   &0   &1    &0    &0    &0   &0    &0  &1    &0  &1 &2   &2    &0   &0                &0   &1  &0  \\ 
\hline RVTAU &1   &0  &13   &0     &0     &1    &0    &0   &0   &0     &0    &0    &0   &0    &0  &0    &0  &0  &0    &0     &0   &0                &0   &0   &0  \\ 
\hline CLCEP &1   &0  &0     &171  &6    &0     &0    &0   &0   &0     &0    &0    &0   &0    &0  &0    &0  &0  &0    &0    &0   &0   &0   &0   &0  \\
\hline PTCEP &1    &2 &0     &14    &17   &0     &0    &0   &0   &0     &0    &0    &0   &0    &0  &0    &0  &0  &0    &0     &1  &0      &0   &0   &0  \\  
\hline DMCEP &0    &0  &0     &0     &0     &92   &1   &0   &0   &0     &0    &0    &0   &0    &0  &0    &0  &2 &0    &0     &0   &0          &0   &0   &0  \\ 
\hline RRAB  &0    &1 &0     &1    &1    &0     &118 &0   &0   &1    &0    &0    &0   &0    &0  &0    &0  &0  &0    &0     &0   &0         &0   &0   &0  \\ 
\hline RRC   &0    &0  &0     &0     &0     &0     &1   &27 &0   &5    &0    &0    &0   &0    &0  &0    &0  &0  &0    &0     &0   &0     &0   &0   &0  \\ 
\hline RRD   &0    &0  &0     &0    &0     &0     &0    &0   &56 &0     &0    &0    &0   &0    &0  &0    &0  &0  &0    &0     &0   &0                &0   &0   &0  \\ 
\hline DSCUT &0    &1 &0     &0     &0     &1    &4   &0   &0   &21   &0    &1   &0   &0    &4 &6   &0  &11&0    &0    &2  &0                &0   &2  &0  \\ 
\hline LBOO  &0    &0  &0     &0     &0     &0     &0    &0   &0   &2    &13  &0    &0   &1   &0  &2   &0  &1 &0    &0     &0   &0                &0   &0   &0  \\ 
\hline BCEP  &0    &0  &0     &0     &0     &0     &0    &1  &0   &24   &0    &26  &0   &0    &5 &1   &0  &3 &0    &1   &0   &0                &0   &0   &0  \\ 
\hline SPB   &0    &1 &0     &0     &0     &0     &0    &0   &0   &0     &0    &6   &41 &6   &12&10  &11&3 &0    &1   &0   &0               &0   &1  &0  \\ 
\hline GDOR  &0    &0  &0     &0     &0     &0     &0    &0   &0   &31  &0    &11  &2  &23  &7 &19  &8 &9 &1   &5   &0   &0                &0   &0   &0  \\
\hline BE    &0    &0  &0     &0     &0     &0     &0    &0   &0   &0     &0    &0    &0   &0    &2 &1   &0  &0  &0    &1    &0   &0                &0   &0   &0  \\
\hline PVSG  &0    &6 &0     &0     &0    &0     &0    &0   &0   &1    &0    &1   &0   &0    &11&18  &1 &13&1   &3    &3  &0                &0   &0   &0  \\
\hline CP    &0    &0  &0     &0     &0     &0     &0    &0   &0   &46   &0    &11  &1  &5   &6 &5   &42&4 &0    &1   &0   &0                &0   &0   &0  \\
\hline WR    &0   &2 &0     &0     &0     &0     &0    &0   &0   &0     &0    &0    &0   &0    &5 &9   &0  &13&2   &1    &2  &0                &0   &0   &0  \\
\hline TTAU  &0    &1 &0     &1    &0     &0     &2   &0   &0   &0     &0   &0    &0   &0    &0  &2   &0  &1 &8  &0     &0   &0                &0   &0   &0  \\
\hline HAEBE &0    &4 &0     &0     &0     &0    &0    &0   &1  &3    &0    &0    &0   &0    &0  &0    &0  &1 &2   &5    &0   &0                &0   &0   &0  \\
\hline LBV   &0    &3 &0     &2    &0     &1    &1   &0   &0   &0     &0    &0    &0   &0    &2 &2   &0  &0  &0    &0    &13 &0                &0   &3  &0  \\ 
\hline ELL   &0    &0  &0     &0     &0     &0     &0    &0   &0   &3    &0    &1   &2  &0    &0  &0    &1 &0  &0    &0     &0   &16              &0   &0   &0  \\    
\hline EA    &0    &0  &0     &0     &0     &0     &0    &0   &0   &1    &0    &0    &0   &0    &0  &0    &0  &0  &0    &0     &0   &0                &154&26 &1 \\ 
\hline EB    &0    &0  &0     &1    &0     &0     &0    &0   &0   &0     &0    &1   &1  &0    &0  &0    &0  &0  &0    &0     &0   &0                &12 &98&3 \\ 
\hline EW    &0   &0   &0     &0     &0     &0     &0    &1  &0   &0     &0    &0    &0   &0    &0  &0    &0  &0  &0    &0     &0   &0                &3  &16 &55\\ 
\hline \hline TOT   &144&42 &13   &195  &24   &95   &129 &29 &57 &139  &13  &58  &47 &35  &57&76  &63&63&17  &21   &21 &16
             &169&147&59 \\
\hline CC(\%)&97 &45 &100  &88   &71   &97   &91  &93 &98 &15   &100 &45  &87 &66  &4 &24  &67&21&47  &24   &62&100&91 &67 &93\\

\end{tabular}

\end{minipage}
\end{sidewaystable*}

\begin{table*}

\caption{The confusion matrix for the Gaussian mixture method using $14$
variability classes and $28$ classification attributes. The last but one line
lists the total number of light curves (TOT) to define every class. The last
line lists the correct classification rate (CC) for every class separately. The
average correct classification rate is about $92\%$.}
\label{table4}
\scriptsize
\centering
\renewcommand{\arraystretch}{2}
\begin{tabular}{l||c|c|c|c|c|c|c|c|c|c|c|c|c|c}

             &MIRA&SR&CLCEP&DMCEP&RRAB&RRC&RRD&DSCUT&BCEP&SPB&GDOR&EA &EB &EW\\ 
\hline \hline MIRA  &140 &0  &3    &0     &0    &0   &0   &0     &0    &0   &0    &0   &0   &0  \\
\hline SR    &0    &36&0     &0     &0    &0   &0   &1    &0    &0   &0    &0   &0   &0  \\ 
\hline CLCEP &4   &1 &187  &0     &0    &0   &0   &0     &0    &0   &0    &0   &0   &0  \\ 
\hline DMCEP &0    &0  &0     &95   &0    &0   &0   &0     &0    &0   &0    &0   &0   &0  \\ 
\hline RRAB  &0    &0  &2    &0     &124 &0   &0   &2    &0    &0   &0    &0   &0   &0  \\ 
\hline RRC   &0    &0  &0     &0     &0    &29 &0   &4    &0    &0   &0    &0   &0   &0  \\ 
\hline RRD   &0    &0  &0     &0     &0    &0   &57 &0     &0    &0   &0    &0   &0   &0  \\ 
\hline DSCUT &0    &5 &3    &0     &5   &0   &0   &90   &0    &0   &0    &0   &0   &0  \\ 
\hline BCEP  &0    &0  &0     &0     &0    &0   &0   &30   &57  &0   &0    &0   &0   &0  \\ 
\hline SPB   &0    &0  &0     &0     &0    &0   &0   &0     &0    &47 &0    &0   &0   &0  \\ 
\hline GDOR  &0    &0  &0     &0     &0    &0   &0   &11    &1   &0   &35  &0   &0   &0  \\ 
\hline EA    &0    &0  &0     &0     &0    &0   &0   &1    &0    &0   &0    &161&17 &0  \\ 
\hline EB    &0    &0  &0     &0     &0    &0   &0   &0     &0    &0   &0    &5  &121&0  \\ 
\hline EW    &0    &0  &0     &0     &0    &0   &0   &0     &0    &0   &0    &3  &9  &59\\
\hline \hline TOT   &144 &42&195  &95   &129 &29 &57 &139  &58  &47 &35  &169&147&59\\
\hline CC(\%)&97  &86&96   &100  &96  &100&100&65   &98  &100&100 &95 &82 &100\\

\end{tabular}

\end{table*}

\subsection{Machine learning techniques}

Selecting a methodology amongst several possible choices is in itself a
statistical problem. Here we only summarize the results of a complete study
comparing the approaches listed in Sect.\,3.2, the details of which will be
published in a specialized journal in the area of Pattern Recognition.

As explained in Sect.\,3.2, one reason models can fail is that they are not
flexible enough to describe the decision boundaries required by the data (the
bias error). Another reason is because the training set, due
to its finite number of samples, is never a perfect representation of the real probability
densities (otherwise one would work directly with them and not with
examples). Since learning algorithms are inevitably bound to use the training
set to construct the model, any deficiency or lack of faithfulness in their
representation of the probability densities will translate into errors. The
bias-variance trade-off explained above is somehow a way to prevent the learning
algorithm from adjusting itself too tightly to the training data (a problem
known as overfitting) because its ability to generalize depends critically on
it. Finally, irrespective of all of the above, we cannot avoid dealing with confusion
regions, i.e., regions of parameter space where different classes coexist.

For the machine learning technique, we selected the combination of 10 sorted
runs of 10-fold cross validation experiments together with the standard t-test
(Demsar 2006). This combination assures small bias, a reduced variance (due to
the repetition of the cross validation experiments) and replicability, an issue
of special importance since these analyses will be iterated as the training set
will be completed with new instances for the poorly represented classes and new
attributes from projects such as CoRoT, Kepler and Gaia.

In the following, we have split the results for single stage and sequential
classifiers. It should be born in mind that the misclassification rates used
in the following sections include entries in the confusion matrices which relate
eclipsing binary subtypes. These are amongst the largest
contributions to the overall misclassification rate and are due to a poor
definition of the subtypes as argued in \cite{Sarro:2006} and as is widely known. In
future applications of the classifiers (i.e.\ for CoRoT data) the specialized
classifier presented in \cite{Sarro:2006} and its classification scheme will be
used. Therefore, the misclassification rates quoted below are too pessimistic by
an estimated 2\%.

\subsubsection{Single stage classifiers}

Table\,\ref{cm-sstage-bnn} shows the confusion matrix for the Bayesian model
averaging of artificial neural networks. This methodology produces an average
correct classification rate of 70\%. For comparison, the second best
single stage classifier measured by this figure is the 3-dependent Bayesian
classifier with an overall rate of success of 66\%. 

According to the t-test run applied to the ten sorted runs of 10-fold cross
validation, the probability of finding this difference under the null hypothesis
is below 0.05\%. However, this difference (equivalent to 73 more instances
classified correctly by the ensemble of neural networks) has to be put into the
context of a more demanding computational requirement of the method (several
hours training time in a single 2.66 GHz processor) compared to the almost
instantaneous search for the Bayesian network. For comparison, the classical
C4.5 algorithm \citep{quinlan93} attains only slightly worse performances
(averages of 65.2) at the expense of a more costly determination of the optimal
parameters and greater variance with respect to the training sample.

Support Vector Machines obtain much poorer results (of the order of 50\% correct
identifications). We searched the parameter space as closely as possible
given the computational needs of a cross validation experiment with 30
classes. The best combination found is not able to compete with other
approaches. It is always possible that we missed an island of particularly good
performance in the grid search but the most plausible explanation for the
seemingly poor results is that SVMs are not optimized for multiclass
problems. These are typically dealt with by reducing them to many two-class
problems, but most implementations assume a common value of the parameters
(complexity and radial basis exponent in our case) for all boundaries.

\begin{sidewaystable*}
\begin{minipage}[c][220mm]{\textwidth}
\caption{The confusion matrix for the Bayesian model averaging of
artificial neural networks. The last but one line lists the total
number of light curves (TOT) to define every class. The last line
lists the correct classification rate (CC) for every class separately
as measured by 10 fold cross validation.}
\label{cm-sstage-bnn}
\tiny
\renewcommand{\tabcolsep}{1.1mm}
\renewcommand{\arraystretch}{1.7}
\begin{tabular}{l||c|c|c|c|c|c|c|c|c|c|c|c|c|c|c|c|c|c|c|c|c|c|c|c|c}
 & MIRA & SR & RVTAU & CLCEP & PTCEP & DMCEP & RRAB & RRC & RRD & DSCUT & LBOO & BCEP & SPB & GDOR & BE & PVSG & CP & WR & TTAU & HAEBE & LBV & ELL & EA & EB & EW\\
\hline
\hline
MIRA &  141 &    8 &    1 &    1 &    0 &    0 &    0 &    0 &    0 &    0 &    0 &    0 &    0 &    0 &    1 &    0 &    0 &    0 &    1 &    1 &    2 &    0 &    0 &    0 &    0 \\
\hline
SR &    3 &   16 &    2 &    0 &    2 &    0 &    0 &    0 &    0 &    1 &    0 &    0 &    0 &    0 &    1 &    7 &    0 &    1 &    4 &    3 &    5 &    0 &    0 &    0 &    0 \\
\hline
RVTAU &    0 &    0 &    4 &    2 &    1 &    0 &    0 &    0 &    0 &    0 &    0 &    0 &    0 &    0 &    0 &    0 &    0 &    0 &    0 &    0 &    0 &    0 &    0 &    1 &    0 \\
\hline
CLCEP &    0 &    1 &    3 &  190 &   18 &    0 &    0 &    0 &    0 &    0 &    0 &    0 &    0 &    0 &    0 &    0 &    0 &    0 &    2 &    1 &    0 &    0 &    0 &    0 &    1 \\
\hline
PTCEP &    1 &    0 &    1 &    0 &    0 &    0 &    0 &    0 &    0 &    0 &    0 &    0 &    0 &    0 &    0 &    0 &    0 &    0 &    1 &    0 &    0 &    0 &    0 &    0 &    0 \\
\hline
DMCEP &    0 &    1 &    0 &    0 &    0 &   93 &    5 &    0 &    0 &    0 &    0 &    0 &    0 &    0 &    0 &    0 &    0 &    0 &    0 &    0 &    1 &    0 &    0 &    1 &    0 \\
\hline
RRAB &    0 &    0 &    1 &    0 &    2 &    0 &  121 &    0 &    0 &    1 &    0 &    0 &    0 &    0 &    0 &    0 &    0 &    0 &    0 &    0 &    0 &    0 &    0 &    0 &    0 \\
\hline
RRC &    0 &    0 &    0 &    0 &    0 &    0 &    0 &   23 &    0 &    4 &    0 &    0 &    0 &    0 &    0 &    0 &    0 &    0 &    0 &    0 &    0 &    0 &    0 &    0 &    0 \\
\hline
RRD &    0 &    0 &    0 &    0 &    0 &    0 &    0 &    0 &   57 &    0 &    0 &    0 &    0 &    0 &    0 &    0 &    0 &    2 &    0 &    0 &    0 &    0 &    0 &    0 &    0 \\
\hline
DSCUT &    0 &    1 &    0 &    0 &    0 &    0 &    3 &    3 &    0 &  105 &    9 &   23 &    0 &    5 &    6 &   10 &   10 &    8 &    0 &    3 &    0 &    2 &    0 &    0 &    2 \\
\hline
LBOO &    0 &    0 &    0 &    0 &    0 &    0 &    0 &    0 &    0 &    2 &    1 &    0 &    0 &    0 &    1 &    0 &    0 &    2 &    0 &    0 &    0 &    0 &    0 &    0 &    0 \\
\hline
BCEP &    0 &    0 &    0 &    0 &    0 &    0 &    0 &    1 &    0 &    7 &    0 &   24 &    1 &    0 &    0 &    0 &    2 &    1 &    0 &    0 &    0 &    0 &    0 &    0 &    0 \\
\hline
SPB &    0 &    0 &    0 &    0 &    0 &    0 &    0 &    0 &    0 &    0 &    0 &    4 &   27 &    2 &    6 &    3 &   12 &    0 &    0 &    1 &    0 &    3 &    0 &    0 &    0 \\
\hline
GDOR &    0 &    0 &    0 &    0 &    0 &    0 &    0 &    0 &    0 &    5 &    1 &    0 &    4 &   15 &    8 &    3 &    1 &    3 &    1 &    1 &    0 &    0 &    0 &    0 &    0 \\
\hline
BE &    0 &    2 &    0 &    0 &    0 &    0 &    0 &    0 &    0 &    0 &    0 &    0 &    4 &    2 &    8 &    7 &    1 &    5 &    0 &    2 &    5 &    0 &    0 &    0 &    0 \\
\hline
PVSG &    0 &    4 &    0 &    0 &    0 &    0 &    0 &    0 &    0 &    1 &    0 &    1 &    2 &    4 &   16 &   16 &    1 &   12 &    2 &    6 &    2 &    0 &    0 &    0 &    0 \\
\hline
CP &    0 &    0 &    0 &    1 &    0 &    0 &    0 &    0 &    0 &    3 &    2 &    4 &    6 &    3 &    4 &   11 &   33 &    5 &    0 &    0 &    0 &    1 &    0 &    0 &    0 \\
\hline
WR &    0 &    1 &    0 &    0 &    0 &    0 &    0 &    0 &    0 &    4 &    0 &    0 &    0 &    3 &    4 &   14 &    1 &   18 &    2 &    0 &    3 &    0 &    1 &    0 &    0 \\
\hline
TTAU &    0 &    2 &    1 &    0 &    1 &    0 &    0 &    0 &    0 &    0 &    0 &    1 &    0 &    0 &    0 &    0 &    0 &    0 &    4 &    0 &    0 &    0 &    0 &    0 &    0 \\
\hline
HAEBE &    0 &    2 &    0 &    0 &    0 &    0 &    0 &    0 &    0 &    0 &    0 &    0 &    0 &    0 &    0 &    1 &    0 &    0 &    0 &    1 &    0 &    0 &    0 &    0 &    0 \\
\hline
LBV &    0 &    1 &    0 &    0 &    0 &    0 &    0 &    0 &    0 &    0 &    0 &    0 &    0 &    0 &    1 &    3 &    0 &    2 &    0 &    0 &    3 &    0 &    0 &    0 &    0 \\
\hline
ELL &    0 &    0 &    0 &    1 &    0 &    0 &    0 &    0 &    0 &    0 &    0 &    0 &    2 &    1 &    0 &    0 &    2 &    0 &    0 &    0 &    0 &   10 &    0 &    0 &    0 \\
\hline
EA &    0 &    1 &    0 &    0 &    0 &    0 &    0 &    0 &    0 &    3 &    0 &    0 &    1 &    0 &    0 &    0 &    0 &    1 &    0 &    0 &    0 &    0 &  152 &   22 &    2 \\
\hline
EB &    0 &    2 &    0 &    0 &    0 &    0 &    0 &    0 &    0 &    0 &    0 &    1 &    0 &    0 &    0 &    1 &    0 &    0 &    0 &    1 &    0 &    0 &   16 &  113 &    9 \\
\hline
EW &    0 &    0 &    0 &    0 &    0 &    2 &    0 &    2 &    0 &    1 &    0 &    0 &    0 &    0 &    0 &    0 &    0 &    1 &    0 &    0 &    0 &    0 &    0 &   10 &   45 \\
\hline
\hline
 TOT & 145 &  42 &  13 & 195 &  24 &  95 & 129 &  29 &  57 & 139 &  13 &  58 &  47 &  35 &  57 &  76 &  63 &  62 &  17 &  21 &  21 &  16 & 169 & 147 &  59 \\
\hline
 CC (\%) & 97.2 & 38.1 & 30.8 & 97.4 &  0.0 & 97.9 & 93.8 & 79.3 & 100.0 & 75.5 &  7.7 & 41.4 & 57.4 & 42.9 & 14.0 & 21.1 & 52.4 & 29.0 & 23.5 &  4.8 & 14.3 & 62.5 & 89.9 & 76.9 & 76.3 \\

\end{tabular}
\end{minipage}
\end{sidewaystable*}

\subsubsection{Sequential classifiers}

One of the most relevant characteristics of the stellar variability
classification problem is the rather high number of classes dealt with. Trying
to construct a single stage classifier for such a large number of different classes
implies a trade off between the optimal values of the model parameters in
different regions of attribute space.  We constructed an optimal multistage
classifier in the perspective of dividing the classification problem into several
stages, during each of which a particular subset of the classes is separated from the
rest.

We have selected four subgroups, one for each of the stages of the
classifier. The choice was based on the internal similarities between instances
in a group (intra cluster distances) and the separations between different
groups. The four groups are eclipsing binaries (EA, EB, EW), Cepheids (CLCEP,
PTCEP, RVTAU, DMCEP), long period variables (MIRA, SR) and RR-Lyrae stars (RRAB,
RRC, RRD). These groups are characterized by having significant entries in the
confusion matrices for elements in each group and small contributions to these
matrices across groups. We have trained sequential classifiers in the sense that
the subsequent classifiers are not trained with the classes identified
first. For example, if the first stage classifier is trained to separate
eclipsing variables from the others, the second classifier will not have
eclipsing variables in its training set. This way, given an instance, we can
construct the complete class probability table as a product of conditional
probabilities.

The experiments consists of performing 10 runs of 10-fold cross validation for
each stage with SVMs, Bayesian $k$-dependent networks and Bayesian neural
network averages. The order in which the groups are filtered is altered in order
to test the 24 possible permutations. Each stage is preceded by a feature
selection process that selects the optimal subset of features for each
particular problem (as opposed to the single feature selection step of single
stage classifiers). The results of the experiments consist of several confusion
matrices of dimension 2 for each 2 class problem, and several other confusion
matrices for the classification of instances within these main groups. These
latter matrices do not depend on the order of the assignment of groups to
stages. With only one exception, all statistical tests were inconclusive in the
sense of not providing enough evidence for the rejection of the null hypothesis
(having a threshold of 99.95\%) that the classifiers have equal performance. The
only exception is the eclipsing binaries classifier, where BAANN clearly
outperforms all other methods. In all other cases the similarities in
performance are remarkable.

Table\,\ref{seq1} shows the BAANN confusion matrices for the different
classification stages, while Tables\,\ref{spec1} and \ref{spec4} show the
corresponding matrices for the internal classification problem of each group and
the remaining classes not assigned to any group. Finally, Table\,\ref{h-baann}
shows the combined confusion matrix constructed by multiplying conditional
probabilities. For example, the probability of an instance being a classical
Cepheid (CLCEP) is the probability of not being an eclipsing binary (first
stage) times the probability of belonging to the Cepheids group (second stage)
times the probability of being a classical Cepheid (specialized classifier).
The average correct classification rate is about 66\% for this classifier.

\begin{table*}
\caption{The confusion matrix for the Bayesian model averaging of artificial
neural networks and the two class problem.  The last but one line lists the
total number of light curves (TOT) to define every class. The last line lists
the correct classification rate (CC) for every class separately as measured by
10-fold cross validation. Separation between: A: eclipsing binaries (ECL) and
all other types; B: Cepheids (CEP) and all other types; C: long period variables
(LPV) and all other types except ECL and CEP; D: RR Lyrae stars (RR) from all
other types except ECL, CEP and LPV.  }
\label{seq1}
\centering
\renewcommand{\arraystretch}{1.2}
\begin{tabular}{lcc|lcc|lcc|lcc}
\hline
\multicolumn{3}{c|}{A} & \multicolumn{3}{c|}{B} & 
\multicolumn{3}{c|}{C} & \multicolumn{3}{c}{D} \\
\hline
  & ECL & REST &  & CEP & REST &  & LPV & REST &   & RR & REST \\
 \hline
 ECL & 374 & 2  & CEP & 302 & 18 &  LPV & 164 & 18 & RR & 205 & 8 \\ 
 REST & 1 & 1377 & REST & 25 & 1034 &  REST & 23 & 847 &  REST & 10 & 642 \\
 TOT & 375 & 1379 & TOT & 327 & 1052 &  TOT & 187 & 865 &  TOT & 215 & 650 \\
 CC (\%) & 99.73 & 99.85 &  CC (\%) & 92.35 & 98.28 &  CC (\%) & 
87.70 & 97.91 &  CC (\%) & 95.34 & 98.76 \\
 \hline
\end{tabular}
\end{table*}

\begin{table*}
\caption{The confusion matrix for the Bayesian model averaging of artificial
neural networks.  The last but one line lists the total number of light curves
(TOT) to define every class. The last line lists the correct classification rate
(CC) for every class separately as measured by 10-fold cross validation.
Separation between: A: Cepheids; B: long period variables; C: RR Lyrae stars.}
\label{spec1}
\centering
\renewcommand{\arraystretch}{1.2}
\begin{tabular}{lcccc|lcc|lccc}
\hline
\multicolumn{5}{c|}{A} & \multicolumn{3}{c|}{B} &
\multicolumn{4}{c}{C}  \\
\hline
 &  CLCEP &  PTCEP &  RVTAU &  DMCEP &  &  MIRA &  SR &  &  RRAB &  RRC &  RRD\\
\hline
CLCEP &  190 &  17 &  2 &  0 & MIRA &  139 &  6 & RRAB &  126 &  5 &  1\\
PTCEP &  4 &  3 &  3 &  0 & SR &  6 &  36 &RRC &  3 &  23 &  0\\
RVTAU &  1 &  2 &  6 &  0 &  TOT  &  145 &  42 &RRD &  0 &  1 &  56\\
DMCEP &  0 &  2 &  2 &  95 &   CC(\%)  &  95.86 &  85.71 & TOT  &  129 &  29 &
57 \\
TOT  &  195 &  24 &  13 &  95 & & & & CC(\%)  &  97.67 &  79.31 &  98.24\\
CC(\%)  &  97.43 &  12.50 &  46.15 &  100.00 & & & & &  & &   \\
\hline
\end{tabular}
\end{table*}


\begin{sidewaystable*}
\begin{minipage}[c][220mm]{\textwidth}
\caption{The confusion matrix for the Bayesian model averaging of artificial
neural networks for the variables not assigned to any group. The last but one
line lists the total number of light curves (TOT) to define every class. The
last line lists the correct classification rate (CC) for every class separately
as measured by 10-fold cross validation.}
\label{spec4}
\centering
\tiny
\renewcommand{\arraystretch}{2}
\begin{tabular}{l||c|c|c|c|c|c|c|c|c|c|c|c|c|c|c|c|c}
  &   PVSG  &   BE  &   BCEP  &   CP  &   DSCUT  &   ELL  &   GDOR  &   HAEBE  &   HMXB  &   LBOO  &   LBV  &   SPB  &   SXPHE  &   TTAU  &   WR  &   FUORI  &   SDBV\\
\hline
\hline PVSG  &   20  &   18  &   1  &   2  &   3  &   0  &   0  &   5  &   4  &   0  &   5  &   2  &   0  &   1  &   15  &   0  &   0\\
\hline BE  &   8  &   5  &   0  &   1  &   0  &   0  &   1  &   3  &   3  &   0  &   3  &   3  &   0  &   0  &   4  &   0  &   0\\
\hline BCEP  &   1  &   0  &   24  &   1  &   9  &   2  &   1  &   0  &   0  &   0  &   0  &   1  &   0  &   0  &   1  &   0  &   0\\
\hline CP  &   11  &   8  &   5  &   34  &   5  &   1  &   4  &   1  &   0  &   2  &   0  &   7  &   0  &   0  &   3  &   0  &   0\\
\hline DSCUT  &   8  &   6  &   22  &   9  &   109  &   1  &   4  &   2  &   0  &   6  &   1  &   0  &   5  &   0  &   8  &   0  &   0\\
\hline ELL  &   0  &   0  &   0  &   2  &   0  &   10  &   0  &   0  &   0  &   0  &   1  &   1  &   0  &   0  &   1  &   0  &   0\\
\hline GDOR  &   3  &   10  &   0  &   1  &   3  &   0  &   16  &   0  &   0  &   0  &   0  &   2  &   0  &   1  &   4  &   0  &   0\\
\hline HAEBE  &   0  &   1  &   0  &   0  &   0  &   0  &   0  &   1  &   0  &   0  &   1  &   0  &   0  &   2  &   0  &   0  &   0\\
\hline HMXB  &   0  &   0  &   0  &   0  &   0  &   0  &   0  &   0  &   0  &   0  &   0  &   0  &   0  &   0  &   0  &   0  &   0\\
\hline LBOO  &   0  &   0  &   0  &   1  &   2  &   0  &   1  &   1  &   0  &   2  &   0  &   0  &   0  &   0  &   0  &   0  &   0\\
\hline LBV  &   6  &   3  &   0  &   0  &   1  &   0  &   0  &   1  &   0  &   0  &   7  &   0  &   0  &   1  &   1  &   2  &   0\\
\hline SPB  &   3  &   5  &   4  &   12  &   0  &   2  &   3  &   1  &   1  &   0  &   0  &   30  &   0  &   1  &   1  &   0  &   0\\
\hline SXPHE  &   0  &   0  &   0  &   0  &   1  &   0  &   0  &   1  &   0  &   0  &   0  &   0  &   1  &   0  &   0  &   0  &   0\\
\hline TTAU  &   2  &   0  &   2  &   0  &   2  &   0  &   0  &   5  &   0  &   0  &   1  &   0  &   0  &   9  &   0  &   0  &   0\\
\hline WR  &   14  &   1  &   0  &   0  &   4  &   0  &   5  &   0  &   1  &   3  &   2  &   1  &   1  &   2  &   24  &   1  &   0\\
\hline FUORI  &   0  &   0  &   0  &   0  &   0  &   0  &   0  &   0  &   0  &   0  &   0  &   0  &   0  &   0  &   0  &   0  &   0\\
\hline SDBV  &   0  &   0  &   0  &   0  &   0  &   0  &   0  &   0  &   0  &   0  &   0  &   0  &   0  &   0  &   0  &   0  &   6\\
\hline \hline TOT   &   76  &   57  &   58  &   63  &   139  &   16  &   35  &   21  &   9  &   13  &   21  &   47  &   7  &   17  &   62  &   3  &   6\\
\hline CC(\%)   &   26.31  &   8.77  &   41.37  &   53.96  &   78.41  &   62.50  &   45.71  &   4.76  &   0  &   15.38  &   33.33  &   63.82  &   14.28  &   52.94  &   38.70  &   0  &   100.00\\

\end{tabular}
\end{minipage}
\end{sidewaystable*}

\begin{sidewaystable*}
\begin{minipage}[c][220mm]{\textwidth}
\caption{The complete confusion matrix for the Bayesian model averaging of
artificial neural networks. The last but one line lists the total number of light
curves (TOT) to define every class. The last line lists the correct
classification rate (CC) for every class separately as measured by 10-fold cross
validation.}
\label{h-baann}
\tiny
\renewcommand{\tabcolsep}{0.6mm}
\renewcommand{\arraystretch}{1.5}
\begin{tabular}{l||c|c|c|c|c|c|c|c|c|c|c|c|c|c|c|c|c|c|c|c|c|c|c|c|c|c|c}
& MIRA & SR & RVTAU & CLCEP & PTCEP & DMCEP & RRAB & RRC & RRD & DSCUT & LBOO & BCEP & SPB & GDOR & BE & PVSG & CP & WR & TTAU & HAEBE & LBV & ELL & ECL & SXPHE & SDBV & FUORI & XB\\ 
\hline\hline
MIRA & 126 & 5 & 3 & 2 & 0 & 0 & 0 & 0 & 0 & 0 & 0 & 0 & 0 & 0 & 1 & 0 & 0 & 1 & 0 & 0 & 2 & 0 & 0 & 0 & 0 & 0 & 0\\ 
\hline SR & 5 & 18 & 7 & 2 & 3 & 0 & 0 & 0 & 0 & 0 & 1 & 1 & 0 & 0 & 7 & 4 & 1 & 3 & 6 & 2 & 6 & 0 & 0 & 0 & 0 & 1 & 1\\ 
\hline RVTAU & 3 & 2 & 0 & 1 & 1 & 0 & 0 & 0 & 0 & 1 & 0 & 0 & 0 & 0 & 0 & 0 & 0 & 0 & 0 & 0 & 0 & 0 & 0 & 0 & 0 & 0 & 0\\ 
\hline CLCEP & 0 & 1 & 1 & 176 & 15 & 1 & 0 & 0 & 0 & 0 & 0 & 0 & 0 & 0 & 0 & 1 & 1 & 0 & 2 & 0 & 0 & 2 & 0 & 0 & 0 & 0 & 0\\ 
\hline PTCEP & 0 & 1 & 1 & 4 & 2 & 0 & 0 & 0 & 0 & 0 & 0 & 1 & 0 & 0 & 0 & 0 & 1 & 0 & 0 & 0 & 0 & 0 & 0 & 0 & 0 & 0 & 0\\ 
\hline DMCEP & 0 & 0 & 0 & 0 & 1 & 86 & 6 & 0 & 0 & 2 & 0 & 0 & 1 & 0 & 0 & 2 & 1 & 2 & 2 & 0 & 1 & 2 & 2 & 1 & 0 & 0 & 0\\ 
\hline RRAB & 0 & 0 & 0 & 0 & 1 & 4 & 112 & 5 & 1 & 2 & 0 & 0 & 0 & 0 & 0 & 0 & 0 & 0 & 0 & 0 & 0 & 0 & 0 & 0 & 0 & 0 & 0\\ 
\hline RRC & 0 & 0 & 0 & 0 & 0 & 0 & 1 & 17 & 0 & 4 & 0 & 1 & 0 & 0 & 0 & 0 & 1 & 0 & 0 & 0 & 0 & 0 & 2 & 0 & 0 & 0 & 0\\ 
\hline RRD & 0 & 0 & 0 & 0 & 0 & 1 & 0 & 0 & 54 & 1 & 0 & 0 & 0 & 0 & 1 & 0 & 0 & 0 & 0 & 0 & 0 & 0 & 0 & 0 & 0 & 0 & 0\\ 
\hline DSCUT & 0 & 1 & 0 & 0 & 0 & 0 & 6 & 4 & 1 & 98 & 5 & 20 & 0 & 4 & 5 & 6 & 8 & 8 & 0 & 2 & 0 & 1 & 1 & 3 & 0 & 0 & 0\\ 
\hline LBOO & 0 & 0 & 0 & 0 & 0 & 0 & 0 & 0 & 0 & 2 & 2 & 0 & 0 & 1 & 0 & 0 & 1 & 0 & 0 & 1 & 0 & 0 & 0 & 0 & 0 & 0 & 0\\ 
\hline BCEP & 0 & 0 & 0 & 0 & 0 & 0 & 0 & 0 & 0 & 8 & 0 & 23 & 1 & 1 & 0 & 1 & 1 & 1 & 0 & 0 & 0 & 2 & 0 & 0 & 0 & 0 & 0\\ 
\hline SPB & 0 & 0 & 0 & 0 & 0 & 0 & 0 & 0 & 0 & 0 & 0 & 4 & 29 & 3 & 4 & 3 & 10 & 0 & 1 & 1 & 0 & 2 & 0 & 0 & 0 & 0 & 1\\ 
\hline GDOR & 0 & 0 & 0 & 0 & 0 & 0 & 0 & 0 & 0 & 3 & 0 & 0 & 2 & 16 & 9 & 3 & 0 & 4 & 1 & 0 & 0 & 0 & 0 & 0 & 0 & 0 & 0\\ 
\hline BE & 0 & 2 & 0 & 0 & 0 & 1 & 0 & 0 & 0 & 0 & 0 & 0 & 3 & 1 & 2 & 7 & 1 & 2 & 0 & 2 & 3 & 0 & 1 & 0 & 0 & 0 & 2\\ 
\hline PVSG & 0 & 1 & 0 & 0 & 0 & 0 & 0 & 0 & 0 & 2 & 0 & 1 & 2 & 0 & 16 & 19 & 2 & 15 & 1 & 5 & 5 & 0 & 0 & 0 & 0 & 0 & 4\\ 
\hline CP & 0 & 1 & 0 & 0 & 0 & 0 & 0 & 0 & 0 & 3 & 2 & 4 & 6 & 4 & 7 & 11 & 28 & 3 & 0 & 1 & 0 & 1 & 0 & 0 & 0 & 0 & 0\\ 
\hline WR & 0 & 0 & 0 & 0 & 0 & 0 & 0 & 0 & 0 & 3 & 3 & 0 & 0 & 5 & 1 & 12 & 0 & 22 & 1 & 0 & 0 & 0 & 0 & 1 & 0 & 1 & 1\\ 
\hline TTAU & 2 & 2 & 1 & 4 & 1 & 0 & 0 & 0 & 0 & 0 & 0 & 0 & 0 & 0 & 0 & 1 & 0 & 0 & 3 & 2 & 0 & 0 & 2 & 0 & 0 & 0 & 0\\ 
\hline HAEBE & 6 & 3 & 0 & 0 & 0 & 0 & 0 & 0 & 0 & 0 & 0 & 0 & 0 & 0 & 1 & 0 & 0 & 0 & 0 & 2 & 1 & 0 & 2 & 0 & 0 & 0 & 0\\ 
\hline LBV & 0 & 3 & 0 & 0 & 0 & 1 & 0 & 0 & 0 & 1 & 0 & 0 & 0 & 0 & 0 & 4 & 0 & 0 & 0 & 0 & 1 & 0 & 0 & 0 & 0 & 1 & 0\\ 
\hline ELL & 0 & 0 & 0 & 0 & 0 & 1 & 0 & 1 & 0 & 0 & 0 & 0 & 1 & 0 & 0 & 0 & 2 & 0 & 0 & 0 & 0 & 6 & 1 & 0 & 0 & 0 & 0\\ 
\hline ECL & 2 & 2 & 0 & 6 & 0 & 0 & 4 & 2 & 1 & 6 & 0 & 3 & 1 & 0 & 3 & 2 & 5 & 1 & 0 & 2 & 1 & 0 & 365 & 0 & 3 & 0 & 0\\ 
\hline SXPHE & 0 & 1 & 0 & 0 & 0 & 0 & 0 & 0 & 0 & 0 & 0 & 0 & 0 & 0 & 0 & 0 & 0 & 0 & 0 & 1 & 0 & 0 & 0 & 2 & 0 & 0 & 0\\ 
\hline SDBV & 0 & 0 & 0 & 0 & 0 & 0 & 0 & 0 & 0 & 0 & 0 & 0 & 0 & 0 & 0 & 0 & 0 & 0 & 0 & 0 & 0 & 0 & 0 & 0 & 3 & 0 & 0\\ 
\hline FUORI & 0 & 0 & 0 & 0 & 0 & 0 & 0 & 0 & 0 & 0 & 0 & 0 & 0 & 0 & 0 & 0 & 0 & 0 & 0 & 0 & 0 & 0 & 0 & 0 & 0 & 0 & 0\\ 
\hline XB & 0 & 0 & 0 & 0 & 0 & 0 & 0 & 0 & 0 & 0 & 0 & 0 & 0 & 0 & 0 & 0 & 0 & 0 & 0 & 0 & 0 & 0 & 0 & 0 & 0 & 0 & 0\\ 
\hline\hline TOT & 144 & 43 & 13 & 195 & 24 & 95 & 129 & 29 & 57 & 136 & 13 & 58 & 46 & 35 & 57 & 76 & 63 & 62 & 17 & 21 & 20 & 16 & 376 & 7 & 6 & 3 & 9\\ 
\hline CC(\%) & 87.5 & 41.86 & 0 & 90.26 & 8.333 & 90.53 & 86.82 & 58.62 & 94.74 & 72.06 & 15.38 & 39.66 & 63.04 & 45.71 & 3.509 & 25 & 44.44 & 35.48 & 17.65 & 9.524 & 5 & 37.5 & 97.07 & 28.57 & 50 & 0 & 0\\

\end{tabular}
\end{minipage}
\end{sidewaystable*}


\section{Conclusions and Future Work}

We presented a uniform description of the most important stellar variability
classes currently known. Our description is based on light curve parameters from
well-known member stars for which high quality data are available. The
parameters are derived using Fourier analysis and harmonic fitting methods, and can be calculated on short timescales. The class descriptions obtained form the
basis for a supervised classification method which produces class probabilities
given a set of time series attributes. It is shown that our class descriptions
are accurate enough to separate the monoperiodic variables, some of the
multiperiodic variables, and eclipsing binaries. An obvious improvement to these
capabilities will come from the addition of color information to the class
definitions. This will be discussed in a subsequent paper, where our methodology
will be applied to the OGLE database.

We obtained overall good classification results. The Gaussian mixture method is
relatively simple and robust, and allows for an easy astrophysical interpretation.
The machine learning algorithms, on the other hand, achieve a lower rate of
misclassifications at the expense of longer training times, reduced
interpretability and a higher level of statistical knowledge of the user.
The following extensions/improvements are planned for the future:
\begin{itemize}
\item Extending the number of variability classes and the number of member stars
when more and better quality light curves become available, e.g.\ those from the
exoplanet field data of the CoRoT mission. In particular, we
will add the classes of stars with transits due to 
exoplanets, main-sequence stars with solar-like
oscillations and stars with magnetic activity.
\item Improve the description of light curves by using methods other than Fourier analysis. Wavelet analysis, e.g., may be more suitable for describing non-periodic variables. Also, additional information derived from the
shape of the power spectrum will be considered.
\item Adapt our codes with the intention to apply them to future large scale
databases, such as those to be assembled by CoRoT, Kepler and Gaia. In
particular, we are now preparing the classification of light curves which will
be measured in the framework of the CoRoT Exoplanet programme.
\item Introduce cost matrices to generate specialized
classifiers. When the goal of a classifier is to generate clean
samples of a given class, it is generally preferred that the
number of false positives be diminished even at the expense of missing some of the
less clear candidates. In these cases, the introduction of cost
matrices in machine learning algorithms allows for the differential
weighting of errors in the training process, resulting in classifiers
specialized in particular tasks.
\end{itemize}
The results presented here are a brief summary of all the experiments and
analyses that were applied to the data. For example, only summaries of the
performance measures of some of the approaches taken were included. The full
statistical analysis and detailed research to characterize the confusion regions
of the classifiers will be published in a specialized journal (Sarro et al., in
preparation). In particular, questions about the general properties of the
subsamples of class $i$ misclassified as $j$, their class probability
distributions and if they constitute a separable subset of class $i$ were not
discussed here, to avoid entering into a highly technical discussion. Given the large
number of classes that constitute this classification problem, it is difficult
to include the answers to these questions for the more than 400 possible
combinations of $i$ and $j$ but, at the same time, they are of paramount
importance for the correct interpretation of classifier results. This analysis
is available from the authors upon request.

It is important to realize that the methodology presented here should be
evaluated in a statistical sense, i.e., one can never be sure that each
individual star is correctly and unambiguously classified. Our method was
specifically designed for databases that are so large that individual inspection
of all the stars is impossible. Of course, such inspection can and should be
done after a first classification with our methods has been achieved, for the
specific class of interest to the user. We also note that our basic methods
described here assume the simplest possible input: single-band photometric time
series. Any additional independent information (color indices or time series,
radial velocity time series, spectral type or $\log g$, etc.) will imply an
improved performance as will be shown in our application to the OGLE database
for which such additional attributes were retrieved through the Virtual
Observatory.  We will judge the performance of the different classifiers
presented here in a future paper by comparing our classifications for the OGLE
stars with published results obtained with extractor-type methods requiring
human interaction.

\begin{acknowledgements}
We are very grateful to the following colleagues for providing us with high
quality light curves: Don Kurtz, Simon Jeffery, Gilles Fontaine, Antonio Kanaan,
Kepler Oliveira, Vik Dhillon, Suzanna Randall, Laure Lef\`{e}vre, Joris De Ridder,
Mario van den Ancker, Bram Acke, Chris Sterken and St\'{e}phane Charpinet. This
work is made possible thanks to support from the Belgian PRODEX programme under
grant PEA C90199 (COROT Mission Data Exploitation II), from the research Council
of Leuven University under grant GOA/2003/04 and from the Spanish Ministerio de
Educaci\'on y Ciencia through grant AYA2004-08067-C03-03 and AYA2005-04286. LMS
wishes to thank M. Garc\'{\i}a for the very useful discussions on the feature
selection problem and JD wishes to thank A. Debosscher for the inspiring discussions.
\end{acknowledgements}

\nocite{Sharma}
\nocite{Jeffery:2005}
\nocite{Aerts:2006psdb}
\nocite{Aerts:2006spb}
\nocite{Lamers}
\nocite{Manfroid:1991}
\nocite{Manfroid:1994}
\nocite{Sterken:1993}
\nocite{Sterken:1995}
\nocite{Debosscher}
\nocite{Kurtz:1989}
\nocite{Kilkenny:1985}
\nocite{Teodorani:1997}
\nocite{Covino:1984}
\nocite{Herbst:1978}
\nocite{Cohen:1976}
\nocite{Breger:1974}
\nocite{Terranegra:1994}
\nocite{VanLeeuwen:1998}
\nocite{VanGenderen:1998a}
\nocite{VanGenderen:1998b}
\nocite{VanGenderen:1999}
\nocite{VanGenderen:1996}
\nocite{VanGenderen:1989a}
\nocite{VanGenderen:1989b}
\nocite{VanGenderen:1989c}
\nocite{Rivinius:2003}
\nocite{Neiner:2002}
\nocite{Neiner:2005}
\nocite{Balona:1995}
\nocite{Vogt:1990}
\nocite{Stankov:2005}
\nocite{Saffe:2005}
\nocite{Kudryavtsev:2003}
\nocite{Romanyuk:1994}
\nocite{Adelman:2002a}
\nocite{Glagolevskij:2003}
\nocite{Glagolevskij:2002}
\nocite{Ziznovsky:1995}
\nocite{Kupka:1994a}
\nocite{Catalano:1998}
\nocite{Zverko:1994}
\nocite{Ryabchikova:1998}
\nocite{Savanov:2001}
\nocite{Adelman:2000}
\nocite{LopezGarcia:1999}
\nocite{Kroll:1993}
\nocite{Shavrina:2001}
\nocite{Adelman:2003}
\nocite{Adelman:1995}
\nocite{Alonso:2003}
\nocite{Bagnulo:2002}
\nocite{North:1995}
\nocite{Adelman:2002b}
\nocite{Adelman:2002c}
\nocite{AlbaceteColombo:2002}
\nocite{Adelman:2000}
\nocite{LopezGarcia:2001}
\nocite{Malanushenko:1994}
\nocite{Kato:1999}
\nocite{Yushchenko:1998}
\nocite{Kupka:1998}
\nocite{Kupka:1994b}
\nocite{LopezGarcia:1994}
\nocite{Catalano:1993}
\nocite{Rodriguez:2000a}
\nocite{Rodriguez:2000b}
\nocite{Morris:1985}
\nocite{Beech:1985}
\nocite{Bakos:1989}
\nocite{Hall:1990}
\nocite{Aslanov:1990}
\nocite{Mantegazza:1995}
\nocite{Henry:1999}
\nocite{Clarke:2005}
\nocite{Abraham:2004}
\nocite{Acke:2005}
\nocite{Sterken:1998}
\nocite{Weis:2003}
\nocite{Lamers:1996}
\nocite{Stothers:1995}
\nocite{Davies:2005}
\nocite{Lamers:1998}
\nocite{Robberto:1998}
\nocite{Stahl:2003}
\nocite{Clark:2005}
\nocite{Guo:2005}
\nocite{McSaveney:2005}
\nocite{Percy:1998}
\nocite{Schmidt:2005}
\nocite{Balog:1997}
\nocite{Vinko:1998}
\nocite{MichalowskaSmak:1965}
\nocite{Demers:1974}
\nocite{Schmidt:2003}
\nocite{LloydEvans:1983}
\nocite{Gonzalez:1996}
\nocite{Petersen:1987}
\nocite{Harris:1985}
\nocite{Elkin:2005}
\nocite{Kurtz:1997a}
\nocite{Handler:2002}
\nocite{Balmforth:2001}
\nocite{Kurtz:2000}
\nocite{Mathys:1996}
\nocite{Cunha:2003}
\nocite{Kurtz:1997b}
\nocite{Baldry:1998}
\nocite{Wozniak:2002}
\nocite{Rao:2005}
\nocite{Gonzalez:1997a}
\nocite{Giridhar:2005a}
\nocite{Zsoldos:1996}
\nocite{DeRuyter:2005}
\nocite{VanWinckel:1998}
\nocite{Shenton:1994a}
\nocite{Maas:2002}
\nocite{Shenton:1994b}
\nocite{Gonzalez:1997b}
\nocite{Giridhar:2005b}
\nocite{Giridhar:1998}
\nocite{Deroo:2005}
\nocite{Gonzalez:1997c}
\nocite{Percy:1997}
\nocite{Cieslinski:2000}
\nocite{Pollard:1997}
\nocite{Fokin:1994}
\nocite{Wahlgren:1992}
\nocite{Shenton:1992}
\nocite{Raveendran:1989}
\nocite{Goldsmith:1987}
\nocite{Evans:1985}
\nocite{Dawson:1979}
\nocite{Preston:1963}
\nocite{Douglas:1996}
\nocite{Cristian:1995}
\nocite{Gal:1995}
\nocite{Mantegazza:1988}
\nocite{Aslan:1976}
\nocite{Simon:1973}
\nocite{Smith:1974}
\nocite{OConnell:1961}
\nocite{McLaughlin:1943}
\nocite{DeCat:2002a}
\nocite{Mathias:2001}
\nocite{Aerts:1999b}
\nocite{Waelkens:1991}
\nocite{North:1994}
\nocite{Chapellier:1996}
\nocite{Clausen:1996}
\nocite{Waelkens:1996}
\nocite{Aerts:1999a}
\nocite{DeRidder:1999}
\nocite{Chapellier:2000}
\nocite{DeCat:2004}
\nocite{DeCat:2002b}
\nocite{Zhou:1999b}
\nocite{Hintz:2004}
\nocite{Blake:2000}
\nocite{Zhou:1999a}
\nocite{Hintz:1997}
\nocite{Nemec:1990}
\nocite{Rodriguez:1993}
\nocite{Rodriguez:1990}
\nocite{Pena:1987}
\nocite{Herbst:1994}
\nocite{Gahm:1975}
\nocite{Weintraub:1989}
\nocite{VanDerHucht:2001}
\nocite{Nugis:2000}
\nocite{Sarro:2006}
\nocite{Eyer:2005a}
\nocite{Eyer:2005b}
\nocite{Eyer:1999}
\nocite{Kepler:2000}
\nocite{Kepler:2003}
\nocite{Lefevre:2005}
\nocite{Andrievsky:2005b}
\nocite{Andrievsky:2005a}
\nocite{Andrievsky:2004}
\nocite{Solano:1997}
\nocite{Pamyatnykh:1998}
\nocite{Breger:1999}
\nocite{Arentoft:1998}
\nocite{Pena:1999}
\nocite{Kiss:2002}
\nocite{Hintz:1998}
\nocite{Kovacs:2000}
\nocite{Eyer:2000}
\nocite{Aerts:1998}
\nocite{Dupret:2005}
\nocite{Kaye:1999}
\nocite{Paunzen:1999}
\nocite{Alvarez:1997}
\nocite{Kotak:2004}
\nocite{Corsico:2001}
\nocite{Mukadam:2004}
\nocite{Bergeron:2004}
\nocite{Fontaine:2003}
\nocite{Kleinman:1998}
\nocite{Metcalfe:2004}
\nocite{Wesemael:1986}
\nocite{Clemens:1993}
\nocite{Vauclair:1992}
\nocite{Daou:1990}
\nocite{Handler:2003}
\nocite{Koester:1983}
\nocite{Handler:2001}
\nocite{Fontaine:1991}
\nocite{Oreiro:2005}
\nocite{Sandage:1993}
\nocite{Bessell:1974}
\nocite{McNamara:1996a}
\nocite{McNamara:1996b}
\nocite{Hintz:2004}
\nocite{Rodriguez:2002}
\nocite{Zhou:1999}
\nocite{Hintz:1997a}
\nocite{Hintz:1997b}
\nocite{Nemec:1990}
\nocite{McNamara:1997a}
\nocite{McNamara:1997b}
\nocite{McNamara:1995}
\nocite{Rodriguez:1993}
\nocite{Quirion:2006}

\bibliographystyle{aa}
\bibliography{references-code}

\begin{thebibliography}{245}
\expandafter\ifx\csname natexlab\endcsname\relax\def\natexlab#1{#1}\fi

\bibitem[{{{\'A}brah{\'a}m} {et~al.}(2004){{\'A}brah{\'a}m}, {K{\'o}sp{\'a}l},
  {Csizmadia}, {Kun}, {Mo{\'o}r}, \& {Prusti}}]{Abraham:2004}
{{\'A}brah{\'a}m}, P., {K{\'o}sp{\'a}l}, {\'A}., {Csizmadia}, S., {et~al.}
  2004, \aap, 428, 89 [FUORI]

\bibitem[{{Acke}(2005)}]{Acke:2005}
{Acke}, B. 2005, Ph.D.~Thesis, Catholic University of Leuven, Belgium [HAEBE]

\bibitem[{{Adelman}(2000)}]{Adelman:2000}
{Adelman}, S.~J. 2000, \aaps, 146, 13 [CP]

\bibitem[{{Adelman}(2002)}]{Adelman:2002a}
{Adelman}, S.~J. 2002, Baltic Astronomy, 11, 475 [CP]

\bibitem[{{Adelman}(2003)}]{Adelman:2003}
{Adelman}, S.~J. 2003, \aap, 401, 357 [CP]

\bibitem[{{Adelman} \& {Meadows}(2002{\natexlab{a}})}]{Adelman:2002b}
{Adelman}, S.~J. \& {Meadows}, S.~A. 2002{\natexlab{a}}, \aap, 390, 1023 [CP]

\bibitem[{{Adelman} \& {Meadows}(2002{\natexlab{b}})}]{Adelman:2002c}
{Adelman}, S.~J. \& {Meadows}, S.~A. 2002{\natexlab{b}}, \aap, 390, 1023 [CP]

\bibitem[{{Adelman} {et~al.}(1995){Adelman}, {Pyper}, {Lopez-Garcia}, \&
  {Caliskan}}]{Adelman:1995}
{Adelman}, S.~J., {Pyper}, D.~M., {Lopez-Garcia}, Z., \& {Caliskan}, H. 1995,
  \aap, 296, 467 [CP]

\bibitem[{{Aerts} {et~al.}(1999{\natexlab{a}}){Aerts}, {De Boeck}, {Malfait},
  \& {De Cat}}]{Aerts:1999b}
{Aerts}, C., {De Boeck}, I., {Malfait}, K., \& {De Cat}, P. 1999{\natexlab{a}},
  \aap, 347, 524 [SPB]

\bibitem[{{Aerts} {et~al.}(2006{\natexlab{a}}){Aerts}, {De Cat}, {Kuschnig},
  {Matthews}, {Guenther}, {Moffat}, {Rucinski}, {Sasselov}, {Walker}, \&
  {Weiss}}]{Aerts:2006spb}
{Aerts}, C., {De Cat}, P., {Kuschnig}, R., {et~al.} 2006{\natexlab{a}}, \apjl,
  642, L165 [SPB]

\bibitem[{{Aerts} {et~al.}(1999{\natexlab{b}}){Aerts}, {De Cat}, {Peeters},
  {Decin}, {De Ridder}, {Kolenberg}, {Meeus}, {Van Winckel}, {Cuypers}, \&
  {Waelkens}}]{Aerts:1999a}
{Aerts}, C., {De Cat}, P., {Peeters}, E., {et~al.} 1999{\natexlab{b}}, \aap,
  343, 872 [SPB]

\bibitem[{{Aerts} {et~al.}(1998){Aerts}, {Eyer}, \& {Kestens}}]{Aerts:1998}
{Aerts}, C., {Eyer}, L., \& {Kestens}, E. 1998, \aap, 337, 790 [GDOR]

\bibitem[{{Aerts} {et~al.}(2006{\natexlab{b}}){Aerts}, {Jeffery}, {Fontaine},
  {Dhillon}, {Marsh}, \& {Groot}}]{Aerts:2006psdb}
{Aerts}, C., {Jeffery}, C.~S., {Fontaine}, G., {et~al.} 2006{\natexlab{b}},
  \mnras, 367, 1317 [SDBV]

\bibitem[{{Albacete-Colombo} {et~al.}(2002){Albacete-Colombo},
  {L{\'o}pez-Garc{\'{\i}}a}, {Levato}, {Malaroda}, \&
  {Grosso}}]{AlbaceteColombo:2002}
{Albacete-Colombo}, J.~F., {L{\'o}pez-Garc{\'{\i}}a}, Z., {Levato}, H.,
  {Malaroda}, S.~M., \& {Grosso}, M. 2002, \aap, 392, 613 [CP]

\bibitem[{{Alonso} {et~al.}(2003){Alonso}, {L{\'o}pez-Garc{\'{\i}}a},
  {Malaroda}, \& {Leone}}]{Alonso:2003}
{Alonso}, M.~S., {L{\'o}pez-Garc{\'{\i}}a}, Z., {Malaroda}, S., \& {Leone}, F.
  2003, \aap, 402, 331 [CP]

\bibitem[{{Alvarez} \& {Mennessier}(1997)}]{Alvarez:1997}
{Alvarez}, R. \& {Mennessier}, M.-O. 1997, \aap, 317, 761 [MIRA]

\bibitem[{{Andrievsky} {et~al.}(2005){Andrievsky}, {Luck}, \&
  {Kovtyukh}}]{Andrievsky:2005b}
{Andrievsky}, S.~M., {Luck}, R.~E., \& {Kovtyukh}, V.~V. 2005, \aj, 130, 1880
  [CLCEP]

\bibitem[{{Arentoft} {et~al.}(1998){Arentoft}, {Kjeldsen}, {Nuspl}, {Bedding},
  {Fronto}, {Viskum}, {Frandsen}, \& {Belmonte}}]{Arentoft:1998}
{Arentoft}, T., {Kjeldsen}, H., {Nuspl}, J., {et~al.} 1998, \aap, 338, 909
  [DSCUT]

\bibitem[{{Aslan}(1976)}]{Aslan:1976}
{Aslan}, Z. 1976, The Observatory, 96, 149 [SR]

\bibitem[{{Aslanov} \& {Khruzina}(1990)}]{Aslanov:1990}
{Aslanov}, A.~A. \& {Khruzina}, T.~S. 1990, Soviet Astronomy, 34, 508 [ELL]

\bibitem[{{Bagnulo} {et~al.}(2002){Bagnulo}, {Landi Degl'Innocenti},
  {Landolfi}, \& {Mathys}}]{Bagnulo:2002}
{Bagnulo}, S., {Landi Degl'Innocenti}, M., {Landolfi}, M., \& {Mathys}, G.
  2002, \aap, 394, 1023 [CP]

\bibitem[{{Bakos} \& {Tremko}(1989)}]{Bakos:1989}
{Bakos}, G.~A. \& {Tremko}, J. 1989, Contributions of the Astronomical
  Observatory Skalnate Pleso, 18, 17 [ELL]

\bibitem[{{Baldry} {et~al.}(1998){Baldry}, {Kurtz}, \& {Bedding}}]{Baldry:1998}
{Baldry}, I.~K., {Kurtz}, D.~W., \& {Bedding}, T.~R. 1998, \mnras, 300, L39
  [ROAP]

\bibitem[{{Balmforth} {et~al.}(2001){Balmforth}, {Cunha}, {Dolez}, {Gough}, \&
  {Vauclair}}]{Balmforth:2001}
{Balmforth}, N.~J., {Cunha}, M.~S., {Dolez}, N., {Gough}, D.~O., \& {Vauclair},
  S. 2001, \mnras, 323, 362 [ROAP]

\bibitem[{{Balog} {et~al.}(1997){Balog}, {Vinko}, \& {Kaszas}}]{Balog:1997}
{Balog}, Z., {Vinko}, J., \& {Kaszas}, G. 1997, \aj, 113, 1833 [PTCEP]

\bibitem[{{Balona}(1995)}]{Balona:1995}
{Balona}, L.~A. 1995, \mnras, 277, 1547 [BE]

\bibitem[{{Beech}(1985)}]{Beech:1985}
{Beech}, M. 1985, \apss, 117, 69 [ELL]

\bibitem[{{Bergeron} {et~al.}(2004){Bergeron}, {Fontaine}, {Bill{\`e}res},
  {Boudreault}, \& {Green}}]{Bergeron:2004}
{Bergeron}, P., {Fontaine}, G., {Bill{\`e}res}, M., {Boudreault}, S., \&
  {Green}, E.~M. 2004, \apj, 600, 404 [DAV]

\bibitem[{{Bessell}(1974)}]{Bessell:1974}
{Bessell}, M.~S. 1974, in IAU Symp. 59: Stellar Instability and Evolution, ed.
  P.~{Ledoux}, A.~{Noels}, \& A.~W. {Rodgers}, 63 [DSCUT, RR]

\bibitem[{Bishop(1995)}]{Bishop}
Bishop, C.~M. 1995, Neural Networks for Pattern Recognition (New York, NY, USA:
  Oxford University Press, Inc.)

\bibitem[{{Blake} {et~al.}(2000){Blake}, {Khosravani}, \&
  {Delaney}}]{Blake:2000}
{Blake}, R.~M., {Khosravani}, H., \& {Delaney}, P.~A. 2000, \jrasc, 94, 124
  [SXPHE]

\bibitem[{Bouckaert(2004)}]{Bouckaert04}
Bouckaert, R.~R. 2004, in ICML '04: Proceedings of the twenty-first
  international conference on Machine learning (New York, NY, USA: ACM Press),
  15

\bibitem[{{Breger}(1974)}]{Breger:1974}
{Breger}, M. 1974, \apj, 188, 53 [HAEBE]

\bibitem[{{Breger} {et~al.}(1999){Breger}, {Pamyatnykh}, {Pikall}, \&
  {Garrido}}]{Breger:1999}
{Breger}, M., {Pamyatnykh}, A.~A., {Pikall}, H., \& {Garrido}, R. 1999, \aap,
  341, 151 [DSCUT]

\bibitem[{{Catalano} {et~al.}(1993){Catalano}, {Leone}, \&
  {Kroll}}]{Catalano:1993}
{Catalano}, F.~A., {Leone}, F., \& {Kroll}, R. 1993, in ASP Conf. Ser. 44: IAU
  Colloq. 138: Peculiar versus Normal Phenomena in A-type and Related Stars,
  ed. M.~M. {Dworetsky}, F.~{Castelli}, \& R.~{Faraggiana}, 605 [CP]

\bibitem[{{Catalano} {et~al.}(1998){Catalano}, {Leone}, \&
  {Kroll}}]{Catalano:1998}
{Catalano}, F.~A., {Leone}, F., \& {Kroll}, R. 1998, \aaps, 129, 463 [CP]

\bibitem[{{Chapellier} {et~al.}(2000){Chapellier}, {Mathias}, {Le Contel},
  {Garrido}, {Le Contel}, \& {Valtier}}]{Chapellier:2000}
{Chapellier}, E., {Mathias}, P., {Le Contel}, J.-M., {et~al.} 2000, \aap, 362,
  189 [SPB]

\bibitem[{{Chapellier} {et~al.}(1996){Chapellier}, {Sadsaoud}, {Valtier},
  {Mathias}, {Garrido}, {Alvarez}, {Sareyan}, {Chauville}, \& {Le
  Contel}}]{Chapellier:1996}
{Chapellier}, E., {Sadsaoud}, H., {Valtier}, J.~C., {et~al.} 1996, \aap, 307,
  91 [SPB]

\bibitem[{{Cieslinski} {et~al.}(2000){Cieslinski}, {Steiner}, {Jablonski}, \&
  {Hickel}}]{Cieslinski:2000}
{Cieslinski}, D., {Steiner}, J.~E., {Jablonski}, F.~J., \& {Hickel}, G.~R.
  2000, \pasp, 112, 642 [RVTAU]

\bibitem[{{Clark} {et~al.}(2005){Clark}, {Larionov}, \&
  {Arkharov}}]{Clark:2005}
{Clark}, J.~S., {Larionov}, V.~M., \& {Arkharov}, A. 2005, \aap, 435, 239 [LBV]

\bibitem[{{Clarke} {et~al.}(2005){Clarke}, {Lodato}, {Melnikov}, \&
  {Ibrahimov}}]{Clarke:2005}
{Clarke}, C., {Lodato}, G., {Melnikov}, S.~Y., \& {Ibrahimov}, M.~A. 2005,
  \mnras, 361, 942 [FUORI]

\bibitem[{{Clausen}(1996)}]{Clausen:1996}
{Clausen}, J.~V. 1996, \aap, 308, 151 [SPB]

\bibitem[{{Clemens}(1993)}]{Clemens:1993}
{Clemens}, J.~C. 1993, Baltic Astronomy, 2, 407 [DAV]

\bibitem[{{Cohen} \& {Schwartz}(1976)}]{Cohen:1976}
{Cohen}, M. \& {Schwartz}, R.~D. 1976, \mnras, 174, 137 [TTAU]

\bibitem[{Cooper \& Herskovits(1992)}]{K2-CH92}
Cooper, G.~F. \& Herskovits, E. 1992, Mach. Learn., 9, 309

\bibitem[{{C{\'o}rsico} {et~al.}(2001){C{\'o}rsico}, {Althaus}, {Benvenuto}, \&
  {Serenelli}}]{Corsico:2001}
{C{\'o}rsico}, A.~H., {Althaus}, L.~G., {Benvenuto}, O.~G., \& {Serenelli},
  A.~M. 2001, \aap, 380, L17 [DAV]

\bibitem[{{Covino} {et~al.}(1984){Covino}, {Terranegra}, {Vittone}, \&
  {Russo}}]{Covino:1984}
{Covino}, E., {Terranegra}, L., {Vittone}, A.~A., \& {Russo}, G. 1984, \aj, 89,
  1868 [HAEBE]

\bibitem[{{Cristian} {et~al.}(1995){Cristian}, {Donahue}, {Soon}, {Baliunas},
  \& {Henry}}]{Cristian:1995}
{Cristian}, V.-C., {Donahue}, R.~A., {Soon}, W.~H., {Baliunas}, S.~L., \&
  {Henry}, G.~W. 1995, \pasp, 107, 411 [SR]

\bibitem[{{Cunha} {et~al.}(2003){Cunha}, {Fernandes}, \&
  {Monteiro}}]{Cunha:2003}
{Cunha}, M.~S., {Fernandes}, J.~M.~M.~B., \& {Monteiro}, M.~J.~P.~F.~G. 2003,
  \mnras, 343, 831 [ROAP]

\bibitem[{{Daou} {et~al.}(1990){Daou}, {Wesemael}, {Fontaine}, {Bergeron}, \&
  {Holberg}}]{Daou:1990}
{Daou}, D., {Wesemael}, F., {Fontaine}, G., {Bergeron}, P., \& {Holberg}, J.~B.
  1990, \apj, 364, 242 [DAV]

\bibitem[{{Davies} {et~al.}(2005){Davies}, {Oudmaijer}, \&
  {Vink}}]{Davies:2005}
{Davies}, B., {Oudmaijer}, R.~D., \& {Vink}, J.~S. 2005, \aap, 439, 1107 [LBV]

\bibitem[{{Dawson}(1979)}]{Dawson:1979}
{Dawson}, D.~W. 1979, \apjs, 41, 97 [RVTAU]

\bibitem[{{De Cat}(2002)}]{DeCat:2002b}
{De Cat}, P. 2002, in ASP Conf. Ser. 259: IAU Colloq. 185: Radial and Nonradial
  Pulsationsn as Probes of Stellar Physics, ed. C.~{Aerts}, T.~R. {Bedding}, \&
  J.~{Christensen-Dalsgaard}, 196 [SPB, BCEP]

\bibitem[{{De Cat} \& {Aerts}(2002)}]{DeCat:2002a}
{De Cat}, P. \& {Aerts}, C. 2002, VizieR Online Data Catalog, 339, 30965 [SPB]

\bibitem[{{De Cat} {et~al.}(2004){De Cat}, {de Ridder}, {Uytterhoeven},
  {Davignon}, {Raskin}, {Cuypers}, {Schoenaers}, {Daszy{\'n}ska-Daszkiewicz},
  {Aerts}, {van Winckel}, {Ausseloos}, {Broeders}, {de Meester},
  {Vanautgaerden}, {van Malderen}, {Vandenbussche}, {Acke}, {Decin}, {Decin},
  {Kolenberg}, {Maas}, {De Ruyter}, {Reyniers}, {Reyniers}, {van Kerckhoven},
  \& {Waelkens}}]{DeCat:2004}
{De Cat}, P., {de Ridder}, J., {Uytterhoeven}, K., {et~al.} 2004, in ASP Conf.
  Ser. 310: IAU Colloq. 193: Variable Stars in the Local Group, ed. D.~W.
  {Kurtz} \& K.~R. {Pollard}, 238 [SPB]

\bibitem[{{De Ridder} {et~al.}(1999){De Ridder}, {Gordon}, {Mulliss}, \&
  {Aerts}}]{DeRidder:1999}
{De Ridder}, J., {Gordon}, K.~D., {Mulliss}, C.~L., \& {Aerts}, C. 1999, \aap,
  341, 574 [SPB]

\bibitem[{{De Ruyter} {et~al.}(2005){De Ruyter}, {van Winckel}, {Dominik},
  {Waters}, \& {Dejonghe}}]{DeRuyter:2005}
{De Ruyter}, S., {van Winckel}, H., {Dominik}, C., {Waters}, L.~B.~F.~M., \&
  {Dejonghe}, H. 2005, \aap, 435, 161 [RVTAU]

\bibitem[{{Debosscher} {et~al.}(2006){Debosscher}, {Aerts}, \&
  {Vandenbussche}}]{Debosscher}
{Debosscher}, J., {Aerts}, C., \& {Vandenbussche}, B. 2006, in Astronomical
  Society of the Pacific Conference Series, ed. C.~{Sterken} \& C.~{Aerts}, 219

\bibitem[{{Demers} \& {Harris}(1974)}]{Demers:1974}
{Demers}, S. \& {Harris}, W.~E. 1974, \aj, 79, 627 [PTCEP]

\bibitem[{Demsar(2006)}]{Demsar06}
Demsar, J. 2006, JMLR, 7, 1

\bibitem[{{Deroo} {et~al.}(2005){Deroo}, {Reyniers}, {van Winckel}, {Goriely},
  \& {Siess}}]{Deroo:2005}
{Deroo}, P., {Reyniers}, M., {van Winckel}, H., {Goriely}, S., \& {Siess}, L.
  2005, \aap, 438, 987 [RVTAU]

\bibitem[{{Dhillon} \& {Marsh}(2001)}]{Dhillon:2001}
{Dhillon}, V. \& {Marsh}, T. 2001, New Astronomy Review, 45, 91

\bibitem[{{Douglas} \& {Henson}(1996)}]{Douglas:1996}
{Douglas}, P. \& {Henson}, G.~D. 1996, International Amateur-Professional
  Photoelectric Photometry Communications, 64, 51 [SR]

\bibitem[{{Dupret} {et~al.}(2005){Dupret}, {Grigahc{\`e}ne}, {Garrido}, {De
  Ridder}, {Scuflaire}, \& {Gabriel}}]{Dupret:2005}
{Dupret}, M.-A., {Grigahc{\`e}ne}, A., {Garrido}, R., {et~al.} 2005, \mnras,
  360, 1143 [GDOR]

\bibitem[{{Elkin} {et~al.}(2005){Elkin}, {Riley}, {Cunha}, {Kurtz}, \&
  {Mathys}}]{Elkin:2005}
{Elkin}, V.~G., {Riley}, J.~D., {Cunha}, M.~S., {Kurtz}, D.~W., \& {Mathys}, G.
  2005, \mnras, 358, 665 [ROAP]

\bibitem[{{ESA}(1997)}]{ESA:1997}
{ESA}. 1997, VizieR Online Data Catalog, 1239, 0

\bibitem[{{Evans}(1985)}]{Evans:1985}
{Evans}, T.~L. 1985, \mnras, 217, 493 [RVTAU]

\bibitem[{{Eyer}(2005)}]{Eyer:2005a}
{Eyer}, L. 2005, in ESA SP-576: The Three-Dimensional Universe with Gaia, ed.
  C.~{Turon}, K.~S. {O'Flaherty}, \& M.~A.~C. {Perryman}, 513

\bibitem[{{Eyer} \& {Aerts}(2000)}]{Eyer:2000}
{Eyer}, L. \& {Aerts}, C. 2000, \aap, 361, 201 [GDOR]

\bibitem[{{Eyer} \& {Bartholdi}(1999)}]{Eyer:1999}
{Eyer}, L. \& {Bartholdi}, P. 1999, \aaps, 135, 1

\bibitem[{{Eyer} \& {Blake}(2005)}]{Eyer:2005b}
{Eyer}, L. \& {Blake}, C. 2005, \mnras, 358, 30

\bibitem[{Fayyad \& Irani(1993)}]{Fayyadirani93}
Fayyad \& Irani. 1993, in Multi-Interval Discretization of Continuous-Valued
  Attributes for Classification Learning, 1022--1027

\bibitem[{{Fokin}(1994)}]{Fokin:1994}
{Fokin}, A.~B. 1994, \aap, 292, 133 [RVTAU]

\bibitem[{{Fontaine} {et~al.}(2003){Fontaine}, {Bergeron}, {Bill{\`e}res}, \&
  {Charpinet}}]{Fontaine:2003}
{Fontaine}, G., {Bergeron}, P., {Bill{\`e}res}, M., \& {Charpinet}, S. 2003,
  \apj, 591, 1184 [DAV]

\bibitem[{{Fontaine} {et~al.}(1991){Fontaine}, {Bergeron}, {Brassard},
  {Wesemael}, {Vauclair}, {Kawaler}, {Grauer}, \& {Winget}}]{Fontaine:1991}
{Fontaine}, G., {Bergeron}, P., {Brassard}, P., {et~al.} 1991, \apjl, 378, L49
  [GWVIR]

\bibitem[{{Gahm}(1975)}]{Gahm:1975}
{Gahm}, G.~F. 1975, in IAU Symp. 67: Variable Stars and Stellar Evolution, ed.
  V.~E. {Sherwood} \& L.~{Plaut}, 101 [TTAU, FUORI]

\bibitem[{{G{\'a}l} \& {Szatm{\'a}ry}(1995)}]{Gal:1995}
{G{\'a}l}, J. \& {Szatm{\'a}ry}, K. 1995, International Amateur-Professional
  Photoelectric Photometry Communications, 59, 30 [SR]

\bibitem[{{Giridhar} {et~al.}(1998){Giridhar}, {Lambert}, \&
  {Gonzalez}}]{Giridhar:1998}
{Giridhar}, S., {Lambert}, D.~L., \& {Gonzalez}, G. 1998, \apj, 509, 366
  [RVTAU]

\bibitem[{{Giridhar} {et~al.}(2005{\natexlab{a}}){Giridhar}, {Lambert},
  {Reddy}, {Gonzalez}, \& {Yong}}]{Giridhar:2005a}
{Giridhar}, S., {Lambert}, D.~L., {Reddy}, B.~E., {Gonzalez}, G., \& {Yong}, D.
  2005{\natexlab{a}}, \apj, 627, 432 [RVTAU]

\bibitem[{{Giridhar} {et~al.}(2005{\natexlab{b}}){Giridhar}, {Lambert},
  {Reddy}, {Gonzalez}, \& {Yong}}]{Giridhar:2005b}
{Giridhar}, S., {Lambert}, D.~L., {Reddy}, B.~E., {Gonzalez}, G., \& {Yong}, D.
  2005{\natexlab{b}}, \apj, 627, 432 [RVTAU]

\bibitem[{{Glagolevskii} \& {Chuntonov}(2002)}]{Glagolevskij:2002}
{Glagolevskii}, Y.~V. \& {Chuntonov}, G.~A. 2002, Astrophysics, 45, 408 [CP]

\bibitem[{{Glagolevskij}(2003)}]{Glagolevskij:2003}
{Glagolevskij}, Y.~V. 2003, Astrophysics, 46, 319 [CP]

\bibitem[{{Goldsmith} {et~al.}(1987){Goldsmith}, {Evans}, {Albinson}, \&
  {Bode}}]{Goldsmith:1987}
{Goldsmith}, M.~J., {Evans}, A., {Albinson}, J.~S., \& {Bode}, M.~F. 1987,
  \mnras, 227, 143 [RVTAU]

\bibitem[{{Gonzalez} {et~al.}(1997{\natexlab{a}}){Gonzalez}, {Lambert}, \&
  {Giridhar}}]{Gonzalez:1997a}
{Gonzalez}, G., {Lambert}, D.~L., \& {Giridhar}, S. 1997{\natexlab{a}}, \apj,
  481, 452 [RVTAU]

\bibitem[{{Gonzalez} {et~al.}(1997{\natexlab{b}}){Gonzalez}, {Lambert}, \&
  {Giridhar}}]{Gonzalez:1997b}
{Gonzalez}, G., {Lambert}, D.~L., \& {Giridhar}, S. 1997{\natexlab{b}}, \apj,
  481, 452 [RVTAU]

\bibitem[{{Gonzalez} {et~al.}(1997{\natexlab{c}}){Gonzalez}, {Lambert}, \&
  {Giridhar}}]{Gonzalez:1997c}
{Gonzalez}, G., {Lambert}, D.~L., \& {Giridhar}, S. 1997{\natexlab{c}}, \apj,
  481, 452 [RVTAU]

\bibitem[{{Gonzalez} \& {Wallerstein}(1996)}]{Gonzalez:1996}
{Gonzalez}, G. \& {Wallerstein}, G. 1996, \mnras, 280, 515 [PTCEP]

\bibitem[{{Grankin} {et~al.}(2007){Grankin}, {Melnikov}, {Bouvier}, {Herbst},
  \& {Shevchenko}}]{ROTOR}
{Grankin}, K.~N., {Melnikov}, S.~Y., {Bouvier}, J., {Herbst}, W., \&
  {Shevchenko}, V.~S. 2007, \aap, 461, 183 [TTAU]

\bibitem[{Gunn {et~al.}(1997)Gunn, Brown, \& Bossley}]{SRM}
Gunn, S.~R., Brown, M., \& Bossley, K.~M. 1997, in IDA '97: Proceedings of the
  Second International Symposium on Advances in Intelligent Data Analysis,
  Reasoning about Data (London, UK: Springer-Verlag), 313--323

\bibitem[{{Guo} {et~al.}(2005){Guo}, {Li}, \& {Shan}}]{Guo:2005}
{Guo}, J.-H., {Li}, Y., \& {Shan}, H.-G. 2005, Chinese Journal of Astronony and
  Astrophysics, 5, 245 [LBV]

\bibitem[{Guyon \& Elisseeff(2003)}]{FS-JMLR}
Guyon, I. \& Elisseeff, A. 2003, J. Mach. Learn. Res., 3, 1157

\bibitem[{{Hall}(1990)}]{Hall:1990}
{Hall}, D.~S. 1990, \aj, 100, 554 [ELL]

\bibitem[{{Handler}(2001)}]{Handler:2001}
{Handler}, G. 2001, \mnras, 323, L43 [DBV]

\bibitem[{{Handler} {et~al.}(2003){Handler}, {O'Donoghue}, {M{\"u}ller},
  {Solheim}, {Gonzalez-Perez}, {Johannessen}, {Paparo}, {Szeidl}, {Viraghalmy},
  {Silvotti}, {Vauclair}, {Dolez}, {Pallier}, {Chevreton}, {Kurtz}, {Bromage},
  {Cunha}, {{\O}stensen}, {Fraga}, {Kanaan}, {Amorim}, {Giovannini}, {Kepler},
  {da Costa}, {Anderson}, {Wood}, {Silvestri}, {Klumpe}, {Carlton}, {Miller},
  {McFarland}, {Grauer}, {Kawaler}, {Riddle}, {Reed}, {Nather}, {Winget},
  {Hill}, {Metcalfe}, {Mukadam}, {Kilic}, {Watson}, {Kleinman}, {Nitta},
  {Guzik}, {Bradley}, {Sekiguchi}, {Sullivan}, {Sullivan}, {Shobbrook},
  {Jiang}, {Birch}, {Ashoka}, {Seetha}, {Girish}, {Joshi}, {Dorokhova},
  {Dorokhov}, {Akan}, {Mei{\v s}tas}, {Janulis}, {Kalytis}, {Ali{\v s}auskas},
  {Anguma}, {Kalebwe}, {Moskalik}, {Ogloza}, {Stachowski}, {Pajdosz}, \&
  {Zola}}]{Handler:2003}
{Handler}, G., {O'Donoghue}, D., {M{\"u}ller}, M., {et~al.} 2003, \mnras, 340,
  1031 [DBV]

\bibitem[{{Handler} {et~al.}(2002){Handler}, {Weiss}, {Paunzen}, {Shobbrook},
  {Garrido}, {Guzik}, {Hempel}, {Moalusi}, {Beach}, {Medupe}, {Chagnon},
  {Matthews}, {Reegen}, \& {Granzer}}]{Handler:2002}
{Handler}, G., {Weiss}, W.~W., {Paunzen}, E., {et~al.} 2002, \mnras, 330, 153
  [ROAP]

\bibitem[{{Harris}(1985)}]{Harris:1985}
{Harris}, H.~C. 1985, \aj, 90, 756 [PTCEP]

\bibitem[{{Henry} \& {Kaye}(1999)}]{Henry:1999}
{Henry}, G.~W. \& {Kaye}, A.~B. 1999, Informational Bulletin on Variable Stars,
  4684, 1 [ELL]

\bibitem[{{Herbst} {et~al.}(1994){Herbst}, {Herbst}, {Grossman}, \&
  {Weinstein}}]{Herbst:1994}
{Herbst}, W., {Herbst}, D.~K., {Grossman}, E.~J., \& {Weinstein}, D. 1994, \aj,
  108, 1906 [TTAU]

\bibitem[{{Herbst} {et~al.}(1978){Herbst}, {Racine}, \& {Warner}}]{Herbst:1978}
{Herbst}, W., {Racine}, R., \& {Warner}, J.~W. 1978, \apj, 223, 471 [HAEBE]

\bibitem[{{Hintz} {et~al.}(1997{\natexlab{a}}){Hintz}, {Hintz}, \&
  {Joner}}]{Hintz:1997}
{Hintz}, E., {Hintz}, M.~L., \& {Joner}, M.~D. 1997{\natexlab{a}}, \pasp, 109,
  1073 [SXPHE]

\bibitem[{{Hintz} {et~al.}(1997{\natexlab{b}}){Hintz}, {Hintz}, \&
  {Joner}}]{Hintz:1997b}
{Hintz}, E., {Hintz}, M.~L., \& {Joner}, M.~D. 1997{\natexlab{b}}, \pasp, 109,
  1073 [SXPHE]

\bibitem[{{Hintz} {et~al.}(2004){Hintz}, {Joner}, {Ivanushkina}, \&
  {Pilachowski}}]{Hintz:2004}
{Hintz}, E.~G., {Joner}, M.~D., {Ivanushkina}, M., \& {Pilachowski}, C.~A.
  2004, \pasp, 116, 543 [SXPHE]

\bibitem[{{Hintz} {et~al.}(1997{\natexlab{c}}){Hintz}, {Joner}, {McNamara},
  {Nelson}, {Moody}, \& {Kim}}]{Hintz:1997a}
{Hintz}, E.~G., {Joner}, M.~D., {McNamara}, D.~H., {et~al.} 1997{\natexlab{c}},
  \pasp, 109, 15 [SXPHE]

\bibitem[{{Hintz} {et~al.}(1998){Hintz}, {Joner}, \& {Hintz}}]{Hintz:1998}
{Hintz}, M.~L., {Joner}, M.~D., \& {Hintz}, E.~G. 1998, \aj, 116, 2993 [DSCUT]

\bibitem[{{Jeffery} {et~al.}(2005){Jeffery}, {Aerts}, {Dhillon}, {Marsh}, \&
  {G{\"a}nsicke}}]{Jeffery:2005}
{Jeffery}, C.~S., {Aerts}, C., {Dhillon}, V.~S., {Marsh}, T.~R., \&
  {G{\"a}nsicke}, B.~T. 2005, \mnras, 362, 66 [SDBV]

\bibitem[{{Kato} \& {Sadakane}(1999)}]{Kato:1999}
{Kato}, K.-I. \& {Sadakane}, K. 1999, \pasj, 51, 23 [CP]

\bibitem[{{Kaye} {et~al.}(1999){Kaye}, {Handler}, {Krisciunas}, {Poretti}, \&
  {Zerbi}}]{Kaye:1999}
{Kaye}, A.~B., {Handler}, G., {Krisciunas}, K., {Poretti}, E., \& {Zerbi},
  F.~M. 1999, \pasp, 111, 840 [GDOR]

\bibitem[{{Kepler} {et~al.}(2000){Kepler}, {Mukadam}, {Winget}, {Nather},
  {Metcalfe}, {Reed}, {Kawaler}, \& {Bradley}}]{Kepler:2000}
{Kepler}, S.~O., {Mukadam}, A., {Winget}, D.~E., {et~al.} 2000, \apjl, 534,
  L185 [DAV]

\bibitem[{{Kepler} {et~al.}(2003){Kepler}, {Nather}, {Winget}, {Nitta},
  {Kleinman}, {Metcalfe}, {Sekiguchi}, {Xiaojun}, {Sullivan}, {Sullivan},
  {Janulis}, {Meistas}, {Kalytis}, {Krzesinski}, {Ogoza}, {Zola}, {O'Donoghue},
  {Romero-Colmenero}, {Martinez}, {Dreizler}, {Deetjen}, {Nagel}, {Schuh},
  {Vauclair}, {Ning}, {Chevreton}, {Solheim}, {Gonzalez Perez}, {Johannessen},
  {Kanaan}, {Costa}, {Murillo Costa}, {Wood}, {Silvestri}, {Ahrens}, {Jones},
  {Collins}, {Boyer}, {Shaw}, {Mukadam}, {Klumpe}, {Larrison}, {Kawaler},
  {Riddle}, {Ulla}, \& {Bradley}}]{Kepler:2003}
{Kepler}, S.~O., {Nather}, R.~E., {Winget}, D.~E., {et~al.} 2003, \aap, 401,
  639 [DBV]

\bibitem[{{Kilkenny} {et~al.}(1985){Kilkenny}, {Whittet}, {Davies}, {Evans},
  {Bode}, {Robson}, \& {Banfield}}]{Kilkenny:1985}
{Kilkenny}, D., {Whittet}, D.~C.~B., {Davies}, J.~K., {et~al.} 1985, South
  African Astronomical Observatory Circular, 9, 55 [HAEBE, TTAU]

\bibitem[{{Kiss} {et~al.}(2002){Kiss}, {Derekas}, {Alfaro}, {B{\'{\i}}r{\'o}},
  {Cs{\'a}k}, {Garrido}, {Szatm{\'a}ry}, \& {Thomson}}]{Kiss:2002}
{Kiss}, L.~L., {Derekas}, A., {Alfaro}, E.~J., {et~al.} 2002, \aap, 394, 97
  [DSCUT]

\bibitem[{{Kleinman} {et~al.}(1998){Kleinman}, {Nather}, {Winget}, {Clemens},
  {Bradley}, {Kanaan}, {Provencal}, {Claver}, {Watson}, {Yanagida}, {Nitta},
  {Dixson}, {Wood}, {Grauer}, {Hine}, {Fontaine}, {Liebert}, {Sullivan},
  {Wickramasinghe}, {Achilleos}, {Marar}, {Seetha}, {Ashoka}, {Meistas},
  {Leibowitz}, {Moskalik}, {Krzesinski}, {Solheim}, {Bruvold}, {O'Donoghue},
  {Kurtz}, {Warner}, {Martinez}, {Vauclair}, {Dolez}, {Chevreton}, {Barstow},
  {Kepler}, {Giovannini}, {Augusteijn}, {Hansen}, \& {Kawaler}}]{Kleinman:1998}
{Kleinman}, S.~J., {Nather}, R.~E., {Winget}, D.~E., {et~al.} 1998, \apj, 495,
  424 [DAV]

\bibitem[{{Koester} {et~al.}(1983){Koester}, {Weidemann}, \&
  {Vauclair}}]{Koester:1983}
{Koester}, D., {Weidemann}, V., \& {Vauclair}, G. 1983, \aap, 123, L11 [DBV]

\bibitem[{{Kotak} {et~al.}(2004){Kotak}, {van Kerkwijk}, \&
  {Clemens}}]{Kotak:2004}
{Kotak}, R., {van Kerkwijk}, M.~H., \& {Clemens}, J.~C. 2004, \aap, 413, 301
  [MIRA]

\bibitem[{{Kov{\'a}cs}(2000)}]{Kovacs:2000}
{Kov{\'a}cs}, G. 2000, \aap, 360, L1 [DMCEP]

\bibitem[{{Kovtyukh} {et~al.}(2005){Kovtyukh}, {Andrievsky}, {Belik}, \&
  {Luck}}]{Andrievsky:2005a}
{Kovtyukh}, V.~V., {Andrievsky}, S.~M., {Belik}, S.~I., \& {Luck}, R.~E. 2005,
  \aj, 129, 433 [CLCEP]

\bibitem[{{Kroll}(1993)}]{Kroll:1993}
{Kroll}, R. 1993, in ASP Conf. Ser. 44: IAU Colloq. 138: Peculiar versus Normal
  Phenomena in A-type and Related Stars, ed. M.~M. {Dworetsky}, F.~{Castelli},
  \& R.~{Faraggiana}, 173 [CP]

\bibitem[{{Kudryavtsev} \& {Romanyuk}(2003)}]{Kudryavtsev:2003}
{Kudryavtsev}, D.~O. \& {Romanyuk}, I.~I. 2003, Astrophysics, 46, 234 [CP]

\bibitem[{{Kupka} \& {Piskunov}(1998)}]{Kupka:1998}
{Kupka}, F. \& {Piskunov}, N.~E. 1998, Contributions of the Astronomical
  Observatory Skalnate Pleso, 27, 228 [CP]

\bibitem[{{Kupka} {et~al.}(1994{\natexlab{a}}){Kupka}, {Ryabchikova},
  {Bolgova}, {Kuschnig}, {Weiss}, \& {Le Contel}}]{Kupka:1994a}
{Kupka}, F., {Ryabchikova}, T., {Bolgova}, G., {et~al.} 1994{\natexlab{a}}, in
  Chemically Peculiar and Magnetic Stars, ed. J.~{Zverko} \& J.~{Ziznovsky},
  130 [CP]

\bibitem[{{Kupka} {et~al.}(1994{\natexlab{b}}){Kupka}, {Ryabchikova},
  {Bolgova}, {Kuschnig}, {Weiss}, \& {Le Contel}}]{Kupka:1994b}
{Kupka}, F., {Ryabchikova}, T., {Bolgova}, G., {et~al.} 1994{\natexlab{b}}, in
  Chemically Peculiar and Magnetic Stars, ed. J.~{Zverko} \& J.~{Ziznovsky},
  130 [CP]

\bibitem[{{Kurtz}(1985)}]{Kurtz:1985}
{Kurtz}, D.~W. 1985, \mnras, 213, 773

\bibitem[{{Kurtz} \& {Martinez}(2000)}]{Kurtz:2000}
{Kurtz}, D.~W. \& {Martinez}, P. 2000, Baltic Astronomy, 9, 253 [ROAP]

\bibitem[{{Kurtz} {et~al.}(1997{\natexlab{a}}){Kurtz}, {Martinez}, {Tripe}, \&
  {Hanbury}}]{Kurtz:1997b}
{Kurtz}, D.~W., {Martinez}, P., {Tripe}, P., \& {Hanbury}, A.~G.
  1997{\natexlab{a}}, \mnras, 289, 645 [ROAP]

\bibitem[{{Kurtz} {et~al.}(1989){Kurtz}, {Matthews}, {Martinez}, {Seeman},
  {Cropper}, {Clemens}, {Kreidl}, {Sterken}, {Schneider}, {Weiss}, {Kawaler},
  \& {Kepler}}]{Kurtz:1989}
{Kurtz}, D.~W., {Matthews}, J.~M., {Martinez}, P., {et~al.} 1989, \mnras, 240,
  881 [ROAP]

\bibitem[{{Kurtz} {et~al.}(1997{\natexlab{b}}){Kurtz}, {van Wyk}, {Roberts},
  {Marang}, {Handler}, {Medupe}, \& {Kilkenny}}]{Kurtz:1997a}
{Kurtz}, D.~W., {van Wyk}, F., {Roberts}, G., {et~al.} 1997{\natexlab{b}},
  \mnras, 287, 69 [ROAP]

\bibitem[{{Lamers} {et~al.}(1998{\natexlab{a}}){Lamers}, {Bastiaanse}, {Aerts},
  \& {Spoon}}]{Lamers}
{Lamers}, H.~J.~G.~L.~M., {Bastiaanse}, M.~V., {Aerts}, C., \& {Spoon},
  H.~W.~W. 1998{\natexlab{a}}, \aap, 335, 605 [LBV]

\bibitem[{{Lamers} {et~al.}(1998{\natexlab{b}}){Lamers}, {Bastiaanse}, {Aerts},
  \& {Spoon}}]{Lamers:1998}
{Lamers}, H.~J.~G.~L.~M., {Bastiaanse}, M.~V., {Aerts}, C., \& {Spoon},
  H.~W.~W. 1998{\natexlab{b}}, \aap, 335, 605 [LBV]

\bibitem[{{Lamers} {et~al.}(1996){Lamers}, {Morris}, {Voors}, {van Gent},
  {Waters}, {de Graauw}, {Kudritzki}, {Najarro}, {Salama}, \&
  {Heras}}]{Lamers:1996}
{Lamers}, H.~J.~G.~L.~M., {Morris}, P.~W., {Voors}, R.~H.~M., {et~al.} 1996,
  \aap, 315, L225 [LBV]

\bibitem[{{Lefevre} {et~al.}(2005){Lefevre}, {Moffat}, \&
  {Marchenko}}]{Lefevre:2005}
{Lefevre}, L., {Moffat}, A.~F.~J., \& {Marchenko}, S.~V.~M. 2005, \jrasc, 99,
  130 [WR]

\bibitem[{{Lloyd Evans}(1983)}]{LloydEvans:1983}
{Lloyd Evans}, T. 1983, The Observatory, 103, 276 [PTCEP]

\bibitem[{{Lomb}(1976)}]{Lomb:1976}
{Lomb}, N.~R. 1976, \apss, 39, 447

\bibitem[{{Lopez-Garcia} \& {Adelman}(1994)}]{LopezGarcia:1994}
{Lopez-Garcia}, Z. \& {Adelman}, S.~J. 1994, \aaps, 107, 353 [CP]

\bibitem[{{L{\'o}pez-Garc{\'{\i}}a} \& {Adelman}(1999)}]{LopezGarcia:1999}
{L{\'o}pez-Garc{\'{\i}}a}, Z. \& {Adelman}, S.~J. 1999, \aaps, 137, 227 [CP]

\bibitem[{{L{\'o}pez-Garc{\'{\i}}a} {et~al.}(2001){L{\'o}pez-Garc{\'{\i}}a},
  {Adelman}, \& {Pintado}}]{LopezGarcia:2001}
{L{\'o}pez-Garc{\'{\i}}a}, Z., {Adelman}, S.~J., \& {Pintado}, O.~I. 2001,
  \aap, 367, 859 [CP]

\bibitem[{{Luck} \& {Andrievsky}(2004)}]{Andrievsky:2004}
{Luck}, R.~E. \& {Andrievsky}, S.~M. 2004, \aj, 128, 343 [CLCEP]

\bibitem[{{Maas} {et~al.}(2002){Maas}, {Van Winckel}, \&
  {Waelkens}}]{Maas:2002}
{Maas}, T., {Van Winckel}, H., \& {Waelkens}, C. 2002, \aap, 386, 504 [RVTAU]

\bibitem[{{Malanushenko} {et~al.}(1994){Malanushenko}, {Polosukchina}, \&
  {Weiss}}]{Malanushenko:1994}
{Malanushenko}, V.~P., {Polosukchina}, N.~S., \& {Weiss}, W.~W. 1994, \aaps,
  105, 125 [CP]

\bibitem[{{Manfroid} {et~al.}(1991){Manfroid}, {Sterken}, {Bruch}, {Burger},
  {de Groot}, {Duerbeck}, {Duemmler}, {Figer}, {Hageman}, {Hensberge},
  {Jorissen}, {Madejsky}, {Mandel}, {Ott}, {Reitermann}, {Schulte-Ladbeck},
  {Stahl}, {Steenman}, {Vander Linden}, \& {Zickgraf}}]{Manfroid:1991}
{Manfroid}, J., {Sterken}, C., {Bruch}, A., {et~al.} 1991, \aaps, 87, 481
  [PVSG, LBV]

\bibitem[{{Manfroid} {et~al.}(1994){Manfroid}, {Sterken}, {Cunow}, {de Groot},
  {Jorissen}, {Kneer}, {Krenzin}, {Kruijswijk}, {Naumann}, {Niehues},
  {Sch{\"o}neich}, {Sevenster}, {Vos}, \& {Vogt}}]{Manfroid:1994}
{Manfroid}, J., {Sterken}, C., {Cunow}, B., {et~al.} 1994, European Southern
  Observatory Scientific Report, 14, 1 [PVSG, LBV]

\bibitem[{{Mantegazza}(1988)}]{Mantegazza:1988}
{Mantegazza}, L. 1988, \aap, 196, 109 [SR]

\bibitem[{{Mantegazza} \& {Poretti}(1995)}]{Mantegazza:1995}
{Mantegazza}, L. \& {Poretti}, E. 1995, \aap, 294, 190 [ELL]

\bibitem[{{Mathias} {et~al.}(2001){Mathias}, {Aerts}, {Briquet}, {De Cat},
  {Cuypers}, {Van Winckel}, {Flanders.}, \& {Le Contel}}]{Mathias:2001}
{Mathias}, P., {Aerts}, C., {Briquet}, M., {et~al.} 2001, \aap, 379, 905 [SPB]

\bibitem[{{Mathys} {et~al.}(1996){Mathys}, {Kharchenko}, \&
  {Hubrig}}]{Mathys:1996}
{Mathys}, G., {Kharchenko}, N., \& {Hubrig}, S. 1996, \aap, 311, 901 [ROAP]

\bibitem[{{McLaughlin}(1943)}]{McLaughlin:1943}
{McLaughlin}, D.~B. 1943, Publications of Michigan Observatory, 8, 107 [SR]

\bibitem[{{McNamara}(1997{\natexlab{a}})}]{McNamara:1997a}
{McNamara}, D. 1997{\natexlab{a}}, \pasp, 109, 1221 [SXPHE, DSCUT, RR]

\bibitem[{{McNamara}(1997{\natexlab{b}})}]{McNamara:1997b}
{McNamara}, D. 1997{\natexlab{b}}, \pasp, 109, 1221 [SXPHE]

\bibitem[{{McNamara}(1995)}]{McNamara:1995}
{McNamara}, D.~H. 1995, \aj, 109, 1751 [SXPHE]

\bibitem[{{McNamara} {et~al.}(1996{\natexlab{a}}){McNamara}, {Powell}, \&
  {Joner}}]{McNamara:1996a}
{McNamara}, D.~H., {Powell}, J.~M., \& {Joner}, M.~D. 1996{\natexlab{a}},
  \pasp, 108, 1098 [SXPHE]

\bibitem[{{McNamara} {et~al.}(1996{\natexlab{b}}){McNamara}, {Powell}, \&
  {Joner}}]{McNamara:1996b}
{McNamara}, D.~H., {Powell}, J.~M., \& {Joner}, M.~D. 1996{\natexlab{b}},
  \pasp, 108, 1098 [SXPHE]

\bibitem[{{McSaveney} {et~al.}(2005){McSaveney}, {Pollard}, \&
  {Cottrell}}]{McSaveney:2005}
{McSaveney}, J.~A., {Pollard}, K.~R., \& {Cottrell}, P.~L. 2005, \mnras, 362,
  331 [PTCEP]

\bibitem[{{Metcalfe} {et~al.}(2004){Metcalfe}, {Montgomery}, \&
  {Kanaan}}]{Metcalfe:2004}
{Metcalfe}, T.~S., {Montgomery}, M.~H., \& {Kanaan}, A. 2004, \apjl, 605, L133
  [DAV]

\bibitem[{{Micha{\l}owska-Smak} \& {Smak}(1965)}]{MichalowskaSmak:1965}
{Micha{\l}owska-Smak}, A. \& {Smak}, J. 1965, Acta Astronomica, 15, 333 [PTCEP]

\bibitem[{{Morris}(1985)}]{Morris:1985}
{Morris}, S.~L. 1985, \apj, 295, 143 [ELL]

\bibitem[{{Mukadam} {et~al.}(2004){Mukadam}, {Mullally}, {Nather}, {Winget},
  {von Hippel}, {Kleinman}, {Nitta}, {Krzesi{\'n}ski}, {Kepler}, {Kanaan},
  {Koester}, {Sullivan}, {Homeier}, {Thompson}, {Reaves}, {Cotter},
  {Slaughter}, \& {Brinkmann}}]{Mukadam:2004}
{Mukadam}, A.~S., {Mullally}, F., {Nather}, R.~E., {et~al.} 2004, \apj, 607,
  982 [DAV]

\bibitem[{Neal(1996)}]{Neal}
Neal, R.~M. 1996, Bayesian Learning for Neural Networks (New York: Lecture
  Notes in StatisticsSpringer Verlag)

\bibitem[{{Neiner} {et~al.}(2005){Neiner}, {Floquet}, {Hubert}, {Fr{\'e}mat},
  {Hirata}, {Masuda}, {Gies}, {Buil}, \& {Martayan}}]{Neiner:2005}
{Neiner}, C., {Floquet}, M., {Hubert}, A.~M., {et~al.} 2005, \aap, 437, 257
  [BE]

\bibitem[{{Neiner} {et~al.}(2002){Neiner}, {Hubert}, {Floquet}, {Jankov},
  {Henrichs}, {Foing}, {Oliveira}, {Orlando}, {Abbott}, {Baldry}, {Bedding},
  {Cami}, {Cao}, {Catala}, {Cheng}, {Domiciano de Souza}, {Janot-Pacheco},
  {Hao}, {Kaper}, {Kaufer}, {Leister}, {Neff}, {O'Toole}, {Sch{\"a}fer},
  {Smartt}, {Stahl}, {Telting}, {Tubbesing}, \& {Zorec}}]{Neiner:2002}
{Neiner}, C., {Hubert}, A.-M., {Floquet}, M., {et~al.} 2002, \aap, 388, 899
  [BE]

\bibitem[{{Nemec} \& {Mateo}(1990)}]{Nemec:1990}
{Nemec}, J. \& {Mateo}, M. 1990, in ASP Conf. Ser. 11: Confrontation Between
  Stellar Pulsation and Evolution, ed. C.~{Cacciari} \& G.~{Clementini}, 64
  [SXPHE]

\bibitem[{{North} \& {Adelman}(1995)}]{North:1995}
{North}, P. \& {Adelman}, S.~J. 1995, \aaps, 111, 41 [CP]

\bibitem[{{North} \& {Paltani}(1994)}]{North:1994}
{North}, P. \& {Paltani}, S. 1994, \aap, 288, 155 [SPB]

\bibitem[{{Nugis} \& {Lamers}(2000)}]{Nugis:2000}
{Nugis}, T. \& {Lamers}, H.~J.~G.~L.~M. 2000, \aap, 360, 227 [WR]

\bibitem[{{O'Connell}(1961)}]{OConnell:1961}
{O'Connell}, D.~J.~K. 1961, Ricerche Astronomiche, 6, 353 [SR]

\bibitem[{{Oreiro} {et~al.}(2005){Oreiro}, {Ulla}, {P{\'e}rez Hern{\'a}ndez},
  {MacDonald}, {{\O}stensen}, \& {Rodr{\'{\i}}guez-L{\'o}pez}}]{Oreiro:2005}
{Oreiro}, R., {Ulla}, A., {P{\'e}rez Hern{\'a}ndez}, F., {et~al.} 2005, in ASP
  Conf. Ser. 334: 14th European Workshop on White Dwarfs, ed. D.~{Koester} \&
  S.~{Moehler}, 631 [SDBV]

\bibitem[{{Pamyatnykh} {et~al.}(1998){Pamyatnykh}, {Dziembowski}, {Handler}, \&
  {Pikall}}]{Pamyatnykh:1998}
{Pamyatnykh}, A.~A., {Dziembowski}, W.~A., {Handler}, G., \& {Pikall}, H. 1998,
  \aap, 333, 141 [DSCUT]

\bibitem[{{Paunzen}(1999)}]{Paunzen:1999}
{Paunzen}, E. 1999, \apss, 266, 379 [LBOO]

\bibitem[{{Pe{\~n}a} {et~al.}(1999){Pe{\~n}a}, {Gonz{\'a}lez}, \&
  {Peniche}}]{Pena:1999}
{Pe{\~n}a}, J.~H., {Gonz{\'a}lez}, D., \& {Peniche}, R. 1999, \aaps, 138, 11
  [DSCUT]

\bibitem[{{Pe{\~n}a} {et~al.}(1987){Pe{\~n}a}, {Peniche}, {Gonzalez}, \&
  {Hobart}}]{Pena:1987}
{Pe{\~n}a}, J.~H., {Peniche}, R., {Gonzalez}, S.~F., \& {Hobart}, M.~A. 1987,
  Revista Mexicana de Astronomia y Astrofisica, vol.~14, 14, 429 [SXPHE]

\bibitem[{Pearl(1988)}]{BN}
Pearl, J. 1988, Probabilistic reasoning in intelligent systems: networks of
  plausible inference (San Francisco, CA, USA: Morgan Kaufmann Publishers Inc.)

\bibitem[{{Percy} {et~al.}(1997){Percy}, {Bezuhly}, {Milanowski}, \&
  {Zsoldos}}]{Percy:1997}
{Percy}, J.~R., {Bezuhly}, M., {Milanowski}, M., \& {Zsoldos}, E. 1997, \pasp,
  109, 264 [RVTAU]

\bibitem[{{Percy} \& {Hale}(1998)}]{Percy:1998}
{Percy}, J.~R. \& {Hale}, J. 1998, \pasp, 110, 1428 [PTCEP]

\bibitem[{{Perryman} \& {ESA}(1997)}]{Perryman:1997}
{Perryman}, M.~A.~C. \& {ESA}. 1997, {The HIPPARCOS and TYCHO catalogues.
  Astrometric and photometric star catalogues derived from the ESA HIPPARCOS
  Space Astrometry Mission} (The Hipparcos and Tycho catalogues.~Astrometric
  and photometric star catalogues derived from the ESA Hipparcos Space
  Astrometry Mission, Publisher: Noordwijk, Netherlands: ESA Publications
  Division, 1997, Series: ESA SP Series vol no: 1200, ISBN: 9290923997 (set))

\bibitem[{{Petersen} \& {Andreasen}(1987)}]{Petersen:1987}
{Petersen}, J.~O. \& {Andreasen}, G.~K. 1987, \aap, 176, 183 [PTCEP]

\bibitem[{{Pollard} {et~al.}(1997){Pollard}, {Cottrell}, {Lawson}, {Albrow}, \&
  {Tobin}}]{Pollard:1997}
{Pollard}, K.~R., {Cottrell}, P.~L., {Lawson}, W.~A., {Albrow}, M.~D., \&
  {Tobin}, W. 1997, \mnras, 286, 1 [RVTAU]

\bibitem[{{Ponman}(1981)}]{Ponman:1981}
{Ponman}, T. 1981, \mnras, 196, 583

\bibitem[{{Preston} {et~al.}(1963){Preston}, {Krzeminski}, {Smak}, \&
  {Williams}}]{Preston:1963}
{Preston}, G.~W., {Krzeminski}, W., {Smak}, J., \& {Williams}, J.~A. 1963,
  \apj, 137, 401 [RVTAU]

\bibitem[{Quinlan(1993)}]{quinlan93}
Quinlan, J.~R. 1993, C4.5: programs for machine learning (San Francisco, CA,
  USA: Morgan Kaufmann Publishers Inc.)

\bibitem[{{Quirion} {et~al.}(2006){Quirion}, {Fontaine}, \&
  {Brassard}}]{Quirion:2006}
{Quirion}, P.~., {Fontaine}, G., \& {Brassard}, P. 2006, ArXiv Astrophysics
  e-prints [GWVIR]

\bibitem[{{Rao} \& {Reddy}(2005)}]{Rao:2005}
{Rao}, N.~K. \& {Reddy}, B.~E. 2005, \mnras, 357, 235 [RVTAU]

\bibitem[{{Raveendran}(1989)}]{Raveendran:1989}
{Raveendran}, A.~V. 1989, \mnras, 238, 945 [RVTAU]

\bibitem[{{Rivinius} {et~al.}(2003){Rivinius}, {Baade}, \& {{\v
  S}tefl}}]{Rivinius:2003}
{Rivinius}, T., {Baade}, D., \& {{\v S}tefl}, S. 2003, \aap, 411, 229 [BE]

\bibitem[{{Robberto} \& {Herbst}(1998)}]{Robberto:1998}
{Robberto}, M. \& {Herbst}, T.~M. 1998, \apj, 498, 400 [LBV]

\bibitem[{{Rodr{\'{\i}}guez} {et~al.}(2000{\natexlab{a}}){Rodr{\'{\i}}guez},
  {L{\'o}pez-Gonz{\'a}lez}, \& {L{\'o}pez de Coca}}]{Rodriguez:2000b}
{Rodr{\'{\i}}guez}, E., {L{\'o}pez-Gonz{\'a}lez}, M.~J., \& {L{\'o}pez de
  Coca}, P. 2000{\natexlab{a}}, in ASP Conf. Ser. 210: Delta Scuti and Related
  Stars, ed. M.~{Breger} \& M.~{Montgomery}, 499 [DSCUT]

\bibitem[{{Rodr{\'{\i}}guez} {et~al.}(2000{\natexlab{b}}){Rodr{\'{\i}}guez},
  {L{\'o}pez-Gonz{\'a}lez}, \& {L{\'o}pez de Coca}}]{Rodriguez:2000a}
{Rodr{\'{\i}}guez}, E., {L{\'o}pez-Gonz{\'a}lez}, M.~J., \& {L{\'o}pez de
  Coca}, P. 2000{\natexlab{b}}, \aaps, 144, 469 [DSCUT]

\bibitem[{{Rodr{\'{\i}}guez} {et~al.}(2002){Rodr{\'{\i}}guez},
  {L{\'o}pez-Gonz{\'a}lez}, \& {L{\'o}pez de Coca}}]{Rodriguez:2002}
{Rodr{\'{\i}}guez}, E., {L{\'o}pez-Gonz{\'a}lez}, M.~J., \& {L{\'o}pez de
  Coca}, P. 2002, in ESA SP-485: Stellar Structure and Habitable Planet
  Finding, ed. B.~{Battrick}, F.~{Favata}, I.~W. {Roxburgh}, \& D.~{Galadi},
  317 [SXPHE]

\bibitem[{{Rodriguez} {et~al.}(1990){Rodriguez}, {Rolland}, \& {Lopez de
  Coca}}]{Rodriguez:1990}
{Rodriguez}, E., {Rolland}, A., \& {Lopez de Coca}, P. 1990, \apss, 169, 113
  [SXPHE]

\bibitem[{{Rodriguez} {et~al.}(1993){Rodriguez}, {Rolland}, \& {Lopez de
  Coca}}]{Rodriguez:1993}
{Rodriguez}, E., {Rolland}, A., \& {Lopez de Coca}, P. 1993, \aaps, 100, 571
  [SXPHE]

\bibitem[{{Romanyuk}(1994)}]{Romanyuk:1994}
{Romanyuk}, I.~I. 1994, in Chemically Peculiar and Magnetic Stars, ed.
  J.~{Zverko} \& J.~{Ziznovsky}, 24 [CP]

\bibitem[{{Ryabchikova} {et~al.}(1998){Ryabchikova}, {Piskunov}, {Savanov}, \&
  {Kupka}}]{Ryabchikova:1998}
{Ryabchikova}, T., {Piskunov}, N., {Savanov}, I., \& {Kupka}, F. 1998,
  Contributions of the Astronomical Observatory Skalnate Pleso, 27, 359 [CP]

\bibitem[{{Saffe} {et~al.}(2005){Saffe}, {Levato}, \&
  {L{\'o}pez-Garc{\'{\i}}a}}]{Saffe:2005}
{Saffe}, C., {Levato}, H., \& {L{\'o}pez-Garc{\'{\i}}a}, Z. 2005, Revista
  Mexicana de Astronomia y Astrofisica, 41, 415 [CP]

\bibitem[{Sahami(1996)}]{sahami96learning}
Sahami, M. 1996, in Second International Conference on Knowledge Discovery in
  Databases

\bibitem[{{Sandage}(1993)}]{Sandage:1993}
{Sandage}, A. 1993, \aj, 106, 703 [RR]

\bibitem[{{Sarro} {et~al.}(2006){Sarro}, {S{\'a}nchez-Fern{\'a}ndez}, \&
  {Gim{\'e}nez}}]{Sarro:2006}
{Sarro}, L.~M., {S{\'a}nchez-Fern{\'a}ndez}, C., \& {Gim{\'e}nez}, {\'A}. 2006,
  \aap, 446, 395 [EA, EB, EW]

\bibitem[{{Savanov} {et~al.}(2001){Savanov}, {Kochukhov}, \&
  {Tsymbal}}]{Savanov:2001}
{Savanov}, I.~S., {Kochukhov}, O.~P., \& {Tsymbal}, V.~V. 2001, Astrophysics,
  44, 206 [CP]

\bibitem[{{Scargle}(1982)}]{Scargle:1982}
{Scargle}, J.~D. 1982, \apj, 263, 835

\bibitem[{{Schmidt} {et~al.}(2005){Schmidt}, {Johnston}, {Langan}, \&
  {Lee}}]{Schmidt:2005}
{Schmidt}, E.~G., {Johnston}, D., {Langan}, S., \& {Lee}, K.~M. 2005, \aj, 130,
  832 [PTCEP]

\bibitem[{{Schmidt} {et~al.}(2003){Schmidt}, {Langan}, {Lee}, {Johnston},
  {Newman}, \& {Snedden}}]{Schmidt:2003}
{Schmidt}, E.~G., {Langan}, S., {Lee}, K.~M., {et~al.} 2003, \aj, 126, 2495
  [PTCEP]

\bibitem[{{Sharma}(1996)}]{Sharma}
{Sharma}, S. 1996, Applied Multivariate Techniques (Wiley)

\bibitem[{{Shavrina} {et~al.}(2001){Shavrina}, {Polosukhina}, {Zverko},
  {Mashonkina}, {Khalack}, {{\v Z}i{\v z}{\v n}ovsk{\'y}}, {Hack}, {Tsymbal},
  {North}, \& {Vygonec}}]{Shavrina:2001}
{Shavrina}, A.~V., {Polosukhina}, N.~S., {Zverko}, J., {et~al.} 2001, \aap,
  372, 571 [CP]

\bibitem[{{Shenton} {et~al.}(1992){Shenton}, {Albinson}, {Barrett}, {Davies},
  {Evans}, {Goldsmith}, {Hutchinson}, {Maddison}, \& {Weight}}]{Shenton:1992}
{Shenton}, M., {Albinson}, J.~S., {Barrett}, P., {et~al.} 1992, \aap, 262, 138
  [RVTAU]

\bibitem[{{Shenton} {et~al.}(1994{\natexlab{a}}){Shenton}, {Evans}, {Albinson},
  {Barrett}, {Davies}, {Goldsmith}, {Hutchinson}, {Laney}, \&
  {Maddison}}]{Shenton:1994b}
{Shenton}, M., {Evans}, A., {Albinson}, J.~S., {et~al.} 1994{\natexlab{a}},
  \aap, 292, 102 [RVTAU]

\bibitem[{{Shenton} {et~al.}(1994{\natexlab{b}}){Shenton}, {Monier}, {Evans},
  {Carter}, {Lloyd Evans}, {Marang}, {Parker}, {Rawlings}, {Scott}, \& {van
  Wyk}}]{Shenton:1994a}
{Shenton}, M., {Monier}, R., {Evans}, A., {et~al.} 1994{\natexlab{b}}, \aap,
  287, 866 [RVTAU]

\bibitem[{{Simon} \& {Buscombe}(1973)}]{Simon:1973}
{Simon}, L.~W. \& {Buscombe}, W. 1973, in IAU Symp. 50: Spectral Classification
  and Multicolour Photometry, ed. C.~{Fehrenbach} \& B.~E. {Westerlund}, 33
  [SR]

\bibitem[{{Smith}(1974)}]{Smith:1974}
{Smith}, H.~A. 1974, Journal of the American Association of Variable Star
  Observers (JAAVSO), 3, 20 [SR]

\bibitem[{{Solano} \& {Fernley}(1997)}]{Solano:1997}
{Solano}, E. \& {Fernley}, J. 1997, \aaps, 122, 131 [DSCUT]

\bibitem[{{Soszynski} {et~al.}(2002){Soszynski}, {Udalski}, {Szymanski},
  {Kubiak}, {Pietrzynski}, {Wozniak}, {Zebrun}, {Szewczyk}, \&
  {Wyrzykowski}}]{Soszynski:2002}
{Soszynski}, I., {Udalski}, A., {Szymanski}, M., {et~al.} 2002, Acta
  Astronomica, 52, 369 [RRAB, RRC, RRD]

\bibitem[{{Stahl} {et~al.}(2003){Stahl}, {G{\"a}ng}, {Sterken}, {Kaufer},
  {Rivinius}, {Szeifert}, \& {Wolf}}]{Stahl:2003}
{Stahl}, O., {G{\"a}ng}, T., {Sterken}, C., {et~al.} 2003, \aap, 400, 279 [LBV]

\bibitem[{{Stankov} \& {Handler}(2005)}]{Stankov:2005}
{Stankov}, A. \& {Handler}, G. 2005, VizieR Online Data Catalog, 215, 80193
  [BCEP]

\bibitem[{{Sterken} {et~al.}(1998){Sterken}, {de Groot}, \& {van
  Genderen}}]{Sterken:1998}
{Sterken}, C., {de Groot}, M., \& {van Genderen}, A.~M. 1998, \aap, 333, 565
  [LBV]

\bibitem[{{Sterken} {et~al.}(1993){Sterken}, {Manfroid}, {Anton}, {Barzewski},
  {Bibo}, {Bruch}, {Burger}, {Duerbeck}, {Duemmler}, {Heck}, {Hensberge},
  {Hiesgen}, {Inklaar}, {Jorissen}, {Juettner}, {Kinkel}, {Liu}, {Mekkaden},
  {Ng}, {Niarchos}, {Puttmann}, {Szeifert}, {Spiller}, {van Dijk}, {Vogt}, \&
  {Wanders}}]{Sterken:1993}
{Sterken}, C., {Manfroid}, J., {Anton}, K., {et~al.} 1993, European Southern
  Observatory Scientific Report, 12, 1 [PVSG, LBV]

\bibitem[{{Sterken} {et~al.}(1995){Sterken}, {Manfroid}, {Beele}, {de Koff},
  {Eggenkamp}, {Goecking}, {Jorissen}, {Kaufer}, {Kuss}, {Schoenmakers},
  {Stil}, {van Loon}, {Vink}, {Vrielmann}, \& {Waelde}}]{Sterken:1995}
{Sterken}, C., {Manfroid}, J., {Beele}, D., {et~al.} 1995, \aaps, 113, 31
  [PVSG, LBV]

\bibitem[{{Stothers} \& {Chin}(1995)}]{Stothers:1995}
{Stothers}, R.~B. \& {Chin}, C.-W. 1995, \apjl, 451, L61 [LBV]

\bibitem[{{Teodorani} {et~al.}(1997){Teodorani}, {Errico}, {Vittone},
  {Giovannelli}, \& {Rossi}}]{Teodorani:1997}
{Teodorani}, M., {Errico}, L., {Vittone}, A.~A., {Giovannelli}, F., \& {Rossi},
  C. 1997, \aaps, 126, 91 [FUORI]

\bibitem[{{Terranegra} {et~al.}(1994){Terranegra}, {Chavarria-K.}, {Diaz}, \&
  {Gonzalez-Patino}}]{Terranegra:1994}
{Terranegra}, L., {Chavarria-K.}, C., {Diaz}, S., \& {Gonzalez-Patino}, D.
  1994, \aaps, 104, 557 [HAEBE, FUORI]

\bibitem[{{Udalski} {et~al.}(1999{\natexlab{a}}){Udalski}, {Soszynski},
  {Szymanski}, {Kubiak}, {Pietrzynski}, {Wozniak}, \& {Zebrun}}]{Udalski:1999d}
{Udalski}, A., {Soszynski}, I., {Szymanski}, M., {et~al.} 1999{\natexlab{a}},
  Acta Astronomica, 49, 1

\bibitem[{{Udalski} {et~al.}(1999{\natexlab{b}}){Udalski}, {Soszynski},
  {Szymanski}, {Kubiak}, {Pietrzynski}, {Wozniak}, \& {Zebrun}}]{Udalski:1999b}
{Udalski}, A., {Soszynski}, I., {Szymanski}, M., {et~al.} 1999{\natexlab{b}},
  Acta Astronomica, 49, 223 [CLCEP]

\bibitem[{{van der Hucht}(2001)}]{VanDerHucht:2001}
{van der Hucht}, K.~A. 2001, New Astronomy Review, 45, 135 [WR]

\bibitem[{{van Genderen}(1989)}]{VanGenderen:1989a}
{van Genderen}, A.~M. 1989, \aap, 208, 135 [PVSG, LBV]

\bibitem[{{van Genderen}(1998)}]{VanGenderen:1998b}
{van Genderen}, A.~M. 1998, Journal of Astronomical Data, 4, 10 [PVSG, LBV]

\bibitem[{{van Genderen} {et~al.}(1989{\natexlab{a}}){van Genderen},
  {Bovenschen}, {Engelsman}, {Goudfrooy}, {van Haarlem}, {Hartmann}, {Latour},
  {Ng}, {Prein}, {van Roermund}, {Roogering}, {Steeman}, \&
  {Tijdhof}}]{VanGenderen:1989c}
{van Genderen}, A.~M., {Bovenschen}, H., {Engelsman}, E.~C., {et~al.}
  1989{\natexlab{a}}, \aaps, 79, 263 [PVSG]

\bibitem[{{van Genderen} {et~al.}(1989{\natexlab{b}}){van Genderen},
  {Breukers}, {Houtekamer}, {van Roermund}, {Rottgering}, \&
  {Steeman}}]{VanGenderen:1989b}
{van Genderen}, A.~M., {Breukers}, R.~J.~L.~H., {Houtekamer}, P., {et~al.}
  1989{\natexlab{b}}, \aap, 213, 161 [PVSG]

\bibitem[{{van Genderen} \& {Sterken}(1996)}]{VanGenderen:1996}
{van Genderen}, A.~M. \& {Sterken}, C. 1996, \aap, 308, 763 [PVSG, LBV]

\bibitem[{{van Genderen} \& {Sterken}(1999)}]{VanGenderen:1999}
{van Genderen}, A.~M. \& {Sterken}, C. 1999, \aap, 349, 537 [PVSG]

\bibitem[{{van Genderen} {et~al.}(1998){van Genderen}, {Sterken}, \& {de
  Groot}}]{VanGenderen:1998a}
{van Genderen}, A.~M., {Sterken}, C., \& {de Groot}, M. 1998, \aap, 337, 393
  [PVSG]

\bibitem[{{van Leeuwen} {et~al.}(1998){van Leeuwen}, {van Genderen}, \&
  {Zegelaar}}]{VanLeeuwen:1998}
{van Leeuwen}, F., {van Genderen}, A.~M., \& {Zegelaar}, I. 1998, \aaps, 128,
  117 [PVSG]

\bibitem[{{Van Winckel} {et~al.}(1998){Van Winckel}, {Waelkens}, {Waters},
  {Molster}, {Udry}, \& {Bakker}}]{VanWinckel:1998}
{Van Winckel}, H., {Waelkens}, C., {Waters}, L.~B.~F.~M., {et~al.} 1998, \aap,
  336, L17 [RVTAU]

\bibitem[{Vapnik(1995)}]{Vapnik95}
Vapnik, V.~N. 1995, The nature of statistical learning theory (New York, NY,
  USA: Springer-Verlag New York, Inc.)

\bibitem[{{Vauclair} {et~al.}(1992){Vauclair}, {Belmonte}, {Pfeiffer},
  {Grauer}, {Jimenez}, {Chevreton}, {Dolez}, {Vidal}, \&
  {Herpe}}]{Vauclair:1992}
{Vauclair}, G., {Belmonte}, J.~A., {Pfeiffer}, B., {et~al.} 1992, \aap, 264,
  547 [DAV]

\bibitem[{{Vinko} {et~al.}(1998){Vinko}, {Remage Evans}, {Kiss}, \&
  {Szabados}}]{Vinko:1998}
{Vinko}, J., {Remage Evans}, N., {Kiss}, L.~L., \& {Szabados}, L. 1998, \mnras,
  296, 824 [PTCEP]

\bibitem[{{Vogt} {et~al.}(1990){Vogt}, {Barrera}, \& {Navarro}}]{Vogt:1990}
{Vogt}, N., {Barrera}, L.~H., \& {Navarro}, M. 1990, \apss, 173, 145 [BE]

\bibitem[{{Waelkens}(1991)}]{Waelkens:1991}
{Waelkens}, C. 1991, \aap, 246, 453 [SPB]

\bibitem[{{Waelkens}(1996)}]{Waelkens:1996}
{Waelkens}, C. 1996, \aap, 311, 873 [SPB]

\bibitem[{{Wahlgren}(1992)}]{Wahlgren:1992}
{Wahlgren}, G.~M. 1992, \aj, 104, 1174 [RVTAU]

\bibitem[{{Weintraub} {et~al.}(1989){Weintraub}, {Sandell}, \&
  {Duncan}}]{Weintraub:1989}
{Weintraub}, D.~A., {Sandell}, G., \& {Duncan}, W.~D. 1989, \apjl, 340, L69
  [TTAU, FUORI]

\bibitem[{{Weis}(2003)}]{Weis:2003}
{Weis}, K. 2003, \aap, 408, 205 [LBV]

\bibitem[{{Wesemael} {et~al.}(1986){Wesemael}, {Lamontagne}, \&
  {Fontaine}}]{Wesemael:1986}
{Wesemael}, F., {Lamontagne}, R., \& {Fontaine}, G. 1986, \aj, 91, 1376 [DAV]

\bibitem[{{Wozniak} {et~al.}(2002){Wozniak}, {Udalski}, {Szymanski}, {Kubiak},
  {Pietrzynski}, {Soszynski}, \& {Zebrun}}]{Wozniak:2002}
{Wozniak}, P.~R., {Udalski}, A., {Szymanski}, M., {et~al.} 2002, Acta
  Astronomica, 52, 129

\bibitem[{{Wyrzykowski} {et~al.}(2004){Wyrzykowski}, {Udalski}, {Kubiak},
  {Szymanski}, {Zebrun}, {Soszynski}, {Wozniak}, {Pietrzynski}, \&
  {Szewczyk}}]{Wyrzykowski:2004}
{Wyrzykowski}, L., {Udalski}, A., {Kubiak}, M., {et~al.} 2004, Acta
  Astronomica, 54, 1 [EA, EB, EW]

\bibitem[{{Yushchenko} {et~al.}(1998){Yushchenko}, {Gopka}, {Khokhlova},
  {Musaev}, \& {Bikmaev}}]{Yushchenko:1998}
{Yushchenko}, A.~V., {Gopka}, V.~F., {Khokhlova}, V.~L., {Musaev}, F.~A., \&
  {Bikmaev}, I.~F. 1998, Contributions of the Astronomical Observatory Skalnate
  Pleso, 27, 365 [CP]

\bibitem[{{Zhou} {et~al.}(1999{\natexlab{a}}){Zhou}, {Fu}, \&
  {Jiang}}]{Zhou:1999b}
{Zhou}, A.-Y., {Fu}, J.-N., \& {Jiang}, S.-Y. 1999{\natexlab{a}}, \apss, 268,
  397 [SXPHE]

\bibitem[{{Zhou} {et~al.}(1999{\natexlab{b}}){Zhou}, {Rodr{\'{\i}}guez},
  {Jiang}, {Rolland}, \& {Costa}}]{Zhou:1999a}
{Zhou}, A.-Y., {Rodr{\'{\i}}guez}, E., {Jiang}, S.-Y., {Rolland}, A., \&
  {Costa}, V. 1999{\natexlab{b}}, \mnras, 308, 631 [SXPHE]

\bibitem[{{Zhou} {et~al.}(1999{\natexlab{c}}){Zhou}, {Rodr{\'{\i}}guez},
  {Jiang}, {Rolland}, \& {Costa}}]{Zhou:1999}
{Zhou}, A.-Y., {Rodr{\'{\i}}guez}, E., {Jiang}, S.-Y., {Rolland}, A., \&
  {Costa}, V. 1999{\natexlab{c}}, \mnras, 308, 631 [SXPHE]

\bibitem[{{Ziznovsky} \& {Zverko}(1995)}]{Ziznovsky:1995}
{Ziznovsky}, J. \& {Zverko}, J. 1995, Contributions of the Astronomical
  Observatory Skalnate Pleso, 25, 39 [CP]

\bibitem[{{Zsoldos}(1996)}]{Zsoldos:1996}
{Zsoldos}, E. 1996, \aaps, 119, 431 [RVTAU]

\bibitem[{{Zverko} {et~al.}(1994){Zverko}, {Zboril}, \&
  {Ziznovsky}}]{Zverko:1994}
{Zverko}, J., {Zboril}, M., \& {Ziznovsky}, J. 1994, \aap, 283, 932 [CP]

\end{thebibliography}

\Online



\end{document}